\documentclass[12pt]{iopart}

\expandafter\let\csname equation*\endcsname=\relax 
\expandafter\let\csname endequation*\endcsname=\relax 
\usepackage{amssymb, amsmath}
\usepackage{graphicx}
\usepackage[dvipsnames]{xcolor}
\usepackage{cite}
\usepackage{hyperref}
\hypersetup{colorlinks=true, linkcolor=blue,  urlcolor=magenta,  citecolor=OliveGreen}
\usepackage{footnotebackref}
\usepackage{etoolbox}

\makeatletter
\long\def\@makefntext#1{\parindent 1em\noindent 
 \makebox[1em][l]{\footnotesize\rm$\m@th{\arabic{footnote}}$}%
 \footnotesize\rm #1}
\def\@makefnmark{\hbox{${\arabic{footnote}}\m@th$}}
\def\@thefnmark{\arabic{footnote}}
\makeatother
\makeatletter
\newcommand{\mainmatter}{%
  \setcounter{footnote}{0}%
  \patchcmd{\@makefntext}{\fnsymbol}{\arabic}{}{}%
  \patchcmd{\@thefnmark}{\fnsymbol}{\arabic}{}{}%
  \def\@makefnmark{\textsuperscript{\arabic{footnote}}}%
}
\makeatother

\begin{document}
 
\title{EBWeyl: a Code to Invariantly Characterize Numerical Spacetimes}

\author{Robyn L. Munoz${{^a}}$ and
Marco Bruni${{^{a,b}}}$}

\address{${{^a}}$Institute of Cosmology {\rm \&} Gravitation, University of Portsmouth, Dennis Sciama Building, Burnaby Road, Portsmouth, PO1 3FX, United Kingdom}
\address{${{^b}}$INFN Sezione di Trieste, Via Valerio 2, 34127 Trieste, Italy}

\ead{robyn.munoz@port.ac.uk, marco.bruni@port.ac.uk}

\vspace{10pt}
\begin{indented}
\item[]December 3, 2022
\end{indented}

\begin{abstract}
In order to invariantly characterise spacetimes resulting from cosmological simulations in numerical relativity, we present two different methodologies to compute the electric and magnetic parts of the Weyl tensor, $E_{\alpha\beta}$ and $B_{\alpha\beta}$, from which we construct scalar invariants and the Weyl scalars.
The first method is geometrical, computing these tensors in full from the metric, and the second uses the 3+1 slicing formulation. 
We developed a code for each method and tested them on five analytic metrics, for which we derived $E_{\alpha\beta}$ and $B_{\alpha\beta}$ and the various scalars constructed from them with computer algebra software. 
We find excellent agreement between the analytic and numerical results. 
The slicing code outperforms the geometrical code for computational convenience and accuracy; on this basis we make it publicly available in {\tt github} with the name EBWeyl \cite{R.L.Munoz_2022_ebweyl}. We emphasize that this post-processing code is applicable to any numerical spacetime in any gauge.
\end{abstract}


\maketitle

\mainmatter

\section{Introduction} \label{sec: Introduction}

Enabled and motivated by new computational resources and observational advances, numerical relativity simulations as an alternative to Newtonian simulations have been gaining interest in cosmology in recent years, some in full general relativity using a fluid description of matter \cite{M.Alcubierre_etal_2015, J.C.Aurrekoetxea_etal_2019, E.Bentivegna_M.Bruni_2016, J.Braden_etal_2017, J.Centrella_1980, K.Clough_E.A.Lim_2016, W.E.East_etal_2018, J.T.Giblin_etal_2016, H.Kurki-Suonio_etal_1987, X-X.Kou_2021, H.J.Macpherson_etal_2018, H.J.Macpherson_2022, J.Rekier_etal_2015, F.Staelens_etal_2019, J.M.Torres_etal_2014}, some using a particle description \cite{J.Adamek_etal_2016, W.E.East_etal_2019, C.Barrera-Hinojosa_B.Li_2020_Jan, C.Barrera-Hinojosa_B.Li_2020_Apr, C.Barrera-Hinojosa_etal_2021_Jan, F.Lepori_etal_2022};   approximation are used in certain cases, such as weak field or neglecting transverse-traceless tensor modes, on the assumption that they represent gravitational waves uninfluential for the dynamics.  
This has been pushing our understanding of the significance of relativistic effects on cosmological scales \cite{W.E.East_etal_2018, F.Lepori_etal_2022, C.Guandalin_etal_2020, L.Coates_etal_2020, H.J.Macpherson_A.Heinesen_2021, C.Barrera-Hinojosa_etal_2021_Jan}.

Numerical relativity simulations are based on the 3+1 formalism \cite{R.Arnowitt_etal_2008, M.Alcubierre_2008, M.Shibata_2015}, where time and space are separated. Hence, by construction, obtained results  in general depend on the gauge choice, which in the 3+1 context  corresponds to a choice of lapse and shift, i.e.\ a choice of mapping between a time-slicing and the next one, cf.\ \cite{J.T.Giblin_etal_2019, C.Tian_etal_2020} for a discussion in the context of cosmology. 
In practice, this gauge choice is equivalent to fix coordinates. 
Physical interpretations and simulation comparisons then need to be based on {\it invariants} \cite{W.E.East_etal_2018, W.E.East_etal_2019, J.Adamek_etal_2020}, i.e.\ {\it scalars independent from the coordinate choice}: these are quantities characterising the spacetime \cite{R.D'Inverno_R.Russell-Clark_1971, A.Karlhede_1980, J.Carminati_R.G.McLenaghan_1991, C.B.G.McIntosh_etal_1995, W.B.Bonnor_1995, E.Zakhary_C.B.G.McIntosh_1997, H.Stephani_etal_2003, M.Alcubierre_2008, L.Wylleman_etal_2019, D.Bini_etal_2021} and should be, at least in principle, {\it observable}, i.e.\ measurable quantities \cite{C.Rovelli_1991}.

Friedmann-Lemaitre-Robertson-Walker (FLRW) spacetimes are conformally flat, and therefore the electric and magnetic parts of the Weyl tensor, $E_{\alpha\beta}$ and $B_{\alpha\beta}$ respectively \cite{A.Matte_1953, P.Jordan_etal_1964, S.W.Hawking_1966}, vanish.
As we are going to see in Section~\ref{sec: cov and GI}, tensorial quantities vanishing in the background spacetime are gauge-invariant first-order perturbation variables \cite{J.M.Stewart_M.Walker_1974}. 
Therefore, if we consider linear perturbations of an FLRW spacetime, $E_{\alpha\beta}$ and $B_{\alpha\beta}$ are first-order gauge-invariant variables \cite{S.W.Hawking_1966, G.F.R.Ellis_M.Bruni_1989, M.Bruni_etal_1992}; they are related to the Bardeen potentials \cite{J.M.Bardeen_1980, M.Bruni_etal_1992}, and so are scalars constructed from them. 

$E_{\alpha\beta}$ and $B_{\alpha\beta}$ are of specific interest for their physical meaning: they describe the non-local tidal gravitational fields (overall represented by the Weyl curvature) and they are related to the shear and vorticity of matter \cite{R.Maartens_B.A.Bassett_1998, G.F.R.Ellis_2009, G.F.R.Ellis_etal_2012}.
They can be computed from simulations in numerical relativity, and they are a clear asset in describing the simulated scenario. 
$E_{\alpha\beta}$ and $B_{\alpha\beta}$ are defined with respect to a specific timelike unit vector, which is then (implicitly or explicitly) the time leg of an orthonormal {\it tetrad frame}. It follows from this that in full non-linearity $E_{\alpha\beta}$ and $B_{\alpha\beta}$ are {\it frame dependent}.
Nonetheless, as we are going to show in Section~\ref{sec: invariants}, they can be used to build a full set of invariant scalars that characterises the spacetime, some frame-dependent and some frame-invariant, and these scalar invariants can be used for the Petrov classification described in Section~\ref{sec: Petrov classification}.

Describing the spacetime of a numerical relativity simulation using $E_{\alpha\beta}$ and $B_{\alpha\beta}$ was first considered for colliding black holes \cite{R.Owen_etal_2011}. 
A cosmological application has been studied for universe models containing latices of masses where a significant magnetic part can arise \cite{M.Korzynski_etal_2015, T.Clifton_etal_2017}. 
Furthermore, an approximation where the magnetic part has no divergence has been found to be valid on large scales in simulations of a non-linear matter-dominated spacetime \cite{A.Heinesen_H.J.Macpherson_2022}. 
Indeed gravito-magnetic effects have been gaining interest for their possible implications in cosmology \cite{M.Bruni_etal_2014_Feb, I.Milillo_etal_2015, D.B.Thomas_etal_2015_16Jul, D.B.Thomas_etal_2015_30Jul, C.Barrera-Hinojosa_etal_2021_Jan, C.Barrera-Hinojosa_etal_2021_Dec}.

In this paper, we present two methods to compute $E_{\alpha\beta}$ and $B_{\alpha\beta}$, with the goal of computing them numerically: we call the first ``geometrical" as the computation only requires the spacetime metric, while  we call the second ``slicing" \cite{R.L.Munoz_2022_ebweyl}, as the required variables are those of the 3+1 decomposition of spacetime.
For each of these two methods a code was created and tested on five example spacetimes, four of which are know exact solutions of general relativity; these tests demonstrate the reliability of our codes. 
These spacetimes were specifically chosen because they provide examples from cosmology.
One of the examples we consider is inhomogeneous, it is a generalisation of the dust-only Szekeres models \cite{P.Szekeres_1975} that includes the cosmological constant $\Lambda$ \cite{J.D.Barrow_J.Stein-Schabes_1984}, which we call $\Lambda$-Szekeres. 
As the latter doesn't have a magnetic Weyl part, in order to test the codes on an inhomogeneous spacetime with a non zero $B_{\alpha\beta}$ we have also introduced a conveniently made-up metric.
Our tests and results show that the code based on the slicing method outperforms the other: on this basis we have made the slicing code, which we dub EBWeyl, publicly available at \cite{R.L.Munoz_2022_ebweyl}. 

The paper is structured as follows. Section~\ref{sec: Th framework} presents the theoretical framework; we start from basic definitions, in order to provide a succinct but comprehensive enough summary for cosmologists and numerical relativists.
In Section~\ref{sec: Geometrical method} we describe how the electric and magnetic parts of the Weyl tensor, $E_{\alpha\beta}$ and $B_{\alpha\beta}$, are derived from the Riemann tensor; this establishes the geometrical method used in our first code.
Then, in Section~\ref{sec: Slicing method}, we present an alternative computational method to obtain $E_{\alpha\beta}$ and $B_{\alpha\beta}$ directly based on the 3+1 slicing formulation; this establishes the slicing method used in our second code \cite{R.L.Munoz_2022_ebweyl}. 
Invariants needed for the Petrov classification that can be computed from $E_{\alpha\beta}$ and $B_{\alpha\beta}$ are described in Section~\ref{sec: invariants}, where we also introduce the Weyl scalars. 
In Section~\ref{sec: cov and GI} we elucidate the difference between general scalar invariants of a spacetime and gauge-invariant perturbations of a background spacetime. 
In Section~\ref{sec: Petrov classification} we then describe the Petrov classification and its physical interpretation in terms of $E_{\alpha\beta}$, $B_{\alpha\beta}$ and the Weyl scalars.

Our codes are tested on five spacetimes, these are each presented in Section~\ref{sec: Example spacetimes}. 
Our Python post-processing codes are described in Section~\ref{sec: code description and numerical implementation}.
The usefulness of these codes is demonstrated using the $\Lambda$-Szekeres metric \cite{J.D.Barrow_J.Stein-Schabes_1984} in Section~\ref{sec: Szekeres invariants}, where for this spacetime we compute $E_{\alpha\beta}$ and $B_{\alpha\beta}$, the 3-D and 4-D Ricci scalars, the invariants of Section~\ref{sec: invariants} and the Petrov type. 
Finally, the performance and computing errors are discussed in Section~\ref{sec: Code tests}. 
In Section~\ref{sec: Conclusion} we draw our conclusions. 
In two appendices we demonstrate finite difference limitations and list the analytical expressions we computed with Maple \cite{Maple} and used in this paper.

Throughout this work Greek indices indicate spacetime $\{0,\;...\;3\}$ and Latin indices space $\{1,\;2,\;3\}$. 
We use the conventions $G = c = 1$, $\kappa=8\pi G$ and the signature $\{$-,+,+,+$\}$. 
The notation $T_{(\alpha\beta)}$ and $T_{[\alpha\beta]}$ respectively represent the symmetric and antisymmetric parts of $T_{\alpha\beta}$ with respects to $\alpha$ and $\beta$. 

\section{Theoretical framework} \label{sec: Th framework}

In this section we present the theoretical methods that will be used to build the codes described in Section~\ref{sec: code description and numerical implementation}. These will be applied in Section~\ref{sec: Results} to the test-bed metrics reviewed in Section~\ref{sec: Example spacetimes}.

In relativity, special or general, the physical notions of the observers and that of  the associated reference frames play a crucial role, and as such they are presented from start in introductory textbooks, where in practice most often {\it frame of reference} is used interchangeably with the notion of  {\it coordinate system}. 
However, while for each coordinate system there is an associated tetrad of basis vectors, the reverse is not true\footnote{A vector basis that corresponds to coordinates, $\{\partial/\partial x^\mu\}$, is a coordinate basis or holonomic frame\cite{H.Stephani_etal_2003}}. In addition it is sometime useful, starting from a given tensor,  to define new tensors by projecting  on one or more vectors. 
The value of scalar quantities only depends on the point on the manifold (a spacetime event), and as such it is independent from the choice of coordinate system, as this is a map (charts for mathematicians) of points on the manifold onto $\mathbb{R}^4$ \cite{R.M.Wald_1984,H.Stephani_etal_2003}. 
On the other hand, some scalars are conveniently defined as components of a tensor on a tetrad basis. 
Therefore, at each point on the manifold  these scalars will be independent from the coordinates,  but they will differ if a different tetrad basis is chosen to define them. 

For the sake of clarity, in this paper we use the word \textit{frame} only in reference to a projection on a unit timelike vector, or on a  tetrad of basis vectors, never in reference to coordinate systems. 
Then, {\it scalar invariants} are simply scalar functions, as such independent from the coordinates labeling spacetime points. 
However, a scalar invariant can as well be {\it frame-invariant}, or can be {\it frame-dependent}, as  it can depend from the frame used to define it.
More in general, as we are going to see below for the electric and  magnetic parts of the Weyl tensor, a tensor can be {\it frame-dependent} if it is defined using a specific vector.
Effectively, when a quantity is frame-dependent, it is just a new quantity ``of the same type", just as the electric and magnetic fields are different for different observers, even if the electromagnetic tensor field is the same.

The notions of {\it frame-dependence} or {\it frame-invariance} are quite important here, hence we will explicitly use the notation ${\{n\}}$ when a given quantity is defined by projecting along a given timelike vector $n^{\mu}$; that will define the {\it quantity in the frame} $n^\mu$.
When a combination of frame-dependent quantities is frame-independent, this will be emphasised by dropping the index ${\{n\}}$. 
We use $u^\mu$ to indicate the unit timelike 4-velocity of a fluid, and the index ${\{u\}}$ for associated quantities; these are then defined in the rest frame of the fluid and associated observers. 
For instance, the energy-momentum tensor of a perfect fluid is $T^{\mu\nu}=(\rho^{\{u\}} +p^{\{u\}}) u^\mu u^\nu + p^{\{u\}} g^{\mu\nu}$, where the energy density $\rho^{\{u\}}=T_{\mu\nu} u^\mu u^\nu$  {\it in the matter rest-frame} is the eigenvalue of $T^{\mu\nu}$ and $u^\mu$ its eigenvector, while  in the 3+1 formalism one uses $\rho^{\{n\}}=T_{\mu\nu} n^\mu n^\nu$ in general. 
Both $\rho^{\{u\}}$ and $\rho^{\{n\}}$ are scalar quantities, both independent from the spacetime coordinates, but they are not frame-invariant. 
Similarly, if the electromagnetic field is $F^{\mu\nu}$, in the rest frame associated with $u^\mu$ the electric field is $E^{\{u\}}_\mu=F_{\mu\nu}u^\nu$, while in the frame $n^\mu$ is $E^{\{n\}}_\mu=F_{\mu\nu}n^\nu$.

\subsection{Geometrical method} \label{sec: Geometrical method}

Let's now introduce the method applied in our first code.
From the 4-D metric $g_{\alpha\beta}$ and it's derivatives, the 4-D Riemann tensor ${}^{(4)}R_{\alpha\beta\mu\nu}$ can be constructed. 
While the 4-D Ricci tensor ${}^{(4)}R_{\alpha\beta}$ and scalar ${}^{(4)}R$ are constructed from it's trace: ${}^{(4)}R = g^{\alpha\beta}{}^{(4)}R_{\alpha\beta} = g^{\alpha\beta}g^{\mu\nu}{}^{(4)}R_{\mu\alpha\nu\beta}$, the Weyl tensor is constructed from it's trace-less part, see e.g.\ \cite{M.Alcubierre_2008}\footnote{ In this paper we mostly follow Alcubierre's book notation \cite{M.Alcubierre_2008}.}:
\begin{equation}\label{eq:Weyl}
    C_{\alpha\beta\mu\nu} = {}^{(4)}R_{\alpha\beta\mu\nu} -\left(g_{\alpha[\mu}{}^{(4)}R_{\nu]\beta} - g_{\beta[\mu}{}^{(4)}R_{\nu]\alpha}\right) + \frac{1}{3}g_{\alpha[\mu}g_{\nu]\beta}{}^{(4)}R.
\end{equation}
By projecting with an arbitrary timelike unit vector, say $n^\mu$, the Weyl tensor can be decomposed into its electric and magnetic parts \cite{A.Matte_1953, P.Jordan_etal_1964, S.W.Hawking_1966}:
\begin{equation}\label{eq:EBGeometrical}
    E_{\alpha\mu}^{\{n\}} = n^{\beta} n^{\nu} C_{\alpha\beta\mu\nu},
    \;\;\;\;\;\;\;\;\;
    B_{\alpha\mu}^{\{n\}} = n^{\beta} n^{\nu} C_{\alpha\beta\mu\nu}^*,
\end{equation}
where $C_{\alpha\beta\mu\nu}^*=\frac{1}{2} C_{\alpha\beta\lambda\sigma} \epsilon^{\lambda\sigma}_{\;\;\;\;\mu\nu}$ is the dual of the Weyl tensor and $\epsilon_{\alpha\beta\mu\nu}$ is the Levi-Civita completely antisymmetric tensor\footnote{ It is connected to the Levi-Civita symbol $\varepsilon_{\alpha\beta\mu\nu}$ as $\epsilon_{\alpha\beta\mu\nu} = \sqrt{-g} \varepsilon_{\alpha\beta\mu\nu} = \sqrt{|det(g_{\alpha\beta})|} \varepsilon_{\alpha\beta\mu\nu}$.}. 
Note that these tensors are \textit{frame dependent} in the sense explained above.

The Weyl tensor retains all the symmetries of the Riemann tensor and it is trace-less. $E_{\alpha\beta}^{\{n\}}$ and $B_{\alpha\beta}^{\{n\}}$ are then symmetric, trace-less and covariantly purely spatial, i.e.\ they live on a 3-D space orthonormal to the chosen timelike vector\footnote{ This space is only local if $n^\mu$ is not hypersurface orthogonal, i.e.\ when $\nabla_{[\alpha}n_{\beta]}\not = 0$, as it is the case when the chosen timelike vector is the 4-velocity of a fluid with vorticity, see the Appendix in \cite{G.F.R.Ellis_etal_1989}.}:
\begin{equation}
    n^{\alpha}E_{\alpha\mu}^{\{n\}} = 0,
    \;\;\;\;\;\;\;\;\;
    n^{\alpha}B_{\alpha\mu}^{\{n\}}=0.
\end{equation}
The trace-less characteristic appears by construction, inherited by properties of the Weyl tensor, in particular for $B_{\alpha\mu}^{\{n\}}$ the trace vanishes due to the first Bianchi identity. In the synchronous gauge (where $g_{0i} = \{-1,0,0,0\}$), and with $n^{\mu}=\{1,0,0,0\}$ the specific expression is:
\begin{equation}\label{eq:BianchiIdentity}
{}^{(4)}R_{1023} - {}^{(4)}R_{2013} + {}^{(4)}R_{3012} = 0;
\end{equation}
we will use this explicitly in Section~\ref{sec: Cancellation error}.

\subsection{3+1 slicing} \label{sec: Slicing method}

We now consider the method applied in our second code: this consists of calculating $E_{\alpha\beta}^{\{n\}}$ and $B_{\alpha\beta}^{\{n\}}$ by using the 3+1 formalism \cite{L.Gunnarsen_etal_1995, R.Arnowitt_etal_2008, R.M.Wald_1984, M.Alcubierre_2008, M.Shibata_2015, Y.Choquet-Bruhat_2015}. 
This foliates the spacetime into spatial hypersurfaces with a spatial metric $\gamma_{\alpha\beta}$ and a normal unit timelike vector\footnote{ As such, $n^\alpha$ is hypersurface orthogonal and by construction $\nabla_{[\alpha}n_{\beta]}=0$.} $n^{\alpha}$: $n^\alpha n_\alpha = -1$ and $\gamma_{\alpha\beta}n^\alpha = 0$. 
The time coordinate is chosen such that it is constant on each of the spatial slices, covered by the space coordinates, thereby defining {\it coordinates adapted to the foliation}.
Each time slice is mapped to the next by the lapse function $\alpha$ and the shift vector $\beta^\mu$.
In this foliation-adapted coordinates, $n_\mu=\{-\alpha, 0,0,0\}$ and $n^\mu=\{1/\alpha, -\beta^{i}/\alpha\}$, $\beta^\mu=\{0, \beta^i\}$ and $\beta_\mu=\{\beta_k \beta^k, \beta_j\}$, with $\beta_i=\gamma_{ij}\beta^j$. 

The 4-D metric $g_{\mu\nu}$ and the 3-D metric $\gamma_{\mu\nu}$ are related through $\alpha$ and $\beta^\mu$: 
\begin{equation}
\begin{aligned}
    g^{\mu\nu} & = \gamma^{\mu\nu} - n^{\mu} n^{\nu} = 
    \begin{pmatrix}
    -1/\alpha^2 & \beta^j/\alpha^2 \\
    \beta^i/\alpha^2 & \gamma^{ij}-(\beta^i \beta^j/\alpha^2) \\
    \end{pmatrix},\\
    g_{\mu\nu} & = \gamma_{\mu\nu} - n_{\mu} n_{\nu} =
    \begin{pmatrix}
    -\alpha^2 + \beta_k \beta^k & \beta_j \\
    \beta_i & \gamma_{ij} \\
    \end{pmatrix}.
\end{aligned}
\end{equation}
Inverting these relations show that:
\begin{equation}
    \gamma^{\alpha\beta} = \begin{pmatrix}
    0 & 0 \\
    0 & \gamma^{ij} \\
    \end{pmatrix},
    \;\;\;\;\;\;\;
    \gamma_{\alpha\beta} = \begin{pmatrix}
    \beta_k\beta^k & \beta_i \\
    \beta_j & \gamma_{ij} \\
    \end{pmatrix}.
\end{equation}

The evolution of $\gamma_{\alpha\beta}$ is given by the extrinsic curvature: $K_{\alpha\beta}=-\frac{1}{2}\mathcal{L}_{\mathbf{n}}\gamma_{\alpha\beta}$, where $\mathcal{L}_{\mathbf{n}}$ represents the Lie derivative along $n^\mu$ \cite{R.M.Wald_1984, Y.Choquet-Bruhat_2015, M.Alcubierre_2008}. In practice, defining the projection tensor $P^{\alpha}_{\beta}=\delta^{\alpha}_{\beta}+n^{\alpha}n_{\beta}$ (with $P_{\alpha\beta}=\gamma_{\alpha\beta}$) we can write:
\begin{equation}
    K_{\alpha\beta} = -P^{\mu}_{\alpha}\nabla_{\mu}n_{\beta} = - \nabla_\alpha n_\beta - n_{\alpha}n^{\mu}\nabla_{\mu}n_{\beta}.
\end{equation}
This tensor is symmetric and covariantly purely spatial, i.e.\ it is orthogonal to $n^\alpha$, $n^{\alpha}K_{\alpha\beta} = 0$. However, while $K^{00}=K^{0i}=0$ in the coordinate system adapted to the foliation, $K_{00}$ and $K_{0i}$ are different from zero in general when the shift is not zero; we can write:
\begin{equation}\label{eq:giveKtimecomponents}
    K^{\alpha\beta} = \begin{pmatrix}
    0 & 0 \\
    0 & K^{ij} \\
    \end{pmatrix}, 
    \;\;\;\;\;\;\;
    K_{\alpha\beta} = \begin{pmatrix}
    \beta_{i}\beta_{j}K^{ij} & \beta_{i}K^{i}_{k} \\
    \beta_{j}K_{l}^{j} & K_{kl} \\
    \end{pmatrix},
\end{equation}
where $K_{i}^{j} = \gamma_{ik}K^{kj}$ and $K_{ij} = \gamma_{ik}\gamma_{jl}K^{kl}$.

The intrinsic 3-D Riemann curvature ${}^{(3)}R_{\alpha\beta\mu\nu}$ of the slices is algebraically related to the 4-D Riemann tensor ${}^{(4)}R_{\lambda\sigma\omega\eta}$ and the extrinsic curvature $K_{\alpha\beta}$ by 
the Gauss equation \cite{M.Shibata_2015}
\begin{equation}\label{eq:Gauss}
    {}^{(3)}R_{\alpha\beta\mu\nu} = \gamma^{\lambda}_{\alpha}\gamma^{\sigma}_{\beta}\gamma^{\omega}_{\mu}\gamma^{\eta}_{\nu} {}^{(4)}R_{\lambda\sigma\omega\eta} + K_{\alpha\nu}K_{\beta\mu} - K_{\alpha\mu}K_{\beta\nu},
\end{equation}
while the Codazzi equation \cite{M.Shibata_2015}
\begin{equation}\label{eq:Codazzi}
    D_{\alpha}K_{\beta\mu}-D_{\beta}K_{\alpha\mu} = -\gamma_{\alpha}^{\nu}\gamma_{\beta}^{\lambda}\gamma_{\mu}^{\sigma}{}^{(4)}R_{\nu\lambda\sigma\eta}n^{\eta},
\end{equation}
relates the 4-D curvature to the spatial covariant derivative $D_{\alpha}$ (associated with $\gamma_{\alpha\beta}$) of the extrinsic curvature. 
To express $E_{ij}^{\{n\}}$ with this formulation one starts with the Gauss equation \eqref{eq:Gauss}, this is then contracted with $\gamma^{\beta\nu}$ and rearranged to find $n^{\beta} n^{\nu} {}^{(4)}R_{\alpha\beta\mu\nu}$. 
The resulting expression is introduced into Eq.~(\ref{eq:EBGeometrical}), then the remaining 4-D Ricci terms are substituted using Einstein's equation, 
\begin{equation}
    {}^{(4)}R_{\alpha\beta}-\Lambda g_{\alpha\beta} = \kappa \left(T_{\alpha\beta}-\frac{1}{2}Tg_{\alpha\beta}\right)
\end{equation}
and its contraction, such that:
\begin{equation}\label{eq:Ebefore}
E^{ij\{n\}} = {}^{(3)}R^{ij} + K^{ij}K - K^{ik}K_{k}^{j} - \frac{2}{3}\gamma^{ij} \left( \Lambda+\kappa\rho^{\{n\}} \right) - \frac{\kappa}{2} \left( S^{ij\{n\}} - \frac{1}{3}\gamma^{ij} S^{\{n\}} \right),
\end{equation}
where $K=K^{\alpha}_{\alpha}=\gamma_{ij}K^{ij}$ is the trace of the extrinsic curvature. Here we have explicitly introduced the cosmological constant $\Lambda$ and $\rho^{\{n\}}=T_{\alpha\beta}n^\alpha n^\beta$ is the energy density in the frame associated with $n^\alpha$. The spatial stress tensor $S_{\alpha\beta}^{\{n\}}$ in the frame $n^\mu$ is obtained from the stress-energy tensor $T_{\alpha\beta}$ by $S_{\alpha\beta}^{\{n\}}=\gamma^{\mu}_{\alpha}\gamma^{\nu}_{\beta}T_{\mu\nu}$,  and it's trace is $S^{\{n\}} = S^{\alpha \{n\}}_{\alpha} = \gamma^{\alpha\beta}T_{\alpha\beta} = T + \rho^{\{n\}}$. 

Note that in the coordinates adapted to the foliation, $E^{\alpha\beta\{n\}}$ is purely spatial. Indeed from Eq.~(\ref{eq:EBGeometrical}) $E^{\alpha\mu\{n\}}=\alpha^2 C^{\alpha 0 \mu 0}$, so that the antisymmetric nature of the Weyl tensor implies that $E^{\alpha\mu\{n\}}$ can only have spatial components. One can then define $E_{i}^{j\{n\}} = \gamma_{ik}E^{kj\{n\}}$ and $E_{ij}^{\{n\}} = \gamma_{ik}\gamma_{jl}E^{kl\{n\}}$. In lowering the indices of $E^{\alpha\beta\{n\}}$ with $\gamma_{\alpha\beta}$, however, we see that the temporal components of $E_{\alpha\beta}^{\{n\}}$ do not vanish when the shift is non zero. $E^{\alpha\beta\{n\}}$ and $E_{\alpha\beta}^{\{n\}}$ can be written in terms of the shift and the space components, as in Eq.~(\ref{eq:giveKtimecomponents}).

Starting from Eq.~(\ref{eq:Ebefore}), it is the Hamiltonian constraint
\begin{equation}
    {}^{(3)}R + K^2 - K^{kl}K_{kl} - 2\Lambda - 2\kappa\rho^{\{n\}} = 0
\end{equation}
that ensures that $E^{ij\{n\}}$ remains trace-less: $E_{\alpha}^{\alpha \{n\}} = g_{\alpha\beta}E^{\alpha\beta\{n\}} = \gamma_{ij}E^{ij\{n\}} = 0$.
In numerical relativity simulations, this constraint is used as a validity check, therefore although small, it tends to be non-zero. 
This carries into $E^{ij\{n\}}$ when computed with Eq.~(\ref{eq:Ebefore}), where the non-zero trace would then correspond to the violation of the Hamiltonian constraint.
Then, in order to avoid introducing errors in the calculation of $E_{ij}^{\{n\}}$, in particular a non-zero trace, we substitute in Eq.~(\ref{eq:Ebefore}) the term $\Lambda+\kappa\rho^{\{n\}}$ from the Hamiltonian constraint, obtaining:
\begin{equation}\label{eq:ESlicing}
    E_{ij}^{\{n\}} = {}^{(3)}R_{ij} + K_{ij}K - K_{i}^{k}K_{kj} - \frac{1}{3}\gamma_{ij} \left( {}^{(3)}R + K^2 - K^{kl}K_{kl} \right) - \frac{\kappa}{2} \left( S_{ij}^{\{n\}} - \frac{1}{3}\gamma_{ij} S^{\{n\}} \right);
\end{equation}
by definition, this expression remains traceless up to numerical error.

Similarly, $B_{ij}^{\{n\}}$ can be expressed using the Codazzi equation (\ref{eq:Codazzi}). This is contracted with $\gamma^{\alpha\mu}$ to provide $n^\beta {}^{(4)}R_{\nu\beta}$, and it is rearranged to have $n^\beta {}^{(4)}R_{\alpha\beta\mu\nu}$. These two terms can be introduced into the expression for $B_{ij}^{\{n\}}$, Eq.~(\ref{eq:EBGeometrical}), so that:
\begin{equation}\label{eq:BSlicing}
    B_{ij}^{\{n\}} = \epsilon^{kl}_{\;\;\;\;j}\left( D_{k}K_{li} + \frac{1}{2}\gamma_{ik}\left(D_{l}K - D_{m}K_{l}^{m}\right) \right),
\end{equation}
where $\epsilon_{\alpha\beta\mu} = n^{\nu}\epsilon_{\nu\alpha\beta\mu}$ and $n^{\alpha}\epsilon_{\alpha\beta\mu} = 0$. 
At this point the momentum constraint 
\begin{equation}\label{eq: momentum constraint}
    D_{m}K^{m}_{l}-D_{l}K=\kappa J_{l}^{\{n\}},
\end{equation}
is typically inserted to simplify the second term, where $J_{i}^{\{n\}} = -P_{i}^{\mu}n^\nu T_{\mu\nu}$ is the momentum density or energy flux in the frame $n^\mu$. 
Again, to avoid introducing errors into the numerical computation, we abstain from taking this last step. Here $B_{ij}^{\{n\}}$ can be seen to be trace-less because of the anti-symmetry of the Levi-Civita tensor. 
Finally, $B_{\alpha\beta}^{\{n\}}$ can be expressed in terms of the shift and its space components in the same way that $K_{\alpha\beta}$ and $E_{\alpha\beta}^{\{n\}}$ are, as in Eq.~(\ref{eq:giveKtimecomponents}).

\subsection{Spacetime Invariants} \label{sec: invariants}
Spacetime invariants have been traditionally considered to address two main and related problems: {\it i)} to establish if two metrics, presented in seemingly different forms, e.g.\ in different coordinates, actually represent the same spacetime; this {\it equivalence problem} became an important one at the time when there was a proliferation of new exact solutions, and the development of the first computer algebra software was under way; {\it ii)} to classify exact solutions into Petrov types, which we describe in Section~\ref{sec: Petrov classification}. 
The equivalence problem was originally formulated by Cartan and Brans, then reconsidered and addressed, and related to the Petrov classification, by D'Inverno and Russel-Clark \cite{R.D'Inverno_R.Russell-Clark_1971}, see Karlhede for an early review \cite{A.Karlhede_1980}. 
More general sets of invariants were then considered in \cite{J.Carminati_R.G.McLenaghan_1991} and  \cite{E.Zakhary_C.B.G.McIntosh_1997}. 
Recently, the specific equivalence problem for cosmological models has been addressed in \cite{L.Wylleman_etal_2019} ; in \cite{D.Bini_etal_2021} a more refined classification for Petrov type I spacetimes has been proposed. 
For a classical and rather detailed account of invariants and the characterization of spacetimes we refer the reader to \cite{H.Stephani_etal_2003}.

Here our goal is that of defining spacetime invariants that can be used in numerical relativity, in particularly in its application to cosmology, in order to address two issues: {\it i)} defining quantities that are independent from the specific 3+1 gauge used in a given simulation, as such at least in principle related to observables; {\it ii)} compare results obtained in different simulations, often computed in different gauges, in order to establish if (within numerical errors) these results are consistent \cite{J.Adamek_etal_2020}. 

In the following we are going to construct all the needed scalar invariants for spacetime comparison, as well as for the Petrov classification in Section~\ref{sec: Petrov classification}, using $E_{\alpha\beta}^{ \{n\}}$ and $B_{\alpha\beta}^{ \{n\}}$; these, given their definitions in Eq.~(\ref{eq:EBGeometrical}), are frame-dependent, and so are $E^{2\{n\}}\equiv E^{\alpha\beta \{n\}}E_{\alpha\beta}^{ \{n\}}$ and $B^{2\{n\}}\equiv B^{\alpha\beta \{n\}}B_{\alpha\beta}^{ \{n\}}$.
The frame dependence of $E_{\alpha\beta}^{ \{n\}}$ and $B_{\alpha\beta}^{ \{n\}}$ is however strictly nonlinear and, as we are going to discuss in Section~\ref{sec: cov and GI}, it disappears at first order for perturbations of an FLRW spacetime.

$E_{\alpha\beta}^{ \{n\}}$ and $B_{\alpha\beta}^{ \{n\}}$ represent the non-local gravitational tidal effects: if we think of the Riemann curvature tensor as made up the Ricci and Weyl parts, as in Eq.~(\ref{eq:Weyl}), then the Ricci part is directly determined locally (algebraically) by the matter distribution through Einstein equations, while the Weyl part can only be determined once Einstein equations are solved for the metric. 
Also, focusing directly on $E_{\alpha\beta}$ and $B_{\alpha\beta}$, these are determined by the Bianchi identities re-written in their Maxwell-like form as differential equations for  $E_{\alpha\beta}$ and $B_{\alpha\beta}$, sourced by the matter field that appear once Einstein equations are used to substitute for the Ricci tensor, see \cite{S.W.Hawking_1966,G.F.R.Ellis_2009,J.Wainwright_G.F.R.Ellis_1997,G.F.R.Ellis_etal_2012}.

In cosmology it is most often useful to exploit the 4-velocity $u^\mu$ of the matter, often a fluid, which can always be uniquely defined, even for an imperfect fluid where one can opt for either the {\it energy frame} where there is no energy flux, $J_{l}^{\{u\}}=0$ in Eq.~(\ref{eq: momentum constraint}), or the {\it particle frame} where there is no particles flux \cite{M.Bruni_etal_1992}; for a perfect fluid, the energy and particle frames coincide. 
Physically, quantities that results from projecting tensors in the frame defined by $u^\mu$ and the related projector tensor $h^{\mu\nu}=g^{\mu\nu}+u^\mu u^\nu$ are unique, as they are rest-frame quantities, e.g.\ the energy density $\rho^{\{u\}}$. 
The same uniqueness applies in the case where there are different matter fields \cite{P.K.S.Dunsby_etal_1992}, each with its own 4-velocity, as one can always define an average $u^\mu$, say an average energy frame, or project tensorial quantities with respect to a specific $u^\mu$, for instance that of pressureless matter, i.e.\ dust.

On the other hand, in numerical relativity we use a 3+1 decomposition, which refers to $n^\mu$, the normal to the slicing. While one can use a slicing such that $n^\mu= u^\mu$, in general the two do not coincide, as the slicing/gauge is chosen in order to optimise numerical computations, or one wishes to consider a fluid with vorticity $\omega_{\alpha\beta}$, so that $u^{\mu}$ can't be chosen as the normal to the slices, as in this case $u^{\mu}$ will not be hypersurface-orthogonal. 
For these reasons, we are now going to  construct $E^{2\{u\}}$ and $B^{2\{u\}}$ by projecting along the fluid flow $u^{\mu}$. 
Changing the projection vector in the geometrical method is straightforward but the slicing method is built on $n^\mu$ and the resulting 3-metric and extrinsic curvature. Hence our first step is the construction of the Weyl tensor from $E_{\alpha\beta}^{\{n\}}$ and $B_{\alpha\beta}^{\{n\}}$. Assuming that these have been computed from Eq.~(\ref{eq:ESlicing}) and Eq.~(\ref{eq:BSlicing}), we can construct $C_{\alpha\beta\mu\nu}$ \cite{M.Alcubierre_2008}:
\begin{equation}\label{eq:Weyl_from_EandB}
    C_{\alpha\beta\mu\nu} = 2\left(l_{\alpha[\mu}E^{\{n\}}_{\nu]\beta} - l_{\beta[\mu}E^{\{n\}}_{\nu]\alpha}  - n_{[\mu}B^{\{n\}}_{\nu]\lambda}\epsilon^{\lambda}_{\;\;\alpha\beta} - n_{[\alpha}B^{\{n\}}_{\beta]\lambda}\epsilon^{\lambda}_{\;\;\mu\nu}\right),
\end{equation}
with $l_{\mu\nu}=g_{\mu\nu}+2n_{\mu}n_{\nu}$. The Weyl tensor is frame-independent, but we explicitly write the index $\{n\}$, as we use $E_{\alpha\beta}^{\{n\}}$ and $B_{\alpha\beta}^{\{n\}}$ to then obtain $E_{\alpha\beta}^{\{u\}}$ and $B_{\alpha\beta}^{\{u\}}$.
Then, projecting along the fluid flow we get:
\begin{equation}
    E^{\{u\}}_{\alpha\mu} = u^{\beta} u^{\nu} C_{\alpha\beta\mu\nu},
    \;\;\;\;\;\;\;\;\;
    B^{\{u\}}_{\alpha\mu} = u^{\beta} u^{\nu} C_{\alpha\beta\mu\nu}^*,
\end{equation}
and from these we obtain $E^{2\{u\}}$ and $B^{2\{u\}}$ in the fluid frame, c.f.\ \cite{A.R.King_G.F.R.Ellis_1973, M.Bruni_etal_1992, D.Bini_etal_1995}.

While $E^{2\{u\}}$ and $B^{2\{u\}}$ are frame-dependent, $E_{\alpha\beta}^{\{n\}}$ and $B_{\alpha\beta}^{\{n\}}$ in any frame can be used to construct the frame-independent invariant scalars \cite{A.Matte_1953, W.B.Bonnor_1995},
\begin{equation}
    L_B = E^2-B^2,
    \;\;\;\;\;\;\;\;\;
    M = E^{\alpha\beta}B_{\alpha\beta},
\end{equation}
in complete analogy with electromagnetism.
In the case of a purely gravitational waves spacetime, i.e.\ Petrov type N, $L_B=M=0$; these two conditions are also valid for Petrov type III \cite{W.B.Bonnor_1995}. 

Two fundamental  scalar  invariants in the classification of spacetimes are $I\equiv \frac{1}{2}\widetilde{C}_{\alpha\beta\mu\nu}\widetilde{C}^{\alpha\beta\mu\nu}$ and $J\equiv \frac{1}{6}\widetilde{C}_{\alpha\beta\lambda\sigma}\widetilde{C}_{\;\;\;\;\mu\nu}^{\lambda\sigma}\widetilde{C}^{\alpha\beta\mu\nu}$, where we use the complex self dual Weyl tensor $\widetilde{C}_{\alpha\beta\mu\nu} = \frac{1}{4}(C_{\alpha\beta\mu\nu} - i C_{\alpha\beta\mu\nu}^*)$. 
Because these definitions are directly in terms of the Weyl tensor, and do not use any projection, $I$ and $J$ are frame-independent.
From $I$ and $J$, one can define the speciality index \cite{J.Baker_M.Campanelli_2000},
\begin{equation}\label{eq: speciality_index}
    \mathcal{S} = 27 J^2/I^3,
\end{equation}
or simply $\mathcal{D}= I^3 - 27 J^2$, as in \cite{A.Coley_etal_2021}. The spacetime is of special Petrov type, see Section~\ref{sec: Petrov classification}, when $\mathcal{D}=0$ or $\mathcal{S}=1$ \cite{H.Stephani_etal_2003}.
 
Defining  the following complex linear combination of $E_{\alpha\beta}^{\{n\}}$ and $B_{\alpha\beta}^{\{n\}}$ 
\begin{equation}\label{eq:Q}
    Q_{\alpha\beta}^{\{n\}} = E_{\alpha\beta}^{\{n\}} + iB_{\alpha\beta}^{\{n\}}
\end{equation}
and  its complex conjugate $\overline{Q}_{\alpha\beta}^{\{n\}}=E_{\alpha\beta}^{\{n\}}-iB_{\alpha\beta}^{\{n\}}$  we can then express $I$ and $J$
in terms of $E_{\alpha\beta}^{\{n\}}$ and $B_{\alpha\beta}^{\{n\}}$
\cite{H.Stephani_etal_2003, C.B.G.McIntosh_etal_1995, M.Alcubierre_2008}:
\begin{equation}\label{eq: IandJ_withEB}
\begin{matrix}
    I = \overline{Q}_{\alpha\beta}\overline{Q}^{\alpha\beta}/2 = L_B/2-iM, \\
    J = -\overline{Q}^{\alpha}_{\beta}\overline{Q}^{\beta}_{\mu}\overline{Q}^{\mu}_{\alpha}/6 = \left[ - E^{\alpha}_{\beta}\left(E^{\beta}_{\mu}E^{\mu}_{\alpha} - 3 B^{\beta}_{\mu}B^{\mu}_{\alpha}\right)
    -iB^{\mu}_{\alpha}\left(B^{\alpha}_{\beta}B^{\beta}_{\mu} - 3 E^{\alpha}_{\beta}E^{\beta}_{\mu}\right)\right]/6.
\end{matrix}
\end{equation}

An alternative method to compute $I$ and $J$ is with the Newman-Penrose (NP) formalism \cite{M.Alcubierre_2008, M.Shibata_2015}. 
This enables us to compute the Weyl scalars and use them to compute further invariants. 
We first start by defining a set of four independent vectors:
\begin{equation}
    v_{(0)}^{\alpha}=n^{\alpha}, \;\;\;\;
    v_{(1)}^{\alpha}=\delta_{1}^{\alpha}/\sqrt{g_{11}}, \;\;\;\;
    v_{(2)}^{\alpha}=\delta_{2}^{\alpha}/\sqrt{g_{22}}, \;\;\;\;
    v_{(3)}^{\alpha}=\delta_{3}^{\alpha}/\sqrt{g_{33}}. 
\end{equation}
These are made orthonormal with the Gram-Schmidt method to obtain $\textbf{e}_{(\alpha)}$, our orthonormal tetrad basis. 
We start this procedure by choosing $e_{(0)}^\alpha = n^\alpha$. 
We distinguish tetrad indices with parenthesis and these are raised or lowered with the Minkowski metric. They have the properties:
\begin{equation}
    e_{(0)}^{\alpha}e_{\alpha}^{(0)}=-1, \;\;\;\;
    e_{(i)}^{\alpha}e_{\beta}^{(i)}=\delta_{\beta}^{\alpha}, \;\;\;\;
    e_{(i)}^{\alpha}e_{\alpha}^{(j)}=\delta_{(i)}^{(j)}, \;\;\;\;
    e_{(0)}^{\alpha}e_{\alpha}^{(j)}=0.
\end{equation}
They span the metric as $g_{\alpha\beta}=-e_{(0)\alpha}e_{(0)\beta} + \delta^{(i)(j)}e_{(i)\alpha}e_{(j)\beta}$ and $e_{(1)}^1e_{(2)}^2e_{(3)}^3=det(\gamma_{ij})^{-1/2}$.

From these, four complex null vectors are defined \footnote{ Here we use Alcubierre's notation in \cite{M.Alcubierre_2008}, the $l^{\alpha}$ and $k^{\alpha}$ vectors are swapped in comparison to the notation in \cite{H.Stephani_etal_2003}.}:
\begin{equation}
\begin{matrix}
    l^{\alpha} = \left(e_{(0)}^{\alpha} + e_{(1)}^{\alpha}\right)/\sqrt{2}, \;\;\;\;\;
    k^{\alpha} = \left(e_{(0)}^{\alpha} - e_{(1)}^{\alpha}\right)/\sqrt{2}, \\
    m^{\alpha} = \left(e_{(2)}^{\alpha} + ie_{(3)}^{\alpha}\right)/\sqrt{2}, \;\;\;\;\;
    \overline{m}^{\alpha} = \left(e_{(2)}^{\alpha} - ie_{(3)}^{\alpha}\right)/\sqrt{2},
\end{matrix}
\end{equation}
together referred to as a null NP tetrad.
By definition their norm is zero, and while $l_\alpha k^\alpha = -m_\alpha \overline{m}^\alpha = -1$ all other combinations vanish. They span the metric as 
\begin{equation} \label{eq: gnull}
    g_{\alpha\beta}=-2l_{(\alpha}k_{\beta)} +2m_{(\alpha}\overline{m}_{\beta)}.
\end{equation}

Finally, this null tetrad base is used to project the Weyl tensor and obtain the Weyl scalars, defined as:
\begin{equation} \label{eq: psis}
\begin{aligned}
    \Psi_{0} & \equiv C_{\alpha\beta\mu\nu}l^{\alpha}m^{\beta}l^{\mu}m^{\nu} = \overline{Q}_{\alpha\beta}^{\{n\}}m^{\alpha}m^{\beta}\\
    \Psi_{1} & \equiv C_{\alpha\beta\mu\nu}l^{\alpha}k^{\beta}l^{\mu}m^{\nu} = -\overline{Q}_{\alpha\beta}^{\{n\}}m^{\alpha}e_{(1)}^{\beta} / \sqrt{2}\\
    \Psi_{2} & \equiv C_{\alpha\beta\mu\nu}l^{\alpha}m^{\beta}\overline{m}^{\mu}k^{\nu} = \overline{Q}_{\alpha\beta}^{\{n\}}e_{(1)}^{\alpha}e_{(1)}^{\beta} / 2  = -\overline{Q}_{\alpha\beta}^{\{n\}}m^{\alpha}\overline{m}^{\beta}\\
    \Psi_{3} & \equiv C_{\alpha\beta\mu\nu}l^{\alpha}k^{\beta}\overline{m}^{\mu}k^{\nu} = \overline{Q}_{\alpha\beta}^{\{n\}}\overline{m}^{\alpha}e_{(1)}^{\beta} / \sqrt{2}\\
    \Psi_{4} & \equiv C_{\alpha\beta\mu\nu}k^{\alpha}\overline{m}^{\beta}k^{\mu}\overline{m}^{\nu} = \overline{Q}_{\alpha\beta}^{\{n\}}\overline{m}^{\alpha}\overline{m}^{\beta},\\
\end{aligned}
\end{equation}
where in the second equalities, the $\Psi$s are related to $E_{\alpha\beta}^{\{n\}}$ and $B_{\alpha\beta}^{\{n\}}$ by $\overline{Q}_{\alpha\beta}^{\{n\}}$.
Clearly, by construction, the $\Psi$s are frame dependent. To express them as a function of $\overline{Q}_{\alpha\beta}^{\{n\}}$, we use Maple \cite{Maple} to substitute the Weyl tensor with Eq.~(\ref{eq:Weyl_from_EandB}) and make simplifications based on the tetrad and null vector properties, as well as $e_{(0)}^{\alpha}e_{(1)}^{\beta}m^{\mu}\bar{m}^{\nu}\epsilon_{\alpha\beta\mu\nu} = e_{(1)}^{\beta}m^{\mu}\bar{m}^{\nu}\epsilon_{\beta\mu\nu} = -i$, meaning that : $e_{(1)}^{\beta}m^{\mu}\epsilon_{\beta\mu\nu} = -im_{\nu}$ and $e_{(1)}^{\beta}\bar{m}^{\nu}\epsilon_{\beta\mu\nu} = -i\bar{m}_{\mu}$ \cite{M.Shibata_2015}.

Conversely, with the Weyl scalars one can express $\overline{Q}^{\alpha\beta \{n\}}$ by projecting Eq.~(3.58) of \cite{H.Stephani_etal_2003} along $n^\mu$, obtaining  \cite{H.Stephani_etal_2003, A.Barnes_R.R.Rowlingson_1989, C.Cherubini_etal_2004}
\begin{equation}\label{eq: QonT}
\begin{aligned}
    \overline{Q}^{\alpha\beta \{n\}} 
    & = \Psi_{2}e_{C}^{\alpha\beta} \\
    & + \frac{1}{2}(\Psi_{0}+ \Psi_{4})e_{T+}^{\alpha\beta}
    -\frac{i}{2}(\Psi_{0}-\Psi_{4})e_{T\times}^{\alpha\beta}\\
    & -2(\Psi_{1}-\Psi_{3})e_{1}^{(\alpha} e_{2}^{\beta)}
    +2i(\Psi_{1}+\Psi_{3})e_{1}^{(\alpha} e_{3}^{\beta)}.
\end{aligned}
\end{equation}
Thus $\Psi_2$ is the component on the Coulombian basis tensor $e_{C}^{\alpha\beta} = 2e_{1}^{\alpha} e_{1}^{\beta} - e_{2}^{\alpha} e_{2}^{\beta} - e_{3}^{\alpha} e_{3}^{\beta}$, $\Psi_0$ and $\Psi_4$ are the components on the two transverse basis tensors $e_{T+}^{\alpha\beta} = e_{2}^{\alpha} e_{2}^{\beta} - e_{3}^{\alpha} e_{3}^{\beta}$ and $e_{T\times}^{\alpha\beta} = 2e_{2}^{(\alpha} e_{3}^{\beta)}$, and $\Psi_1$ and $\Psi_3$ are the components on the two longitudinal basis tensors $e_{1}^{(\alpha} e_{2}^{\beta)}$ and $e_{1}^{(\alpha} e_{3}^{\beta)}$. 
One can then express $E^{\mu\nu \{n\}}$ and $B^{\mu\nu \{n\}}$ in terms of Weyl scalar components, on the above defined tetrad basis, by using Eq.~(\ref{eq:Q}) and its complex conjugate:
\begin{equation}\label{eq: EBonT}
    \begin{aligned}
        E^{\alpha\beta \{n\}} 
        & = \Re(\Psi_{2}) e_{C}^{\alpha\beta}\\
        & + \frac{1}{2} \Re( \Psi_{0} + \Psi_{4} ) e_{T+}^{\alpha\beta} + \frac{1}{2}\Im(\Psi_{0}-\Psi_{4})e_{T\times}^{\alpha\beta}\\
        & - 2\Re(\Psi_{1}-\Psi_{3})e_{1}^{(\alpha} e_{2}^{\beta)} - 2\Im(\Psi_{1}+\Psi_{3})e_{1}^{(\alpha} e_{3}^{\beta)}, \\
        B^{\alpha\beta \{n\}} 
        & = -\Im(\Psi_{2})e_{C}^{\alpha\beta}\\
        & - \frac{1}{2} \Im( \Psi_{0}+ \Psi_{4} ) e_{T+}^{\alpha\beta} + \frac{1}{2}\Re(\Psi_{0}-\Psi_{4})e_{T\times}^{\alpha\beta}\\
        & + 2\Im(\Psi_{1}-\Psi_{3})e_{1}^{(\alpha} e_{2}^{\beta)} - 2\Re(\Psi_{1}+\Psi_{3})e_{1}^{(\alpha} e_{3}^{\beta)}.
    \end{aligned}
\end{equation}
Note that $\overline{Q}^{\alpha\beta \{n\}}$, $E^{\alpha\beta \{n\}}$ and $B^{\alpha\beta \{n\}}$ are defined in terms of a generic $n^\mu$ frame, therefore the expressions in Eq.~(\ref{eq: QonT}) and Eq.~(\ref{eq: EBonT}) are valid in any orthonormal frame with timelike vector $e_{(0)}^\mu=n^\mu$. 
In this sense, these expressions are frame-invariant.
They are valid for the general Petrov type expressed in a generic frame. 
Even for the general Petrov type, it is possible to choose a {\it transverse frame} where $\Psi_1=\Psi_3=0$ \cite{C.Beetle_L.M.Burko_2002, E.Berti_etal_2005}, then we can see from Eq.\eqref{eq: EBonT} that  both $E^{\mu\nu \{n\}}$ and $B^{\mu\nu \{n\}}$ have a Coulombian component, plus one for each transverse ``polarization".

Then to express $I$ and $J$ in Eq.~(\ref{eq: IandJ_withEB}) in terms of the Weyl scalars, we explicitly use the inverse of the metric Eq.~(\ref{eq: gnull}) to lower indices, e.g.\ $Q^{\alpha}_{\beta} = g^{\alpha\mu}Q_{\mu\beta}$, and using the definition Eq.~(\ref{eq: psis}) we obtain the well know expression:
\begin{equation}
    I = \Psi_{0}\Psi_{4} - 4\Psi_{1}\Psi_{3} +3\Psi_{2}^2, \;\;\;\;
    J = \begin{vmatrix}
    \Psi_{4} & \Psi_{3} & \Psi_{2} \\
    \Psi_{3} & \Psi_{2} & \Psi_{1} \\
    \Psi_{2} & \Psi_{1} & \Psi_{0}
    \end{vmatrix}.
\end{equation}
Further scalar invariants, which are relevant for the Petrov classification, see Section~\ref{sec: Petrov classification}, are defined as\footnote{ Where $K$ should not be confused with the trace of the extrinsic curvature.} \cite{R.Penrose_1960, H.Stephani_etal_2003, R.D'Inverno_R.Russell-Clark_1971}:
\begin{equation}
    K = \Psi_{1}\Psi_{4}^2 - 3\Psi_{4}\Psi_{3}\Psi_{2} + 2\Psi_{3}^3, \;\;\;\;
    L = \Psi_{2}\Psi_{4} - \Psi_{3}^2, \;\;\;\;
    N = 12 L^2 - \Psi_{4}^2 I.
\end{equation}
However, if $\Psi_4=0$ and $\Psi_0\neq 0$ then $\Psi_0$ and $\Psi_4$ need to be interchanged as well as $\Psi_1$ and $\Psi_3$.

We conclude this section by emphasizing again that $E^2$, $B^2$ and the various Weyl scalars are all frame-dependant scalars. Depending on the background, some of these scalars will be gauge-invariant at first order, as will be described in the next section. Then, $L_B$, $M$, $I$, $J$, $\mathcal{S}$ and $\mathcal{D}$ are coordinate and frame-invariant scalars, and $K$, $L$ and $N$ are coordinate-invariant and frame-dependent scalars \cite{D.Bini_etal_2021}.

\subsection{Covariant and Gauge-invariant perturbations and observables} \label{sec: cov and GI}

It is useful at this point to remark links and differences between spacetime coordinate invariance of the above defined scalars, either frame-invariant or not, and gauge-invariance in perturbation theory. 
The gauge-dependence of perturbations in relativity maybe confused with a simple lack of invariance under coordinate transformations, especially within a field-theoretical approach\footnote{ Relativistic perturbation theory can be seen as a classical field theory on a fixed  background spacetime, but this approach is not helpful in understanding the gauge dependence of perturbations, especially in going beyond the first order.}, see e.g.\ \cite{S.Weinberg_1972}. 
If this was the case, the gauge-dependence of perturbations of scalar quantities would then be obscure. 
The issue becomes totally clear when a geometrical approach is instead used, as it was first pointed out by Sachs \cite{R.K.Sachs_1964}, see \cite{J.M.Stewart_M.Walker_1974} for first-order perturbations and \cite{M.Bruni_etal_1997, S.Sonego_M.Bruni_1997, M.Bruni_S.Sonego_1999} for non-linear perturbations. 

The crucial point is that in relativistic perturbation theory we deal with two spacetimes: the realistic one, which we wish to describe as a small deviation from an idealised background, and the fictitious background spacetime itself. 
The gauge dependence of perturbations is due to the fact that a gauge choice in this context is a choice of mapping between points of the realistic spacetime and points of the background, so that a passive change of coordinates in the first (a change of labels for a given point) produces a change of points in the background. 
Since the background points also have their own coordinates, the change of points in the background results in what is sometime referred to as an active coordinate transformation (or point transformation) of the perturbation fields on the background. 
The latter is the point of view in \cite{S.Weinberg_1972}, and clearly also affects scalar quantities in general.  

This gauge dependence can then be formalised (at first order) in the Stewart and Walker Lemma \cite{J.M.Stewart_M.Walker_1974}, the essence of which being that for a tensorial quantity $T$ the relation between its perturbations in two different gauges is given by $\delta\tilde{T}=\delta T +{\cal L}_\xi T_o$, where $\xi$ is the vector field generating the said mapping of points in the background at first order and ${\cal L}_\xi T_o$ is the Lie derivative along $\xi$ of $T_o$, the tensor $T$ evaluated in the background. It then immediately follows that a tensorial quantity is gauge-invariant at first order if $T_o=0$. 

Both Teukolsky \cite{S.A.Teukolsky_1973} for the Kerr black hole and Stewart and Walker \cite{J.M.Stewart_M.Walker_1974} for any type D spacetime based their studies of first-order perturbations on Weyl scalars: if the background is of type D, both $\Psi_0$ and $\Psi_4$ vanish at zero order if the right frame is chosen (see next section for further details), hence they are gauge-invariant perturbations\footnote{ Things are more complicated for more general black holes \cite{P.Pani_2013}.}. 
The approach to perturbation theory where tensorial quantities that vanish in the background are directly used as perturbation variables may be called covariant and gauge-invariant \cite{G.F.R.Ellis_M.Bruni_1989}. 
 
In the context of cosmology, the  covariant and gauge-invariant approach was first partly used by Hawking \cite{S.W.Hawking_1966}, as the shear and vorticity of the matter 4-velocity vanish in an FLRW background, together with the electric and magnetic Weyl tensors. This was then extended in \cite{G.F.R.Ellis_M.Bruni_1989} by defining fully nonlinear variables characterising inhomogeneities in the matter density field, as well as in the pressure and expansion fields, i.e.\ covariantly defined spatial gradients that as such vanish in the homogeneous FLRW background and, therefore, are gauge-invariant at first order, following the Stewart and Walker Lemma \cite{J.M.Stewart_M.Walker_1974}. 

Clearly, not all perturbations of interest can be directly characterised by a tensor field that vanishes in the background, notably perturbations of the metric. Nonetheless, first-order gauge-invariant variables can be constructed as linear combinations of gauge-dependent quantities, such as the metric components and velocity perturbations, as first proposed in  \cite{U.H.Gerlach_U.K.Sengupta_1978}, then fully developed for perturbations of an FLRW background by Bardeen \cite{J.M.Bardeen_1980} and extended by Kodama and Sasaki \cite{H.Kodama_M.Sasaki_1984} to the multi-fluid and scalar field cases. 

Bardeen's approach is such that gauge-invariant variables only acquire a physical meaning in a specific perturbation gauge, or at least the specification of a slicing. It is clear, however, that  physical results can't depend on the mathematical approach used, and the two approaches are equivalent\footnote{ More precisely, a full equivalence with Bardeen's original variables is obtained under minimal and reasonable assumptions, in essence those required for  a harmonic expansion on the homogeneous and isotropic 3-space of the  FLRW background, see also \cite{J.M.Stewart_1990}. }, as shown in \cite{M.Bruni_etal_1992} and \cite{P.K.S.Dunsby_etal_1992}, cf.\ also \cite{S.W.Goode_1989},  where the physical meaning of Bardeen-like variables is  elucidated through the use of the covariant variables. In essence, Bardeen-like variables naturally appear when  the covariant nonlinear variables are fully expanded at first-order, e.g.\ Bardeen's potentials appear in the expansion of the electric Weyl tensor; Bardeen's evolution equations for the gauge-invariant  variable are also recovered in the same process, see \cite{M.Bruni_etal_1992} section 5. 

Gauge transformations and conditions for gauge invariance of perturbations generalising the Stewart and Walker Lemma  to an arbitrary higher order were derived in \cite{M.Bruni_etal_1997, S.Sonego_M.Bruni_1997}, soon leading to  applications in cosmology \cite{S.Matarrese_etal_1998, S.Mollerach_S.Matarrese_1997, R.Maartens_etal_1999}
and the theory of black holes perturbations \cite{M.Campanelli_C.O.Lousto_1999, A.Garat_R.H.Price_2000}, cf.\ \cite{R.J.Gleiser_etal_2000}, see 
\cite{C.G.Tsagas_etal_2008, K.A.Malik_D.Wands_2008, K.A.Malik_D.R.Matravers_2008} and \cite{K.D.Kokkotas_B.G.Schmidt_1999, E.Berti_etal_2009, E.Berti_etal_2018, A.Pound_B.Wardell_2021} for reviews, and e.g.\  \cite{C.A.Clarkson_2004, N.Bartolo_etal_2004, D.H.Lyth_etal_2005, K.Nakamura_2007, H.Noh_J-C.Hwang_2004, B.Osano_etal_2007, G.W.Pettinari_etal_2013, M.Bruni_etal_2014_Mar, E.Villa_C.Rampf_2015, H.A.Gressel_M.Bruni_2018, N.Loutrel_etal_2021, J.L.Ripley_etal_2021, M.H-Y.Cheung_etal_2022, K.Mitman_etal_2022} for more recent applications. The theory of gauge dependence of relativistic perturbations was then  extended to higher order  with two and more parameters in \cite{M.Bruni_etal_2003, C.F.Sopuerta_etal_2003}, again with applications in cosmology \cite{C.Pitrou_etal_2015, A.Talebian-Ashkezari_etal_2018, S.R.Goldberg_etal_2016} and
sources of gravitational waves \cite{A.Passamonti_etal_2005, A.Passamonti_etal_2006, A.Passamonti_etal_2007, C.F.Sopuerta_N.Yunes_2009, M.Lenzi_C.F.Sopuerta_2021}, cf.\ also \cite{P.Pani_2013} and references therein.

In summary, measurable quantities in relativity \cite{C.Rovelli_1991}, commonly referred to as observables, should be gauge-invariant, and the invariant scalars discussed in the previous section are gauge-invariant observables in that sense, i.e.\ they are coordinate independent. 
The electric and magnetic Weyl tensors $E^{\mu\nu \{u\}}$ and $B^{\mu\nu \{u\}}$ are first-order gauge-invariant variables on an FLRW background in the sense of perturbation theory explained above. 
Since the Weyl tensor $C_{\alpha\beta\mu\nu}$ itself is a first-order gauge-invariant variable on the conformally-flat FLRW background, at first order $E^{\mu\nu \{u\}}$ and $B^{\mu\nu \{u\}}$ are actually built by contraction of $C_{\alpha\beta\mu\nu}$ with the background fluid 4-velocity $u^\mu$. 
Then, the electric and magnetic Weyl tensors are also frame-invariant at first order \cite{M.Bruni_etal_1992}, if we define them with respect to a frame $n^\mu=u^\mu+V^\mu$, where $V^\mu$ is first-order deviation from the background $u^\mu$. 
Indeed, contracting $C_{\alpha\beta\mu\nu}$ with  $n^\mu=u^\mu+V^\mu$ gives, at first order, the same result than contracting with $u^\mu$,  i.e.\ $E^{\mu\nu \{n\}}\simeq E^{\mu\nu \{u\}} +{\cal O}(2)$ and $B^{\mu\nu \{n\}}\simeq B^{\mu\nu \{u\}} +{\cal O}(2)$. 
By their very definitions, the invariant scalars of the previous section are nonlinear. 
If expanded in perturbations, they may well have  a deviation from the background value that is only second or higher order, see \cite{J.Baker_M.Campanelli_2000} and \cite{C.Cherubini_etal_2004, C.Cherubini_etal_2005} for examples using the speciality index $\mathcal{S}$. 
As pointed out in \cite{C.Cherubini_etal_2004}, however, this has more to do with their definition than anything else, as one can always define a monotonic function of a scalar invariant that is first order, e.g.\ $\mathcal{S}_{norm}=\sqrt{|1-\mathcal{S}|}$ in considering $\mathcal{S}$ for perturbations of a Petrov type D spacetime. 
The point is that, even intuitively, we expect a {\it general} perturbation of a given background spacetime - typically of some special Petrov type, to be Petrov type general, even at first order; in other words, the first-order perturbations should separate  the principal null directions that coincide in the background, see  \cite{C.Cherubini_etal_2004} for a discussion.

How about observables as usually intended by cosmologists? Both the scalar invariants of the previous section and the gauge-invariant perturbations discussed above are {\it local quantities}, but in cosmology  observers can't go in a galaxy far far away \cite{StarWars} and measure $E^{\mu\nu \{u\}}$ and $B^{\mu\nu \{u\}}$ there: rather we have to link  points (spacetime events) on the past light-cone of the observer, points where local observables are defined,  with the  point of the observers themselves, i.e.\ ray tracing is key, see \cite{J.Adamek_etal_2020} for a concrete example in numerical relativistic cosmology and an application to simulation comparisons  and \cite{M.Grasso_etal_2021, M.Grasso_E.Villa_2021, T.Buchert_etal_2022, H.J.Macpherson_2022} and references therein for a general discussion. 
Nonetheless, although the issue of defining observables is  simple for first-order perturbations but more involved at second and higher order, first-order gauge invariance, i.e.\ the vanishing  in the background of the value of the observable tensorial quantity, plays a crucial role, see \cite{M.Bruni_S.Sonego_1999} and \cite{J.Yoo_R.Durrer_2017} for a recent and extended discussion.

In conclusion, $E^{\mu\nu}$ and $B^{\mu\nu}$ and the scalars built from  them can be useful for the physical interpretation of simulations and for the comparison of different codes, as well as for comparison of the fully nonlinear results of a simulation  with the result of perturbation theory.  Although our codes can compute $E^{\mu\nu}$ and $B^{\mu\nu}$ in any spacetime and in any frame, for the case of cosmology and especially for comparison with perturbation theory results  $E^{\mu\nu \{u\}}$ and $B^{\mu\nu \{u\}}$, computed in the matter rest frame, seem particularly useful \cite{L.Wylleman_etal_2019}.

\subsection{Petrov classification} \label{sec: Petrov classification}

Spacetimes can be classified according to their Weyl tensor, Ricci tensor, energy-momentum tensor, some special vector fields and symmetries \cite{ H.Stephani_etal_2003}. The Weyl tensor classification leads to the definition of the different Petrov types \cite{A.Z.Petrov_2000} and because it can be obtained invariantly it has become more significant \cite{H.Stephani_etal_2003}. There are six different types going from the general one to the most special: I, II, D, III, N, and O. There are multiple interrelated methods to determine the classification, which we now briefly summarise; we refer the reader to \cite{H.Stephani_etal_2003}, c.f.\ \cite{D.Bini_etal_2021} for a recent account.
\begin{itemize}
    \item Via the \textbf{Q} matrix. This is the tensor $Q_{\alpha\beta}$, Eq.~(\ref{eq:Q}), expressed with respect to an arbitrary orthonormal basis. This matrix has 3 complex eigenvalues and whether or not they are distinct will establish the Petrov type \cite{H.Stephani_etal_2003, A.Barnes_2014, D.Bini_etal_2021}.
   \item Via the principal spinors (or Debever spinors). The Weyl tensor can be expressed as a combination of these four spinors, and the Petrov type is related to whether or not these are independent or aligned \cite{J.Plebanski_A.Krasinski_2006}.
    \item Via the principal null directions that can be found using the Weyl scalars  \cite{H.Stephani_etal_2003, M.Alcubierre_2008, S.Chandrasekhar_1992, M.Shibata_2015}. Depending on the null tetrad base, certain Weyl scalars vanish\footnote{ The five complex Weyl scalars are just a different representation of the ten components of the Weyl tensor in 4-D. Even in the general case, therefore, coordinates or frame transformations can be used to make four of these ten components vanish, or two complex Weyl scalars.}. In the frame that maximises the number of vanishing scalars, those scalars will determine the Petrov type. Starting from a generic null base, a frame rotation can be chosen such that the new $\Psi_0$ vanishes. This is done by solving a $4^{th}$ order complex polynomial and the number of distinct roots, and whether or not they coincide, will determine the Petrov type and the principal null directions. Indeed, the more roots coincide, the more Weyl scalars can be made to vanish with further transformations. 
    \item Via the $I$, $J$, $K$, $L$, and $N$ invariants. Finding the roots of the polynomial described above is not a trivial task. So, based on the discriminant of the polynomial, these invariants are constructed \cite{R.D'Inverno_R.Russell-Clark_1971} and whether or not they vanish will establish the number of distinct roots and therefore the Petrov type, see the flow diagram in Figure 9.1 of \cite{H.Stephani_etal_2003}. For all special Petrov types $\mathcal{S}=1$ \cite{J.Baker_M.Campanelli_2000}, see Eq.~(\ref{eq: speciality_index}), or $\mathcal{D}=I^3-27J^2=0$ \cite{A.Coley_etal_2021}.
\end{itemize}

The physical interpretation of the different Petrov types has been described by Szekeres in \cite{P.Szekeres_1965} using a thought-device, the {\it gravitational compass}, measuring tidal effects, i.e.\ using the geodesic deviation equation. This physical interpretation is then based on looking at which of the Weyl scalars are non-zero in each case and on their specific distortion effects:
$\Psi_0$ and $\Psi_4$ generate a transverse geodesic deviation, while $\Psi_1$ and $\Psi_3$ generate a longitudinal tidal distortion; the real part of $\Psi_2$ represents the tidal distortion associated with a Coulomb-type field associated with a central mass (the only one that would be present in a Newtonian gravitational field); its imaginary part, if present, is associated with frame dragging. 
All of this becomes perfectly clear if one thinks of the Weyl tensor as the combination of $E_{\alpha\beta}$ and $B_{\alpha\beta}$ in Eq.~(\ref{eq:Weyl_from_EandB}), with  $E_{\alpha\beta}$ and $B_{\alpha\beta}$ expressed as in Eq.~(\ref{eq: EBonT}): they contain all of the effects mentioned above, notably $E_{\alpha\beta}$ contains the real part of $\Psi_2$, and $B_{\alpha\beta}$ its imaginary part.
For an FLRW spacetime linearly perturbed with only scalar perturbations, $B_{\alpha\beta}$ is zero and $E_{\alpha\beta}$ corresponds to the second derivatives of a linear combination of the Bardeen potentials that plays the role of the Newtonian gravitational potential\footnote{ In analogy to the relation between the Newtonian gravitational potential and the Newtonian tidal field.} \cite{M.Bruni_etal_1992}, sometimes also called the Weyl potential \cite{A.Lewis_A.Challinor_2006}.
Then, the physical interpretation of the different Petrov types is as follows.  
\begin{itemize}
    \item Type O is conformally flat, i.e.\ all Weyl scalars vanish and there are no tidal fields other than those associated with the Ricci curvature, e.g.\ like in FLRW spacetimes. 
    \item The Petrov type N is associated with plane waves \cite{F.A.E.Pirani_1957}, as the null tetrad can be chosen so that only $\Psi_4$ (or $\Psi_0$) is not zero; the tidal field associated to  $\Psi_0$ (or $\Psi_4$) is purely transverse and,   indeed, in the gauge-invariant perturbative formalism of Teukolsky \cite{S.A.Teukolsky_1973}, cf.\ also \cite{J.M.Stewart_M.Walker_1974}, gravitational wave perturbations of black holes are represented by $\Psi_4$. 
    \item In type III the null tetrad rotations allow to make all Weyl scalars but $\Psi_3$ zero: since $\Psi_3$ gives rise to a longitudinal tidal effect, this is  a strange case of spacetimes with pure longitudinal tidal fields. The sole perfect fluid solution known \cite{H.Stephani_etal_2003} has been found in \cite{J.A.Allnutt_1981}.
    \item In type D spacetimes, a null frame can be found such that only $\Psi_2$ is the non-zero Weyl scalar. This is the case of the Schwarzschild and Kerr spacetimes, where the real part of  $\Psi_2$ represents the Coulomb-type tidal field  and the imaginary part (vanishing for Schwarzschild) is associated with frame dragging. This is often referred to as the Kinnersley frame \cite{S.A.Teukolsky_1973, W.Kinnersley_1969}.
    \item In type II spacetimes the scalars $\Psi_2$ and  $\Psi_0$ can be made non-zero by appropriate rotations: these spacetimes can be seen as the superposition of an outgoing wave and a Coulomb-type field. A perfect fluid example of this Petrov type  was found by \cite{W.B.Bonnor_W.Davidson_1985} as a special case of the Robinson–Trautman metrics \cite{I.Robinson_A.Trautman_1962}.
    \item For Petrov type I, a standard choice is to have  $\Psi_1$, $\Psi_2$ and $\Psi_3$  non-zero \cite{H.Stephani_etal_2003,D.Bini_etal_2021}, but the alternative choice $\Psi_0$,$\Psi_2$ and $\Psi_4$ non-zero is also possible. This latter choice of the NP null tetrad can be called transverse \cite{C.Beetle_L.M.Burko_2002}. In the context of black hole perturbation theory this transverse tetrad can be called quasi-Kinnersley, as there are $\Psi_0$ and $\Psi_4$ perturbations on the Kinnersley background, as in  \cite{S.A.Teukolsky_1973} and \cite{J.M.Stewart_M.Walker_1974}. In numerical relativity applied to isolated sources of gravitational waves the search for this quasi-Kinnersley frame is a non-trivial task associated with the goal of properly extracting gravitational waves, see \cite{A.Nerozzi_etal_2005, A.Nerozzi_etal_2006} and Refs.\ therein.  However, a type I spacetime doesn't necessarily contain gravitational waves; a noteworthy example is the spacetime of stationary rotating neutron stars. In this case, in the quasi-Kinnersley transverse frame $\Psi_0$ and $\Psi_4$ can be interpreted as transverse (but stationary) tidal field deviation from the  Kerr geometry, see \cite{E.Berti_etal_2005}.
\end{itemize}

In summary, the process of making some of the Weyl scalars vanishing by rotations of the NP null tetrad can lead to ambiguities, as there is a certain set of degrees of freedom for each type leading to a set number of non-vanishing Weyl scalars, hence there is a certain freedom of choosing which scalar to cancel out. For instance, one may see that for type II it is also possible to have $\Psi_2$ and $\Psi_3$ instead of $\Psi_2$ and $\Psi_4$ non zero \cite{S.Chandrasekhar_1992}. Nonetheless, each Weyl scalar has a precise interpretation as a specific type of tidal field on the basis of the geodesic deviation equation and the associated gravitational compass \cite{P.Szekeres_1965}.  In general, the geodesic deviation equation is linear in the Riemann tensor (and therefore in its Weyl plus Ricci decomposition), hence it allows a superposition of the tidal effects associated with each Weyl scalar. However, this decomposition is not unique, as it differs for different observers associated with the different possible tetrad bases. This just means that different observers would measure different tidal fields, even if the Petrov type - and the corresponding intrinsic nature of the tidal field - would be invariant. In the following, we will focus - except for the tilted case in Section \ref{sec: Ex Bianchi VI} - on the class of observers comoving with matter and associated with its 4-velocity $u^\mu$, i.e.\ the timelike leg of the tetrad carried by these observers.

\section{Test-bed spacetimes} \label{sec: Example spacetimes}

In this section, we summarise the spacetimes that we use to test our two codes. These are all exact solutions of General Relativity, except for the test metric of Section~\ref{sec: Ex test metric}. Since our codes are motivated by cosmological applications, most of the solutions are homogeneous but we also consider one inhomogeneous solution, a $\Lambda$-Szekeres spacetime. However, it has no magnetic part of the Weyl tensor, we then have created this test metric that presents an inhomogeneous spacetime with an electric and magnetic part of the Weyl tensor. These metrics were chosen for their potential challenge to the codes, indeed by order of presentation, two of these spacetimes have a sinusoidal dependence on the space coordinates, the next is polynomial, and the last two are exponential. They are all provided to the codes which then compute $R$, $E^2$, and $B^2$ that we compare to the analytical solutions to establish code performance. Then for the $\Lambda$-Szekeres spacetime, in particular, we show what those variables and the scalar invariants look like, and we verify the resulting Petrov classification.

\subsection{The $\Lambda$-Szekeres models of Barrow and Stein-Schabes} \label{sec: Ex Szekeres}

\begin{figure}
    \centering
    \includegraphics[width=0.5\linewidth]{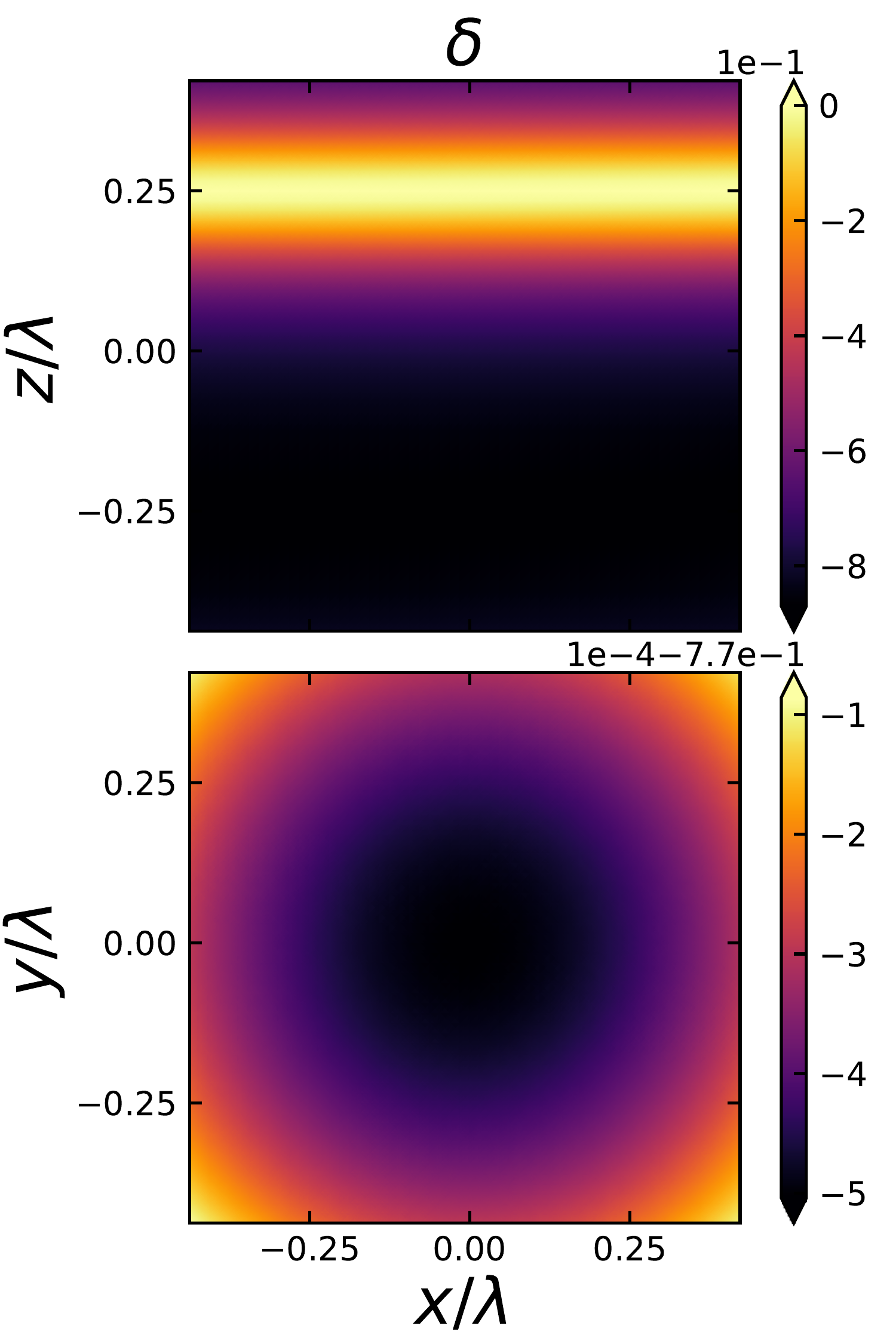}
    \caption{Space distribution of the density contrast $\delta$ in Eq.~(\ref{eq:szekdelta}) indicated by color coding (the top plot ranges from -0.8 to 0 and the bottom plot ranges from -0.7705 to -0.7701) for the $\Lambda$-Szekeres spacetime on the $z$-$x$ and $y$-$x$ planes (with $y/\lambda=0$ and $z/\lambda=0$ respectively) of a data box of size $\lambda=20$Mpc with $64^3$ grid points, at redshift $230$.}
    \label{fig: Szekeres delta}
\end{figure}

The first spacetime we consider is a cosmological solution generalising the FLRW metric to include a nonlinear inhomogeneity, it is the $\Lambda$-Szekeres \cite{P.Szekeres_1975, S.W.Goode_J.Wainwright_1982} model with dust and cosmological constant $\Lambda$ first considered by Barrow and Stein-Schabes in \cite{J.D.Barrow_J.Stein-Schabes_1984}. Following the representation of Szekeres models introduced in \cite{S.W.Goode_J.Wainwright_1982}, this model can be presented as a nonlinear exact perturbation of a flat (zero curvature) $\Lambda$CDM  background \cite{N.Meures_M.Bruni_2011_Jan, N.Meures_M.Bruni_2011_Jun} where the dust fluid represents Cold Dark Matter (CDM). In Cartesian-like coordinates  the line element is:
\begin{equation}\label{eq: SZmetric}
    ds^2 = - dt^2 + a^2(t) ( dx^2 + dy^2 + Z^2(t, x, y, z) dz^2 ).
\end{equation}
This model is well known to be Petrov type D and has no magnetic part of the Weyl tensor. It is a well-understood exact inhomogeneous cosmological solution, making it an interesting example for code testing, cf.\ \cite{M.Grasso_E.Villa_2021}. 

It turns out \cite{S.W.Goode_J.Wainwright_1982, N.Meures_M.Bruni_2011_Jan, N.Meures_M.Bruni_2011_Jun} that the scale factor $a(t)$ in Eq.~\eqref{eq: SZmetric} satisfies the Friedmann equations for the flat $\Lambda$CDM model, with background matter density  $\bar{\rho}^{\{u\}}$; they  are:
\begin{equation}\label{eq: SzekeresBackground}
    a(t) = \left(\frac{\Omega_{m0}}{\Omega_{\Lambda 0}}\right)^{1/3}\sinh{\left(\frac{3 t H_0}{2}\sqrt{\Omega_{\Lambda 0}}\right)}^{2/3},
    \;\;\;\;\;\;\;\;\;\;\;\;
    \bar{\rho}^{\{u\}} = \frac{3H_{0}^{2} \Omega_{m0}}{\kappa}{a}^{-3}.
\end{equation}

$H_0$, $\Omega_{\Lambda 0}$ and $\Omega_{m0}$ are, respectively, the Hubble constant and the energy density parameters for $\Lambda$ and the matter measured today; we have set, with no loss of generality,  $a(t_0)=1$, where $t_0$ is the current age of the universe depending on $H_0$ and $\Omega_{m0}$. For these parameters, in the result section, we chose to work with the measurements of the Planck collaboration (2018) \cite{Planck_CMB_2018}.

The inhomogeneous matter density is $\rho^{\{u\}} = \bar{\rho}^{\{u\}}(1+\delta)$, where the density contrast $\delta$ can be written as:
\begin{equation}\label{eq:szekdelta}
    \delta = -F(t,z)/Z(t, x, y, z).
\end{equation}
The function $Z(t,x,y,z)$ represents inhomogeneity in the metric Eq.~\eqref{eq: SZmetric} and can be written as:
\begin{equation}\label{eq: Szekeres_Z}
    Z(t, x, y, z) = 1 + F(t,z) + \beta_{+}(z)(x^2+y^2) \frac{3}{4} H^{2}_0 \Omega^{1/3}_{\Lambda 0} \Omega^{2/3}_{m 0}.
\end{equation}

Remarkably, as far as its time-dependence is concerned,  the nonlinear perturbation $F(t,z)$ satisfies exactly the same linear second order differential equation satisfied by $\delta$ in linear perturbation theory \cite{S.W.Goode_J.Wainwright_1982, N.Meures_M.Bruni_2011_Jan, N.Meures_M.Bruni_2011_Jun}, cf.\ \cite{M.Bruni_etal_2014_Mar}. Because of this, $F$ is in general composed of a growing and decaying mode. The latter is usually neglected in cosmological structure formation theory, while the former is sourced by the curvature perturbations related to ${}^{(3)}R$ \cite{M.Bruni_etal_2014_Mar}, cf.\ also \cite{M.Bruni_etal_2014_Sep} for a different approximation leading to the same equations. Therefore for our test we only consider the growing mode:
\begin{equation}\label{eq: Szekeres_F}
    F(\tau,z) = \beta_+(z)\frac{3}{5} \cosh{(\tau)}\sinh{(\tau)}^{2/3} {}_{2}F_{1}\left(\frac{5}{6},\; \frac{3}{2};\; \frac{11}{6};\; -\sinh{(\tau)}^2 \right),
\end{equation}
where $\tau = t\sqrt{\frac{3\Lambda}{4}}$, and ${}_{2}F_{1}$ is a hypergeometric function, see \ref{sec: analyticsolution_szekeres}. Then we chose the spatial perturbation to be $\beta_{+}(z) = 10^3 (1-\sin{(2\pi z/\lambda)})$, $\lambda$ being the data box size. Given these choices, the density contrast $\delta$ in Eq.~(\ref{eq:szekdelta}) is negative and it is illustrated in Fig.~(\ref{fig: Szekeres delta}), where we can see the influence of the sinusoidal distribution on the $x-z$ plane, and the paraboloid structure on the $x-y$ plane. This spacetime's invariants and its Petrov type will be discussed in Section~\ref{sec: Szekeres invariants}.

\subsection{A non-diagonal inhomogeneous test metric} \label{sec: Ex test metric}

In order to have an inhomogeneous example with a non-vanishing magnetic part of the Weyl tensor, we introduce a spacetime with the following line element:
\begin{equation}\label{eq: testmetric}
    ds^2 = - dt^2 + t A(z) \delta_{ij} dx^i dx^j + 2 dx (dy + dz),
\end{equation}
where $A(z)$ is an arbitrary function, that for practical purposes we assume to be positive.
This can be found as a solution to Einstein's equations by the method of reverse engineering the metric, where one starts with the metric, and then finds the corresponding energy-momentum tensor with Einstein's field equations. 
In the analytical computations of this spacetime, we find it to be of Petrov type I, see \ref{sec: analyticsolution_testmetric}.

However, since the determinant of this metric is $g = A(z)t[2-A(z)^2 t^2]$, one can see that this spacetime is only valid for a certain domain in time, also depending on $A(z)$. 
Additionally, the resulting energy-momentum tensor doesn't have any particular physical meaning, so we are not referring to this spacetime as a solution to Einstein's equations. 
All we need to test our code is a specific form of the metric.
In this light, although $A(z)$ is an arbitrary function, we define it for the purpose of the test as: $ A(z) = 2.3 + 0.2 \sin{(2 \pi z /\lambda)}$ so we can use periodic boundary conditions, with $\lambda$ the box size. Then, in the frame associated with $n^\mu$, we obtain the (rather fictitious) non-perfect fluid energy-momentum tensor $T_{\alpha\beta}$ from Einstein's equations, see \ref{sec: analyticsolution_testmetric}.

\subsection{Bianchi II Collins-Stewart} \label{sec: Ex Bianchi II}

The Collins and Stewart Bianchi II $\gamma$-law perfect fluid homogeneous solution \cite{C.B.Collins_J.M.Stewart_1971, J.Wainwright_G.F.R.Ellis_1997} has the spatial metric
\begin{equation}\label{eq: BianchiIIspatialmetric}
    \gamma_{ij}=\begin{pmatrix}
    t^{(2-\gamma)/\gamma} & t^{(2-\gamma)/\gamma}(sz/2\gamma) & 0 \\
    t^{(2-\gamma)/\gamma}(sz/2\gamma) & t^{(2+\gamma)/2\gamma}+t^{(2-\gamma)/\gamma}\left(sz/2\gamma\right)^2 & 0 \\
    0 & 0 & t^{(2+\gamma)/2\gamma}
    \end{pmatrix},
\end{equation}
with the constant $s^2 = (2 - \gamma)(3\gamma-2)$; the synchronous comoving gauge and Cartesian-like coordinates are assumed. 
The perfect fluid has energy density $ \rho^{\{u\}} = (6-\gamma)/4\kappa t^2\gamma^2$, and pressure following the $\gamma$-law: $p = (\gamma-1)\rho^{\{u\}}$, so $\gamma=1$ for dust and $\gamma=4/3$ for radiation. 
In the latter case ${}^{(4)}R=0$, in both cases this spacetime is of Petrov type D, see \ref{sec: analyticsolution_BianchiII}. 
This is our sole example showing the spacial metric having a polynomial dependence on the space coordinates.

\subsection{Bianchi VI tilted model} \label{sec: Ex Bianchi VI}

Assuming the synchronous gauge and Cartesian-like coordinates, the Rosquist and Jantzen Bianchi VI tilted $\gamma$-law perfect fluid homogeneous solution with vorticity\footnote{ See footnote 3 and 4, emphasising that when vorticity is present a spacial hypersurface can not be constructed, hence a comoving frame can not be used and only tilted spacial hypersurfaces can be constructed with $n^\alpha \neq u^\alpha$.} \cite{K.Rosquist_R.T.Jantzen_1985, H.Stephani_etal_2003}, has the spacial metric:
\begin{equation}\label{eq: BianchiVImetric}
    \gamma_{ij}=\begin{pmatrix}
    (1+m^2)(kt)^2    & mkt^{1+s-q}e^{x}  & 0   \\
    mkt^{1+s-q}e^{x} & t^{2(s-q)}e^{2x} & 0   \\
    0                & 0                 & t^{2(s+q)}e^{-2x}
    \end{pmatrix},
\end{equation}
with the constants:
\begin{equation}
\begin{matrix}
    s = (2-\gamma)/(2\gamma), \\
    m^2 = -32q^2s/(s-q-1)^2(3s+3q-1), \\
    q = (6-5\gamma)(2-\gamma+2\sqrt{(9\gamma-1)(\gamma-1)})/2\gamma(35\gamma-36), \\
    k^2 = -(3s+3q-1)/(s+3q-1)(3s^2+(6q-1)s-q^2-q).
\end{matrix}
\end{equation}
With this definition of $q$, $\gamma$ is limited to the domain $]6/5, 1.7169...[$ \cite{H.Stephani_etal_2003}. For our test, we use $\gamma=1.22$ and although this solution is described by a perfect fluid following the $\gamma$-law in a tilted frame, $T_{\alpha\beta}$ used in the slicing code was computed from Einstein's equations for a non-perfect fluid in the $n^\alpha$ frame. An other relevant note for the code testing, is that the space dependence of the metric is exponential. Using Maple \cite{Maple}, we find that this spacetime is of Petrov type I, see \ref{sec: analyticsolution_BianchiVI}.

\subsection{Bianchi IV vacuum plane wave} \label{sec: Ex Bianchi IV plane wave}

The final spacetime we consider is the Harvey and Tsoubelis Bianchi IV vacuum plane wave homogeneous solution \cite{A.Harvey_D.Tsoubelis_1977, A.Harvey_etal_1979, J.Wainwright_G.F.R.Ellis_1997} with spatial metric:
\begin{equation}\label{eq: planewavemetric}
    \gamma_{ij}=\begin{pmatrix}
    t^2 & 0 & 0 \\
    0 & te^{x} & te^{x}(x+\log{t}) \\
    0 & te^{x}(x+\log{t}) & te^{x}((x+\log{t})^2+1)
    \end{pmatrix}.
\end{equation}
Again, the synchronous comoving gauge and Cartesian-like coordinates are assumed.  
The plane wave represented by this model makes it a very interesting example: it is easy to check with Maple \cite{Maple}, see \ref{sec: analyticsolution_PlaneWave}, that $E^{2\{u\}} = B^{2\{u\}} = 1/2t^4$ and that the Petrov type is N \cite{W.B.Bonnor_1995}.
This gives an additional point of comparison for $E^{2\{u\}}$ and $B^{2\{u\}}$, otherwise, as we are in vacuum, ${}^{(4)}R=0$.

\section{Description of the codes and Numerical implementation} \label{sec: code description and numerical implementation}

A Python post-processing code has been developed for each computational method in Section~\ref{sec: Th framework}: the geometrical and slicing methods \cite{R.L.Munoz_2022_ebweyl}. 

\subsection{Geometrical code}

In the code using the geometrical approach, the 4-D Riemann tensor is calculated from its definition in terms of the derivatives of the metric $g_{\alpha\beta}$. 
Because of the added complexity in computing time derivatives, this code has been developed only for the synchronous gauge, $g_{0\alpha}=\{-1,\;0,\;0,\;0\}$.
In practice, assuming that this post-processing code is applied to data produced by a numerical simulation in this gauge, then the metric is directly given by $\gamma_{ij}$.

The first spatial derivatives of the metric are computed with a centred finite difference (FD) scheme where the boundary points are obtained using a periodic boundary condition when applicable (here only for the test metric case Section~\ref{sec: Ex test metric}), otherwise a combination of forward and backward schemes are used. 
As the centred scheme has lower relative error than either the forward or backward scheme, the points along the edges affected by this boundary choice are cut off, see \ref{sec: append FDtest}.
These considerations are of no concern when applying these codes to cosmological simulation results, as in this case the boundary  conditions commonly used are periodic. 

Then, the first time derivative of the metric in the synchronous gauge coincides with the extrinsic curvature, $K_{ij}=-\frac{1}{2}\partial_t\gamma_{ij}$, and therefore can directly be retrieved from the data of the underlying simulation.

Finally, to compute second derivatives of the metric, spatial derivatives of all of the above are computed with the same scheme applied for the first spatial derivatives, and time derivatives are computed with a backward scheme. 

Then, having all the necessary derivatives, the 4-D Christoffel symbols and their derivatives are computed to obtain the 4-D Riemann tensor. From this, the 4-D Ricci tensor and Ricci scalar ${}^{(4)}R$ are constructed and $E_{\alpha\beta}$ and $B_{\alpha\beta}$ are computed using Eq.~(\ref{eq:Weyl}) and Eq.~(\ref{eq:EBGeometrical}). The outputs of this code are ${}^{(4)}R$, $E^2$, and $B^2$; these are used in the examples in Section~\ref{sec: Results}.

The FD schemes are all of $4^{th}$ order, then to increase accuracy we also implement the option to use $6^{th}$ order schemes \cite{B.Fornberg_1988}, and to have Riemann symmetries enforced: $R_{\alpha\beta\mu\nu} = - R_{\beta\alpha\mu\nu} = - R_{\alpha\beta\nu\mu} = R_{\mu\nu\alpha\beta}$.

\subsection{Slicing code: EBWeyl}
In EBWeyl \cite{R.L.Munoz_2022_ebweyl}, the code using the slicing approach, the metric $g_{\alpha\beta}$ provides the 3+1 variables needed to compute ${}^{(3)}R$. Then with $K_{\alpha\beta}$ and $T_{\alpha\beta}$,  $E_{\alpha\beta}$ and $B_{\alpha\beta}$ are computed with Eq.~(\ref{eq:ESlicing}) and Eq.~(\ref{eq:BSlicing}). No time derivatives are needed and the spatial derivatives are obtained with the same scheme used in the geometrical code. 

EBWeyl is essentially a module with functions and classes providing FD tools and computations of tensorial expressions. 
In the {\tt github} repository there is an example Jupyter Notebook demonstrating how to use it for the Bianchi IV vacuum plane wave spacetime in Section~\ref{sec: Ex Bianchi IV plane wave}. 
The user needs to provide $g_{\alpha\beta}$, and $K_{\alpha\beta}$ as numerical numpy arrays to the class called Weyl, this will automatically define the 3+1 terms, then the class's functions will compute the expressions of Section~\ref{sec: Slicing method} and \ref{sec: invariants}, as demonstrated in the Jupyter Notebook \cite{R.L.Munoz_2022_ebweyl}. 
Note that in EBWeyl the electric part of the Weyl tensor is defined as a 3+1 quantity, i.e.\ as in Eq.~(\ref{eq:ESlicing}); therefore the energy-momentum tensor $T_{\alpha\beta}$ also needs to be provided to get $E^{\{n\}}_{ij}$.
For example, in vacuum, as in Section~\ref{sec: Ex Bianchi IV plane wave}, this simply means that $T_{\alpha\beta}$ should be provided as a array of zeros.
Finally, we emphasise that although the examples of this paper are all cosmological and we always use the synchronous gauge, EBWeyl is general enough to be applied to any spacetime in any gauge. 

\subsection{Application of the codes}
Both these codes were applied to the example spacetimes in Section~\ref{sec: Example spacetimes} in order to establish which code is most suitable for cosmological numerical relativity simulations. 
The discussion in the next Section will demonstrate preference towards the slicing code EBWeyl, therefore 
this code has been made  publicly available in \cite{R.L.Munoz_2022_ebweyl}. 

\section{Results} \label{sec: Results}

Here we present two forms of tests. Firstly, we demonstrate applications of these codes to the $\Lambda$-Szekeres spacetime in Section~\ref{sec: Ex Szekeres} \cite{P.Szekeres_1975, J.D.Barrow_J.Stein-Schabes_1984, N.Meures_M.Bruni_2011_Jan, N.Meures_M.Bruni_2011_Jun}. We compute ${}^{(4)}R$ with the geometrical code and we compute ${}^{(3)}R$, $B^2$, $E^2$ and the invariants of Section~\ref{sec: invariants} with the slicing code. With the invariants, we then check that this spacetime is of Petrov type D. This process is then applicable to any numerical spacetime where the Petrov type is not known.

Secondly, we show the numerical error, and convergence, on computing $R$, $E^2$, and $B^2$ for each code on each example spacetime of Section~\ref{sec: Example spacetimes}. As we identify different types of numerical errors, each is addressed individually showing how reliable these codes are.

To do these tests using the metrics of Section~\ref{sec: Example spacetimes} we generate 3-D data boxes of $N^3$ points where the $x$, $y$, and $z$ coordinates vary, such that at each data point is associated with a numerical metric tensor computed from the analytical metric. We additionally associate a numerical extrinsic curvature and stress tensor with each of these points. The provided data has then been generated exactly at a singular arbitrary time for the slicing code, and multiple times, with a small time step, for the geometrical code. These numerical arrays are provided to the two codes where the outputs can be plotted, as in Section~\ref{sec: Szekeres invariants}, or compared to the expected solution as in Section~\ref{sec: Code tests}. This comparison is done by computing the average relative difference between the code outputs, say $v$, and the analytical solution, $v_{th}$: $\mathbb{E}\left(|v/v_{th} - 1|\right)$. These solutions are derived analytically using Maple, see \ref{sec: analyticsolution}, and provided as numerical arrays for comparison.

\subsection{Invariants for the $\Lambda$-Szekeres spacetime} \label{sec: Szekeres invariants}

\begin{figure}
    \centering
    \includegraphics[width=0.87\linewidth]{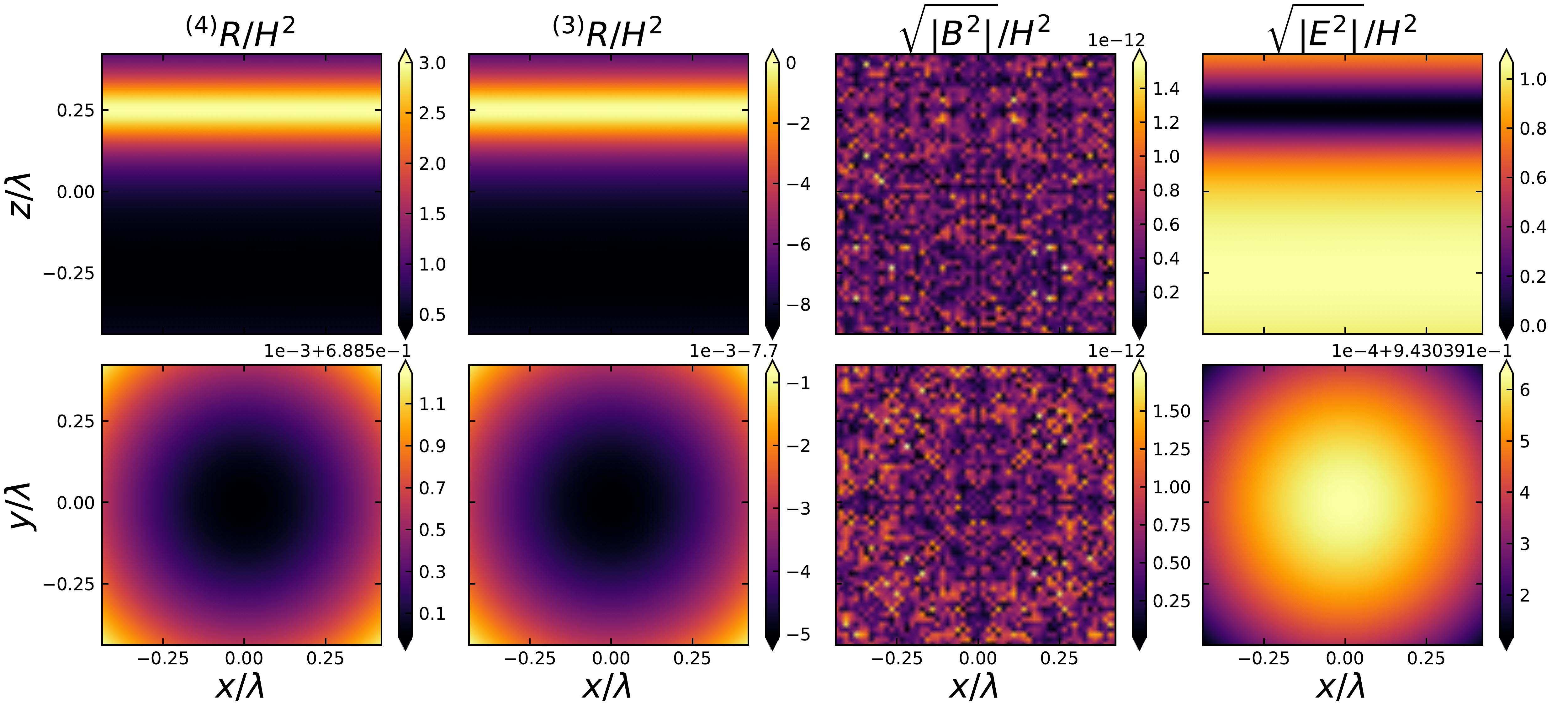}
    \includegraphics[width=0.87\linewidth]{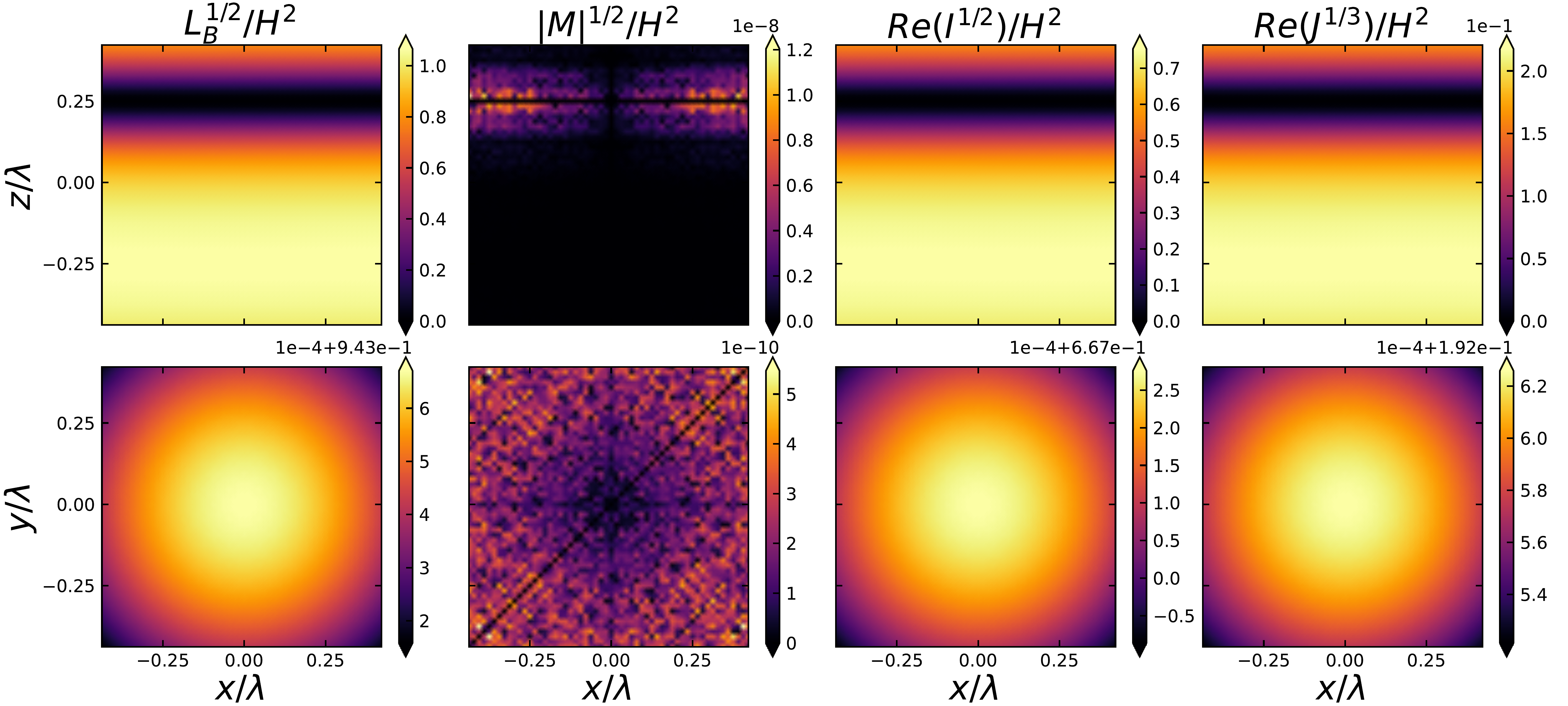}
    \includegraphics[width=0.87\linewidth]{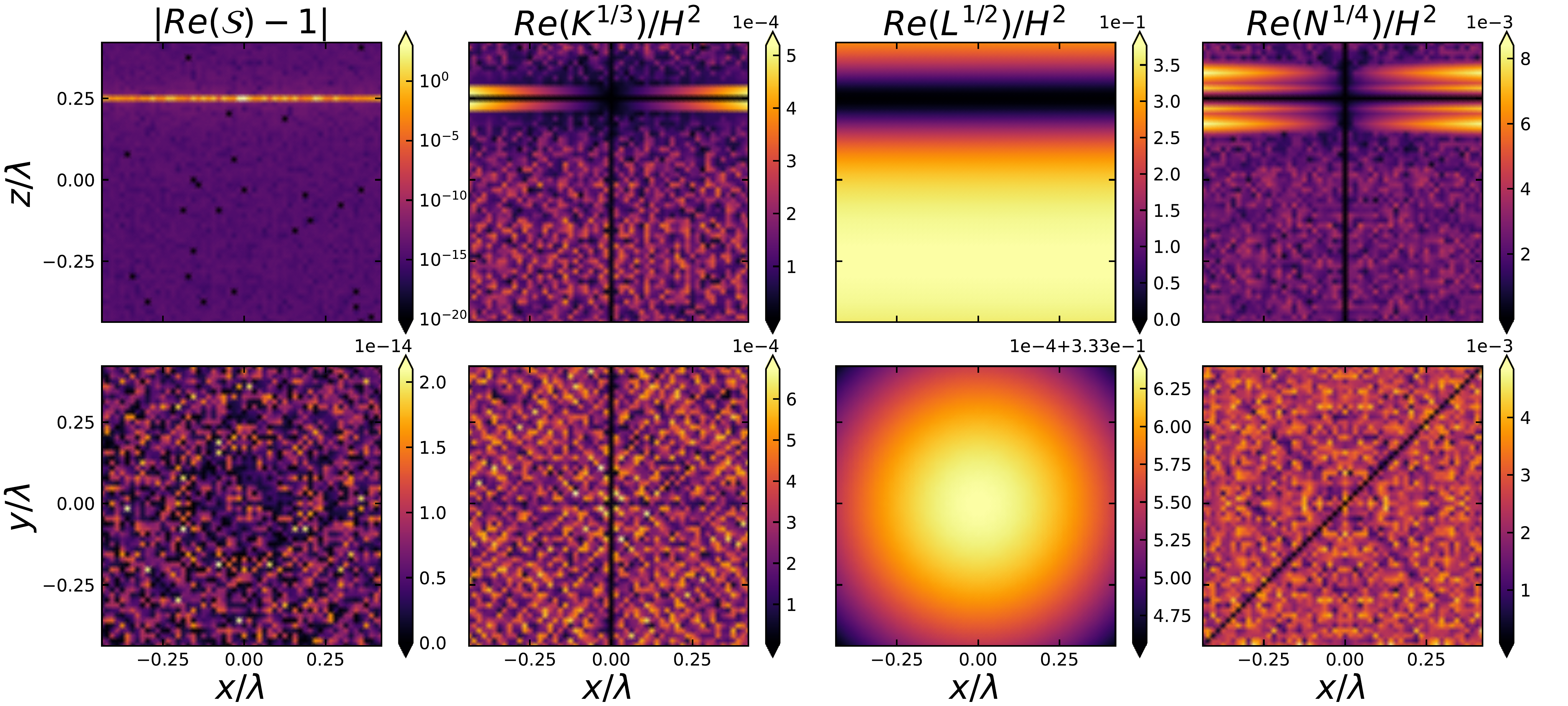}
    \caption{For the $\Lambda$-Szekeres spacetime we show the spatial distribution of the 4-D and 3-D Ricci scalar ${}^{(4)}R$ and ${}^{(3)}R$, the magnitude of the magnetic and electric parts of the Weyl tensor $\sqrt{|B^2|}$ and $\sqrt{|E^2|}$ and the invariant scalars along the $z$-$x$ and $y$-$x$ planes (with $y/\lambda=0$ and $z/\lambda=0$ respectively) of a data box with $64^3$ grid points. These quantities are made dimensionless by dividing by the square of the Hubble scalar $H$. Only the real part of the complex invariants are shown, as the imaginary parts are zero, up to numerical noise.}
    \label{fig: Szekeres invariants}
\end{figure}

The 4-D and 3-D Ricci scalars and the invariants from Section~\ref{sec: invariants} of the $\Lambda$-Szekeres spacetime have been computed and are presented in Fig.~(\ref{fig: Szekeres invariants}). This shows their spatial distribution along the $x-z$ and $x-y$ planes. We present them in homogeneous (first) powers of the Weyl tensor, e.g.\ $I^{1/2}$, and make them dimensionless by dividing by the Hubble scalar $H = \frac{\partial a}{a \partial t}$, cf.\ \cite{J.Wainwright_G.F.R.Ellis_1997}, e.g.\ $I^{1/2}/H^2$. For complex scalars, only the real part is shown, as for the imaginary part we only get numerical noise. The geometrical code was used for ${}^{(4)}R$, and then the slicing code otherwise. ${}^{(3)}R$ shows flatness where the density contrast goes to zero $\delta\rightarrow 0$, Fig.~(\ref{fig: Szekeres delta}), and negative curvature where $\delta<0$. $B^2$ vanishes, so only numerical noise is visible. Contrasting the panels for $E^2$ and ${}^{(3)}R$, we note that where the negative 3-curvature is strongest, $E^2$ is also strongest, and where it is flatter the electric tidal field is weaker. 

The Szekeres spacetime and the Barrow and Stein-Schabes model with $\Lambda$ are of Petrov type D \cite{H.Stephani_etal_2003, A.Barnes_R.R.Rowlingson_1989, N.Meures_M.Bruni_2011_Jun}, meaning that $I\neq 0$, $J\neq 0$, $\mathcal{S}=1$, and $K=N=0$ \cite{H.Stephani_etal_2003, J.Baker_M.Campanelli_2000}. We can check this here, indeed $B_{\alpha\beta}=0$ so $L_B=E^2$, $M=0$, $I=E^2/2$, and $J$ is a combination of $E_{\alpha}^{\beta}$ with $Im(J)=0$. Therefore we can see the similarity of $L_B$ and $Re(I)$ with $E^2$, and then $M$ shows numerical noise but is otherwise null. Hence we see that $I$ and $J$ are not null, except along the $z/\lambda=0.25$ plane. Next, to verify $\mathcal{S}=1$, $|Re(\mathcal{S})-1|$ is plotted, showing that everywhere we have $\mathcal{S}=1$ except at $z/\lambda=0.25$. On this plane, $I=0$, so $\mathcal{S}$ has a singularity plane, being ill-defined for spacetimes other than I, II, and D. Numerically we do not have exactly zero but a small value that makes $\mathcal{S}$ extremely large. Finally, $K$ and $N$ are both shown to present numerical noise, completing all the requirements for us to numerically confirm that indeed, this $\Lambda$-Szekeres metric is of Petrov type D. Except on the $z/\lambda=0.25$ plane where we can see that all invariants vanish, along this coordinate $\beta_+(z=\lambda/4)=0$ and so the spacetime is pure FLRW and of Petrov type O.

In summary, we have shown the potential of this code in deriving various invariants of an analytic spacetime. The same type of analysis can be done on any spacetime generated numerically. We next look into the accuracy of these measurements.

\subsection{Code tests} \label{sec: Code tests}

\begin{figure}
    \centering
    \includegraphics[width=\linewidth]{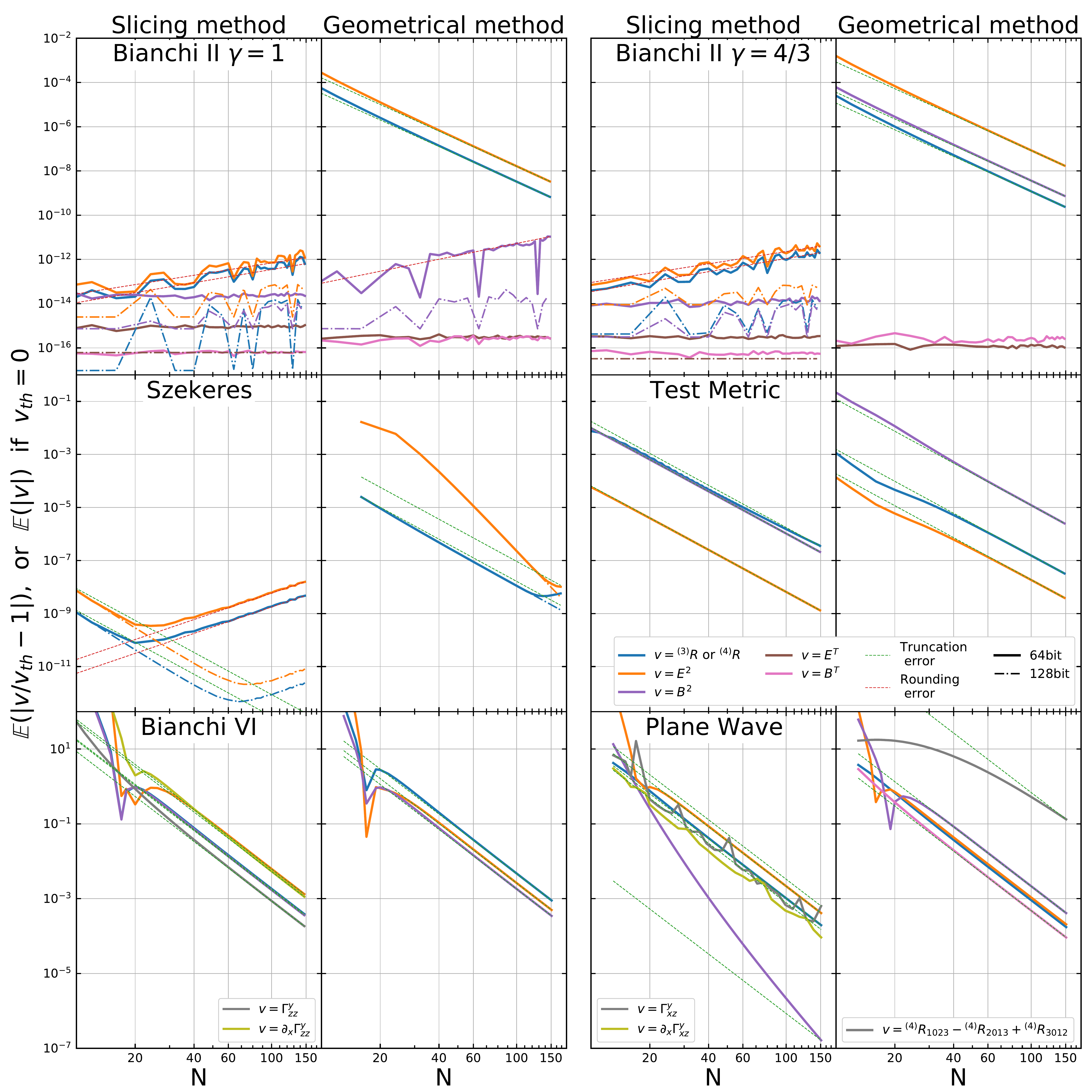}
    \caption{Six panels showing, respectively, the average relative error of the slicing code, left side, and geometrical code, right side, of the Bianchi II (with $\gamma=1$ and $\gamma=4/3$), $\Lambda$-Szekeres, test metric, Bianchi VI, Bianchi IV plane wave spacetimes, applied to data boxes of N${}^3$ grid points, using a $4^{th}$ order FD scheme.}
    \label{fig: Error}
\end{figure}

\begin{figure}
    \centering
    \includegraphics[width=\linewidth]{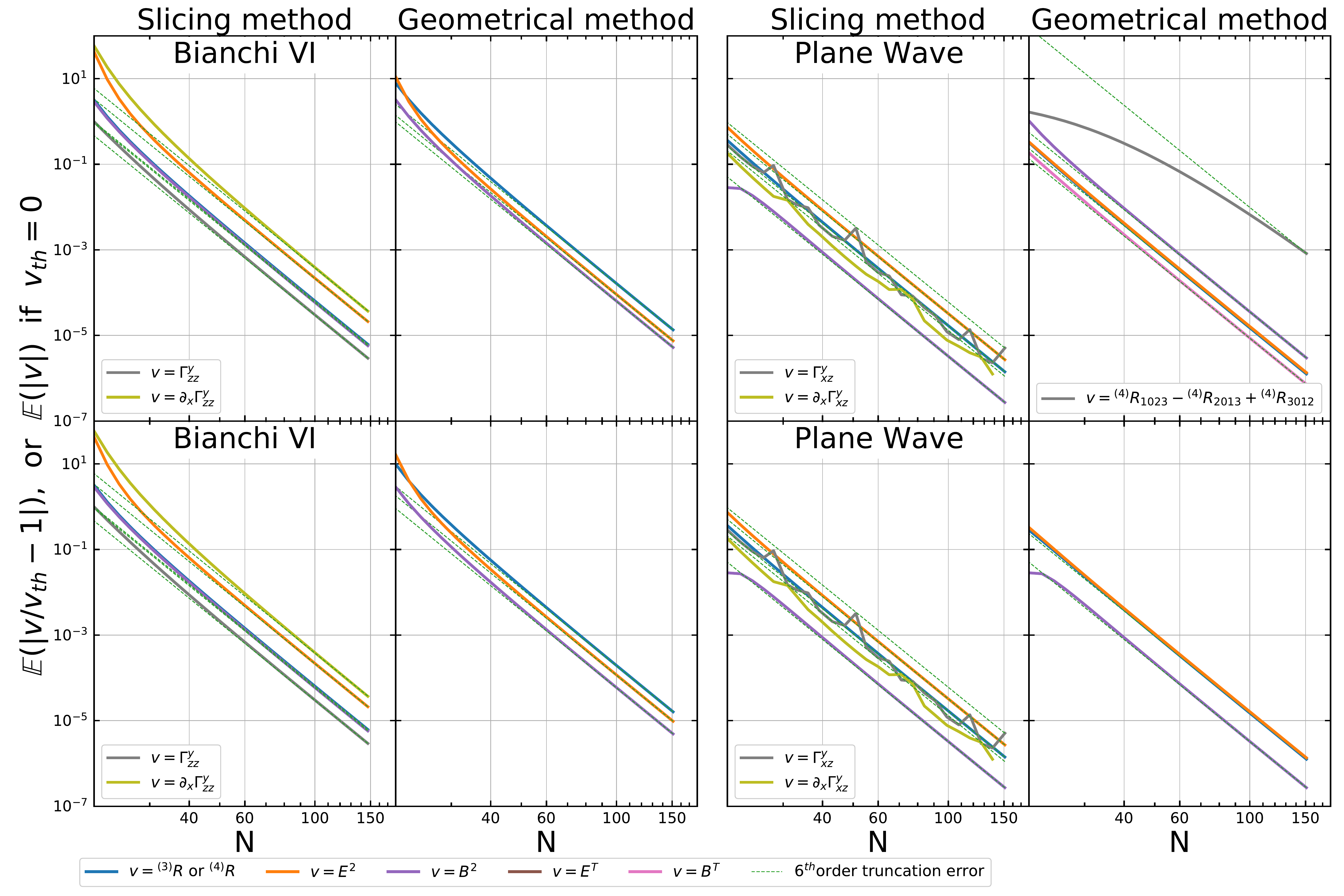}
    \caption{Top row: Two panels showing, respectively, the average relative error of the slicing, left side, and geometrical code, right side, of the Bianchi VI and Bianchi IV plane wave spacetimes with data boxes of N${}^3$ grid points, using a $6^{th}$ order FD scheme. Bottom row: the same with Riemann symmetries enforced. For the plane wave geometrical case, $B^T$ and the Bianchi identity vanish.}
    \label{fig: Erroro6}
\end{figure}

To test our codes, we run them on all the example spacetimes listed in Section~\ref{sec: Example spacetimes} and compare the results to the expected analytical expressions of \ref{sec: analyticsolution}. Fig.~(\ref{fig: Error}) shows the resulting numerical error when computing ${}^{(3)}R$ or ${}^{(4)}R$, $E^2$, $B^2$, $E^T$ and $B^T$ (their trace, which should be zero). If the analytical solution is different from zero, the relative error is shown, otherwise, the value itself is presented. The scalars that are absent from the plot, are omitted because the error is too small to fit in, and so is of lesser interest. All these plots display multiple types of numerical errors so we will address these individually. To complete this analysis we show again in Fig.~(\ref{fig: Erroro6}) the numerical error for the Bianchi VI and plane wave cases, this time using the $6^{th}$ order FD and Riemann symmetry enforcement options. Additionally, the plane wave case has $E^2=B^2$, so Fig.~(\ref{fig: Plane Wave B2vsE2}) shows how accurately each code can reproduce this. 

\subsubsection{Truncation error} \label{sec: Truncation error}
\begin{figure}
    \centering
    \includegraphics[width=0.65\linewidth]{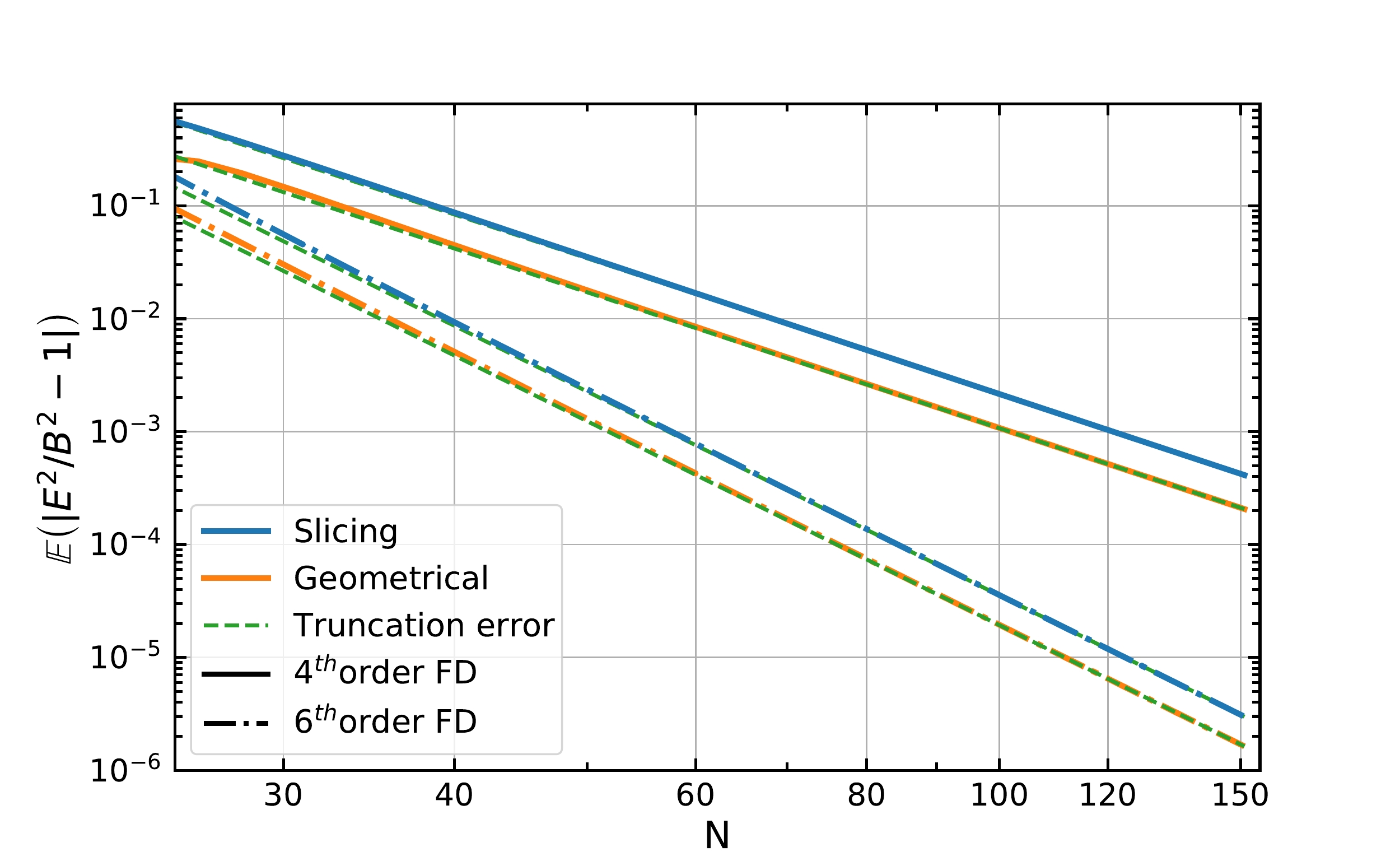}
    \caption{Average relative difference between $B^2$ and $E^2$ for both the slicing and geometrical codes applied to the Bianchi IV plane wave spacetime, where $B^2=E^2$, with data boxes of N${}^3$ grid points. Both $4^{th}$ and $6^{th}$ order FD schemes were used.}
    \label{fig: Plane Wave B2vsE2}
\end{figure}

The derivatives are computed with FD schemes, this introduces truncation errors that decrease as resolution increases. This follows the power law $N^{-o}$, with $N^3$ the number of grid points, and $o$ the order of the FD scheme. In Figs.~(\ref{fig: Error}, \ref{fig: Erroro6}, \ref{fig: Plane Wave B2vsE2}, \ref{fig: FDtest}) this power law is shown with dashed green lines. The codes' capacity to numerically compute derivatives will be determined by the dependence of the metric on the time and space coordinates (time coordinate only for the geometrical method). 
\begin{itemize}
    \item The Bianchi II metric in Eq.~(\ref{eq: BianchiIIspatialmetric}) has a polynomial dependence on $z$ and $t$. The top plots of Fig.~(\ref{fig: Error}) show that the spatial dependence is not an issue as the slicing method is not limited by the truncation error, however, the additional FD for the time derivatives is the limiting factor in the geometrical method. Even the change in the temporal powers of Eq.~(\ref{eq: BianchiIIspatialmetric}) that results from changing the $\gamma$-law index from $1$ to $4/3$ has changed the $B^2$ error to being truncation dominated.
    \item The test metric in Eq.~(\ref{eq: testmetric}) has a sinusoidal dependence on $z$ and linear dependence on $t$. The middle right plots in Fig.~(\ref{fig: Error}) shows that the truncation error is the limiting factor here, the error decreases following $4^{th}$ order convergence, as predicted by the FD order used in this figure. 
    \item The $\Lambda$-Szekeres metric in Eq.~(\ref{eq: SZmetric}) is sinusoidal along $z$ and paraboloidal in the orthogonal direction and its time dependence follows hyperbolic functions. The middle left plots in Fig.~(\ref{fig: Error}) show that the truncation error is a limiting factor, it indeed follows the expected power law, occasionally with an even steeper slope (showing better convergence). More on this figure in the floating point error section.
    \item Both Bianchi VI and Bianchi IV plane wave metrics in Eq.~(\ref{eq: BianchiVImetric}) and Eq.~(\ref{eq: planewavemetric}) have an exponential spatial distribution. The bottom row plots of Fig.~(\ref{fig: Error}) are similar as they both decrease with $4^{th}$ order convergence. For these cases, a Christoffel component of interest, and its derivative, are also displayed to demonstrate the errors introduced by the FD scheme. Being inhomogeneous and zero in certain locations there are bumps in these curves. These Christoffel components are limited by the FD order, so they benefit from the $6^{th}$ order scheme as seen in the top row of Fig.~(\ref{fig: Erroro6}). In this figure, the error in the Christoffel components manages to reach lower values, therefore decreasing the errors in the other terms as they all have $6^{th}$ order convergence. This behaviour is also visible in Fig.~(\ref{fig: Plane Wave B2vsE2}). More on this in the cancellation error section.
\end{itemize}
Whether the spacetime is inhomogeneous ($\Lambda$-Szekeres and test metric) or homogeneous (other spacetimes) does not seem to make much of a difference on the truncation error, only extra bumps along the curves. However, awareness of the metric spatial and temporal dependence is needed to understand the impact of the truncation error. If the space dependence is simple, as is the case of the Bianchi II metric, then the slicing method is preferred. Otherwise, if the space dependence is challenging, as is the case for the Bianchi VI and plane wave cases, the higher order FD method ought to be used for more accurate results, more on this in \ref{sec: append FDtest}.

\subsubsection{Floating point error} \label{sec: Floating point error}
Floating point error or round-off error comes from the limited number of digits stored in the computer memory. It accumulates as the amount of handled numbers and computational steps increases. Consequently, this type of error grows with the resolution, as it is visible in the top plots and middle left plot of Fig.~(\ref{fig: Error}), the increasing slopes follow power laws between $N^0$ and $N^{2.5}$. To ensure this is a floating point error and not a coding error we change the computational precision from 64bit to 128bit (dash-dotted lines). In all cases the amplitude of these dash-dotted lines is smaller, confirming the origin of this error. The $\Lambda$-Szekeres case is an interesting example where the transition from truncation error to floating point error is visible. The precision change decreases the amount of floating point error, therefore, pushing the error transition to happen at a higher resolution. This type of error displays computational limitations, however in all cases, it remains very small, so this does not pose much concern to results obtained with our codes.

\subsubsection{Cancellation error} \label{sec: Cancellation error}
When comparing large numbers to small ones the relative error may mislead the result if the error on the large number is of the same order of magnitude as the small number. Large numbers cancelling each other out will then introduce significant errors in the rest of the computation. A good example of this type of error arises in the computation of the Bianchi identity Eq.~(\ref{eq:BianchiIdentity}). In the Bianchi IV plane wave case, see bottom right in Fig.~(\ref{fig: Error}) and right side of Fig.~(\ref{fig: Erroro6}), each of the Riemann components in Eq.~(\ref{eq:BianchiIdentity}) are $\sim 10^{3}$, say the relative error is $10^{-3}$ from truncation error, then the introduced cancellation error is of $\sim 10^0$. It is then multiplied with smaller numbers and gives the error in the trace $B^T$ (this should be zero and is indeed negligible in all other cases). This error can be related to the truncation error so the cancellation error here decreases as the former gets corrected. This can then be improved by increasing the FD order, as seen from $4^{th}$ order FD Fig.~(\ref{fig: Error}) to $6^{th}$ order Fig.~(\ref{fig: Erroro6}) (top row) where the error in the Bianchi identity and other quantities in the plot significantly decrease. Additionally, this can also be improved by enforcing the symmetries of the Riemann tensor, see bottom row of Fig.~(\ref{fig: Erroro6}), where the Bianchi identity is enforced and $B^T$ vanishes and the $B^2$ error decreases. This additional step does not make a difference in the slicing method or the Bianchi VI case, i.e.\ symmetries of the Riemann or Ricci tensors are not limiting issues in the slicing code or when the magnetic part is small with respect to the electric part.

\begin{figure}
    \centering
    \includegraphics[width=0.65\linewidth]{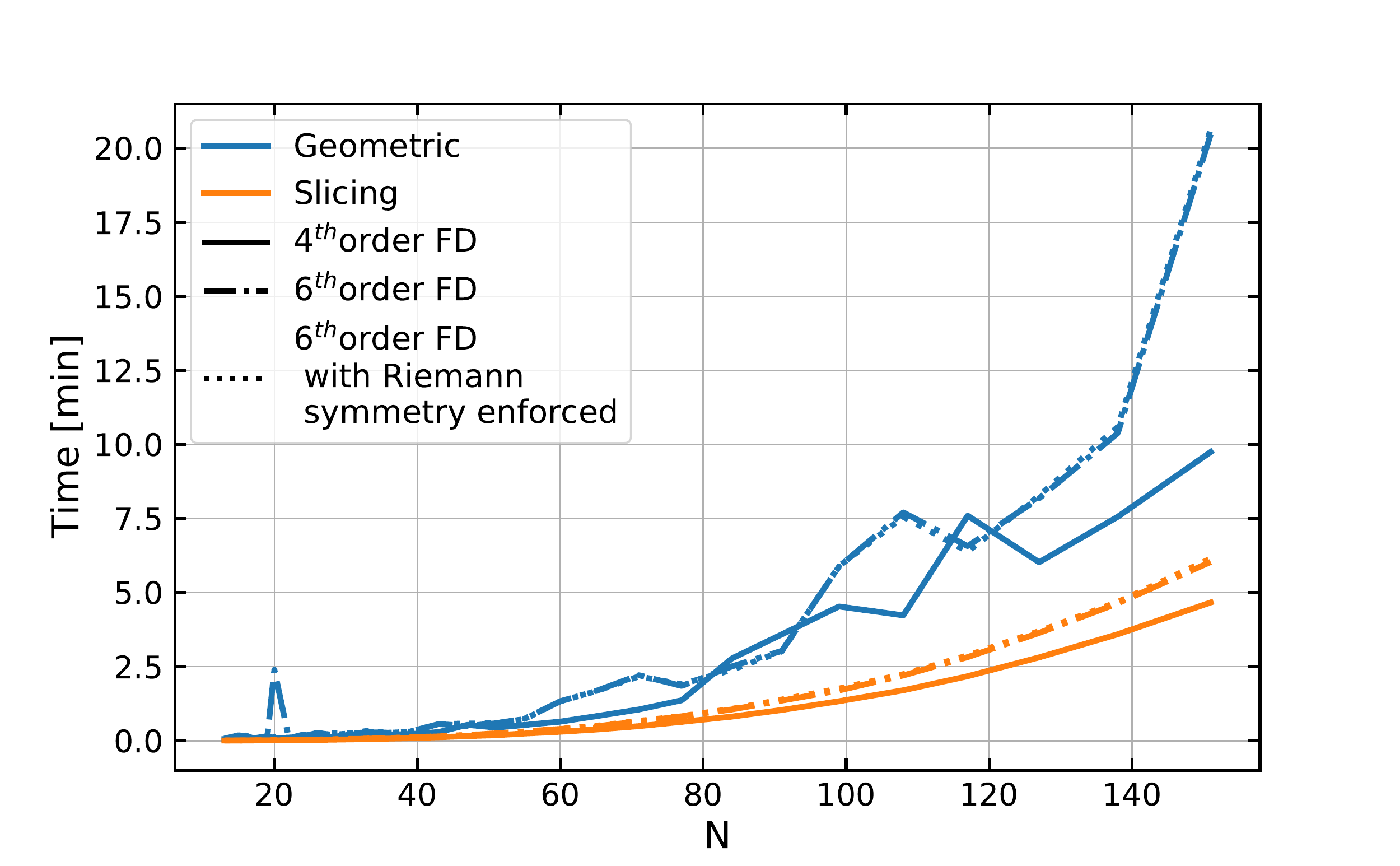}
    \caption{Computing time of both codes applied to data boxes of N${}^3$ grid points at a given time, with both $4^{th}$ and $6^{th}$ order FD and Riemann symmetry enforcing.}
    \label{fig: time}
\end{figure}

\subsubsection{Performance comparison} \label{sec: Performance comparison}
The Bianchi II and $\Lambda$-Szekeres cases show that the slicing method outperforms the geometrical one because of the temporal FD. Those additional steps introduce more truncation errors and significantly increase the computing time of the geometrical method as seen in Fig.~(\ref{fig: time}). To manage time derivatives over the entire data box, data files are appropriately written and read making the computing time curve bumpy. Additionally, the higher order FD needs more time steps, this then significantly increases computing time for the geometrical code, but not for the slicing code. For computational cost and ease of independently treating simulation time steps, the slicing method is therefore preferred here.

However, the results of the two methods are comparable in the test metric, Bianchi VI and Bianchi IV plane wave cases. This is reflected in Fig.~(\ref{fig: Plane Wave B2vsE2}), where $B^2$ and $E^2$ results are compared as they should be equal. The geometric code computes $E^2$ more accurately, and when the $B^2$ error is large it is very close to the slicing $E^2$ error, making the geometrical code perform better in Fig.~(\ref{fig: Plane Wave B2vsE2}). Yet the slicing code is very close and given the points raised above, it remains the preferred method \cite{R.L.Munoz_2022_ebweyl}.

\section{Conclusions} \label{sec: Conclusion}

The Riemann tensor can be expressed in terms of the Ricci tensor and the electric and magnetic parts $E_{\alpha\beta}$ and $B_{\alpha\beta}$ of the Weyl tensor. Here we have presented two methods to compute $E_{\alpha\beta}$ and $B_{\alpha\beta}$ and the Ricci scalars ${}^{(4)}R$ and ${}^{(3)}R$ from numerical relativity simulations, along with further scalar invariants that can be used to invariantly characterise any spacetime and to classify it according to the Petrov type. 
The first method is geometrical as it computes these quantities in full from the metric, and the second, which we dub slicing, uses the 3+1 decomposition of the metric. 
Special care has been taken to not introduce the constraint equations into the expressions in the slicing method. 
However, for the electric part of the Weyl tensor, Einstein's equations were necessary, therefore, this could potentially be a caveat when applying this method to simulation results, possibly introducing extra numerical error.

A post-processing Python code has been developed for each method, they have been applied to the $\Lambda$-Szekeres spacetime in Section~\ref{sec: Szekeres invariants}. We have shown that where $\beta_+$ is strongest we find negative curvature and a strong electric part, then when it is small the curvature tends to flatness and the $E_{\alpha\beta}$ is weak. 
As it is well known, the magnetic part vanishes and the spacetime is of Petrov type D, everywhere but when $\beta_+=0$ where it is of Petrov type O \cite{N.Meures_M.Bruni_2011_Jun}. 
We have verified this with our codes and this is a demonstration of their applicability.

We have tested our two codes on the five different spacetimes introduced in Section~\ref{sec: Example spacetimes}. 
The results, in Section~\ref{sec: Code tests}, show the presence of truncation, floating point and cancellation error depending on the spatial and temporal distribution of the metric. 
In the most challenging cases, we make higher-order FD schemes and Riemann symmetry enforcing available. With all best efforts introduced, in the most difficult case, we can report a relative error of $~10^{-4}$ for a box with $100^3$ points, and the relative error continues to decrease for higher resolution. 
But one should keep in mind that the numerical error we find depends on the considered case, in less challenging scenarios we find smaller errors. 
Then, when applying these codes to simulation results, one would also need to consider the accuracy of the simulation results. 
Should a $4^{th}$ order Runge Kutta scheme be used to evolve a simulation, then one could not expect better than $4^{th}$ order convergence on variables computed with these post-processing codes (even if the $6^{th}$ order FD scheme is used).

For three of the spacetimes we considered, our tests show that both methods have comparable performance, however in the two other ones the slicing method outperforms the geometrical one. 
Then when considering the computing time, the slicing method drastically outperforms the geometrical one. 
This is because of the additional FD scheme required by the geometrical method. 
On the basis of it's capacities demonstrated here, we have made the slicing post-processing code EBWeyl available in {\tt github} \cite{R.L.Munoz_2022_ebweyl}; it is applicable to any spacetime in any gauge.

While these methods and codes were developed for post-processing numerical relativity simulations, in this paper they were solely tested on exact solutions. 
We leave showing the applicability of the EBWeyl code on cosmological simulation results to our next paper. 
Finally, we remark that the use of EBWeyl is not limited to numerical relativity simulations, as it can be applied to any spacetime obtained numerically.

\ack
We thank Kazuya Koyama and Helvi Witek for useful suggestions during the development of this work and Emanuele Berti, Donato Bini, Ian Hawke and Hayley J. Macpherson for comments on the final draft.
RLM is grateful to Ian W. Harry for guidance on the presentation of EBWeyl on {\tt github}, Coleman Krawczyk for help identifying floating point error and to Obinna Umeh for help in using symbolic computing softwares.
MB has been supported by UK STFC Grant No. ST/S000550/1.
RLM has been supported by UK STFC studentship grants  ST/S505651/1 and ST/T506345/1 and by University of Portsmouth funding.
Some numerical computations were done on the Sciama High Performance Compute (HPC) cluster \cite{Sciama} which is supported by the ICG, SEPNet and the University of Portsmouth.
For the purpose of open access, the authors have applied a Creative Commons Attribution (CC BY) licence to any Author Accepted Manuscript version arising. Supporting research data are available on reasonable request from the corresponding author.

\section*{References}
\bibliography{Ref}

\providecommand{\newblock}{}
\begin{thebibliography}{100}
\expandafter\ifx\csname url\endcsname\relax
  \def\url#1{{\tt #1}}\fi
\expandafter\ifx\csname urlprefix\endcsname\relax\def\urlprefix{URL }\fi
\providecommand{\eprint}[2][]{\url{#2}}

\bibitem{R.L.Munoz_2022_ebweyl}
Munoz R~L 2022 {EBW}eyl \urlprefix\url{https://github.com/robynlm/ebweyl}

\bibitem{M.Alcubierre_etal_2015}
Alcubierre M, de~la Macorra A, Diez-Tejedor A and Torres J~M 2015 {\em Physical
  Review D\/} {\bf 92} 063508 {\bf 92} 063508 (\textit{Preprint}
  \eprint{gr-qc/1501.06918})

\bibitem{J.C.Aurrekoetxea_etal_2019}
Aurrekoetxea J~C, Clough K, Flauger R and Lim E~A 2020 {\em Journal of
  Cosmology and Astroparticle Physics\/} {\bf 2020} 030 {\bf 2020} 030
  (\textit{Preprint} \eprint{1910.12547})

\bibitem{E.Bentivegna_M.Bruni_2016}
Bentivegna E and Bruni M 2016 {\em Physical Review Letters\/} {\bf 116} 251302
  {\bf 116} 251302 (\textit{Preprint} \eprint{1511.05124})

\bibitem{J.Braden_etal_2017}
Braden J, Johnson M~C, Peiris H~V and Aguirre A 2017 {\em Physical Review D\/}
  {\bf 96} 023541 {\bf 96} 023541 (\textit{Preprint} \eprint{1604.04001})

\bibitem{J.Centrella_1980}
Centrella J 1980 {\em The Astrophysical Journal\/} {\bf 241} 875--885 {\bf 241}
  875--885

\bibitem{K.Clough_E.A.Lim_2016}
Clough K and Lim E~A 2016 Critical phenomena in non-spherically symmetric
  scalar bubble collapse (\textit{Preprint} \eprint{1602.02568})

\bibitem{W.E.East_etal_2018}
East W~E, Wojtak R and Abel T 2017 {\em Physical Review D\/} {\bf 97} 043509
  {\bf 97} 043509 (\textit{Preprint} \eprint{1711.06681})

\bibitem{J.T.Giblin_etal_2016}
Giblin J~T, Mertens J~B and Starkman G~D 2016 {\em Physical Review Letters\/}
  {\bf 116} 251301 {\bf 116} 251301 (\textit{Preprint} \eprint{1511.01105})

\bibitem{H.Kurki-Suonio_etal_1987}
Kurki-Suonio H, Matzner R~A, Centrella J and Wilson J~R 1987 {\em Physical
  Review D\/} {\bf 35} 435--448 {\bf 35} 435--448

\bibitem{X-X.Kou_2021}
Kou X~X, Mertens J~B, Tian C and Zhou S~Y 2022 {\em Physical Review D\/} {\bf
  105} 123505 {\bf 105} 123505 (\textit{Preprint} \eprint{gr-qc/2112.07626})

\bibitem{H.J.Macpherson_etal_2018}
Macpherson H~J, Lasky P~D and Price D~J 2018 {\em The Astrophysical Journal\/}
  {\bf 865} L4 {\bf 865} L4 (\textit{Preprint} \eprint{1807.01714})

\bibitem{H.J.Macpherson_2022}
Macpherson H~J 2022  (\textit{Preprint} \eprint{2209.06775})

\bibitem{J.Rekier_etal_2015}
Rekier J, Cordero-Carri\'on I and F\"uzfa A 2015 {\em Physical Review D\/} {\bf
  91} 024025 {\bf 91} 024025 (\textit{Preprint} \eprint{1409.3476})

\bibitem{F.Staelens_etal_2019}
Staelens F, Rekier J and F\"uzfa A 2021 {\em General Relativity and
  Gravitation\/} {\bf 53} 38 {\bf 53} 38 (\textit{Preprint}
  \eprint{gr-qc/1912.00677})

\bibitem{J.M.Torres_etal_2014}
Torres J~M, Alcubierre M, Diez-Tejedor A and Núñez D 2014 {\em Physical
  Review D\/} {\bf 90} 123002 {\bf 90} 123002 (\textit{Preprint}
  \eprint{1409.7953})

\bibitem{J.Adamek_etal_2016}
Adamek J, Daverio D, Durrer R and Kunz M 2016 {\em Nature Physics\/} {\bf 12}
  346--349 {\bf 12} 346--349 (\textit{Preprint} \eprint{1509.01699})

\bibitem{W.E.East_etal_2019}
East W~E, Wojtak R and Pretorius F 2019 {\em Physical Review D\/} {\bf 100}
  103533 {\bf 100} 103533 (\textit{Preprint} \eprint{1908.05683})

\bibitem{C.Barrera-Hinojosa_B.Li_2020_Jan}
Barrera-Hinojosa C and Li B 2020 {\em Journal of Cosmology and Astroparticle
  Physics\/} {\bf 2020} 007 {\bf 2020} 007 (\textit{Preprint}
  \eprint{1905.08890})

\bibitem{C.Barrera-Hinojosa_B.Li_2020_Apr}
Barrera-Hinojosa C and Li B 2020 {\em Journal of Cosmology and Astroparticle
  Physics\/} {\bf 2020} 056 {\bf 2020} 056 (\textit{Preprint}
  \eprint{2001.07968})

\bibitem{C.Barrera-Hinojosa_etal_2021_Jan}
Barrera-Hinojosa C, Li B, Bruni M and He J 2021 {\em Monthly Notices of the
  Royal Astronomical Society\/} {\bf 501} 5697--5713 {\bf 501} 5697--5713
  (\textit{Preprint} \eprint{2010.08257})

\bibitem{F.Lepori_etal_2022}
Lepori F, Schulz S, Adamek J and Durrer R 2022  (\textit{Preprint}
  \eprint{2209.10533})

\bibitem{C.Guandalin_etal_2020}
Guandalin C, Adamek J, Bull P, Clarkson C, Abramo L~R and Coates L 2020 {\em
  Monthly Notices of the Royal Astronomical Society\/} {\bf 501} 2547--2561
  {\bf 501} 2547--2561 (\textit{Preprint} \eprint{2009.02284})

\bibitem{L.Coates_etal_2020}
Coates L, Adamek J, Bull P, Guandalin C and Clarkson C 2021 {\em Monthly
  Notices of the Royal Astronomical Society\/} {\bf 504} 3534--3543 {\bf 504}
  3534--3543 (\textit{Preprint} \eprint{2011.12936})

\bibitem{H.J.Macpherson_A.Heinesen_2021}
Macpherson H and Heinesen A 2021 {\em Physical Review D\/} {\bf 104} 023525
  {\bf 104} 023525 (\textit{Preprint} \eprint{2103.11918})

\bibitem{R.Arnowitt_etal_2008}
Arnowitt R, Deser S and Misner C~W 2008 {\em General Relativity and
  Gravitation\/} {\bf 40} 1997–2027 {\bf 40} 1997–2027

\bibitem{M.Alcubierre_2008}
Alcubierre M 2008 {\em Introduction to 3+1 Numerical Relativity\/} (Oxford
  Science Publications)

\bibitem{M.Shibata_2015}
Shibata M 2015 {\em Numerical Relativity\/} (World Scientific Publishing
  Company)

\bibitem{J.T.Giblin_etal_2019}
Giblin J~T, Mertens J~B, Starkman G~D and C T 2019 {\em Physical Review D\/}
  {\bf 99} 023527 {\bf 99} 023527 (\textit{Preprint} \eprint{1810.05203})

\bibitem{C.Tian_etal_2020}
Tian C, Anselmi S, Carney M~F, Giblin J~T, Mertens J~B and Strakman G~D 2020
  {\em Physical Review D\/} {\bf 103} 083513 {\bf 103} 083513
  (\textit{Preprint} \eprint{2010.07274})

\bibitem{J.Adamek_etal_2020}
Adamek J, Barrera-Hinojosa C, Bruni M, Li B, Macpherson H~J and Mertens J~B
  2020 {\em Classical and Quantum Gravity\/} {\bf 37} 154001 {\bf 37} 154001
  (\textit{Preprint} \eprint{2003.08014v2})

\bibitem{R.D'Inverno_R.Russell-Clark_1971}
D'Inverno R and Russell-Clark R 1971 {\em Journal of Mathematical Physics\/}
  {\bf 12} 1258--1263 {\bf 12} 1258--1263

\bibitem{A.Karlhede_1980}
Karlhede A 1980 {\em General Relativity and Gravitation\/} {\bf 12} 693--707
  {\bf 12} 693--707

\bibitem{J.Carminati_R.G.McLenaghan_1991}
Carminati J and McLenaghan R~G 1991 {\em Journal of Mathematical Physics\/}
  {\bf 32} 3135--3140 {\bf 32} 3135--3140

\bibitem{C.B.G.McIntosh_etal_1995}
McIntosh C~B~G, Arianrhod R, Wade S~T and Hoenselaers C 1995 {\em Classical and
  Quantum Gravity\/} {\bf 11} 1555--1564 {\bf 11} 1555--1564

\bibitem{W.B.Bonnor_1995}
Bonnor W~B 1995 {\em Classical and Quantum Gravity\/} {\bf 12} 499--502 {\bf
  12} 499--502

\bibitem{E.Zakhary_C.B.G.McIntosh_1997}
Zakhary E and McIntosh C~B~G 1997 {\em General Relativity and Gravitation\/}
  {\bf 29} 539--581 {\bf 29} 539--581

\bibitem{H.Stephani_etal_2003}
Stephani H, Kramer D, MacCallum M, Hoenselaers C and Herlt E 2003 {\em Exact
  Solutions of Einstein's Field Equations\/} (Cambridge University Press)

\bibitem{L.Wylleman_etal_2019}
Wylleman L, Coley A, McNutt D and Aadne M 2019 {\em Classical and Quantum
  Gravity\/} {\bf 36} 235018 {\bf 36} 235018 (\textit{Preprint}
  \eprint{2007.15915})

\bibitem{D.Bini_etal_2021}
Bini D, Geralico A and Jantzen R~T 2021  (\textit{Preprint}
  \eprint{gr-qc/2111.01283v2})

\bibitem{C.Rovelli_1991}
Rovelli C 1991 {\em Classical and Quantum Gravity\/} {\bf 8} 297 {\bf 8} 297

\bibitem{A.Matte_1953}
Matte A 1953 {\em Canadian Journal of Mathematics\/} {\bf 5} 1--16 {\bf 5}
  1--16

\bibitem{P.Jordan_etal_1964}
Jordan P, Beiglb{\"o}ck W, Bichteler K, Budich W, Kundt W and Tr{\"u}mper M
  1964 Contributions to actual problems of general relativity Airforce Report,
  University of Hamburg

\bibitem{S.W.Hawking_1966}
Hawking S~W 1966 {\em The Astrophysical Journal\/} {\bf 145} 544 {\bf 145} 544

\bibitem{J.M.Stewart_M.Walker_1974}
Stewart J~M and Walker M 1974 {\em Proceedings of the Royal Society of London.
  A. Mathematical and Physical Sciences\/} {\bf 341} 49--74 {\bf 341} 49--74

\bibitem{G.F.R.Ellis_M.Bruni_1989}
Ellis G~F~R and Bruni M 1989 {\em Physical Review D\/} {\bf 40} 1804--1818 {\bf
  40} 1804--1818

\bibitem{M.Bruni_etal_1992}
Bruni M, Dunsby P~K~S and Ellis G~F~R 1992 {\em The Astrophysical Journal\/}
  {\bf 395} 34--53 {\bf 395} 34--53

\bibitem{J.M.Bardeen_1980}
Bardeen J~M 1980 {\em Physical Review D\/} {\bf 22} 1882--1905 {\bf 22}
  1882--1905

\bibitem{R.Maartens_B.A.Bassett_1998}
Maartens R and Bassett B~A 1998 {\em Classical and Quantum Gravity\/} {\bf 15}
  705--717 {\bf 15} 705--717 (\textit{Preprint} \eprint{gr-qc/9704059})

\bibitem{G.F.R.Ellis_2009}
Ellis G~F~R 2009 {\em General Relativity and Gravitation\/} {\bf 41} 581--660
  {\bf 41} 581--660

\bibitem{G.F.R.Ellis_etal_2012}
Ellis G~F~R, Maartens R and MacCallum M~A~H 2012 {\em Relativistic Cosmology\/}
  (Cambridge University Press)

\bibitem{R.Owen_etal_2011}
Owen R, Brink J, Chen Y, Kaplan J~D, Lovelace G, Matthews K~D, Nichols D~A,
  Scheel M~A, Zhang F, Zimmerman A and Thorne K~S 2011 {\em Physical Review
  Letters\/} {\bf 106} 151101 {\bf 106} 151101 (\textit{Preprint}
  \eprint{1012.4869})

\bibitem{M.Korzynski_etal_2015}
Korzyński M, Hinder I and Bentivegna E 2015 {\em Journal of Cosmology and
  Astroparticle Physics\/} {\bf 2015} 25--25 {\bf 2015} 25--25
  (\textit{Preprint} \eprint{1505.05760})

\bibitem{T.Clifton_etal_2017}
Clifton T, Gregoris D and Rosquist K 2017 {\em General Relativity and
  Gravitation\/} {\bf 49} 30 {\bf 49} 30 (\textit{Preprint}
  \eprint{1607.00775})

\bibitem{A.Heinesen_H.J.Macpherson_2022}
Heinesen A and Macpherson H~J 2022 {\em Journal of Cosmology and Astroparticle
  Physics\/} {\bf 2022} 57 {\bf 2022} 57 (\textit{Preprint}
  \eprint{2111.14423})

\bibitem{M.Bruni_etal_2014_Feb}
Bruni M, Thomas D~B and Wands D 2014 {\em Physical Review D\/} {\bf 89} 044010
  {\bf 89} 044010 (\textit{Preprint} \eprint{1306.1562})

\bibitem{I.Milillo_etal_2015}
Milillo I, Bertacca D, Bruni M and Maselli A 2015 {\em Physical Review D\/}
  {\bf 92} 023519 {\bf 92} 023519 (\textit{Preprint} \eprint{gr-qc/1502.02985})

\bibitem{D.B.Thomas_etal_2015_16Jul}
Thomas D~B, Bruni M and Wands D 2015 {\em Monthly Notices of the Royal
  Astronomical Society\/} {\bf 452} 1727--1742 {\bf 452} 1727--1742
  (\textit{Preprint} \eprint{1501.00799})

\bibitem{D.B.Thomas_etal_2015_30Jul}
Thomas D~B, Bruni M, Koyama K, Li B and Zhao G~B 2015 {\em Journal of Cosmology
  and Astroparticle Physics\/} {\bf 07} 051 {\bf 07} 051 (\textit{Preprint}
  \eprint{gr-qc/1503.07204})

\bibitem{C.Barrera-Hinojosa_etal_2021_Dec}
Barrera-Hinojosa C, Li B and Cai Y~C 2021 {\em Monthly Notices of the Royal
  Astronomical Society\/} {\bf 510} 3589--3604 {\bf 510} 3589--3604
  (\textit{Preprint} \eprint{2109.02632})

\bibitem{P.Szekeres_1975}
Szekeres P 1975 {\em Communications in Mathematical Physics\/} {\bf 41} 55--64
  {\bf 41} 55--64

\bibitem{J.D.Barrow_J.Stein-Schabes_1984}
Barrow J~D and Stein-Schabes J 1984 {\em Physics Letters A\/} {\bf 103}
  315--317 {\bf 103} 315--317

\bibitem{Maple}
{Maplesoft, a division of Waterloo Maple Inc} Maple
  \urlprefix\url{https://hadoop.apache.org}

\bibitem{R.M.Wald_1984}
Wald R~M 1984 {\em General Relativity\/} (The University of Chicago Press)

\bibitem{G.F.R.Ellis_etal_1989}
Ellis G~F~R, Bruni M and Hwang J 1990 {\em Physical Review D\/} {\bf 42}
  1035--1046 {\bf 42} 1035--1046

\bibitem{L.Gunnarsen_etal_1995}
Gunnarsen L, Hisa-Aki S and Kei-Ichi M 1995 {\em Classical and Quantum
  Gravity\/} {\bf 12} 133--140 {\bf 12} 133--140

\bibitem{Y.Choquet-Bruhat_2015}
Choquet-Bruhat Y 2015 {\em Introduction to General Relativity, Black Holes, and
  Cosmology\/} (Oxford University Press)

\bibitem{J.Wainwright_G.F.R.Ellis_1997}
Wainwright J and Ellis G~F~R 1997 {\em Dynamical Systems in Cosmology\/}
  (Cambridge University Press)

\bibitem{P.K.S.Dunsby_etal_1992}
Dunsby P~K~S, Bruni M and Ellis G~F~R 1992 {\em The Astrophysical Journal\/}
  {\bf 395} 54--73 {\bf 395} 54--73

\bibitem{A.R.King_G.F.R.Ellis_1973}
King A~R and Ellis G~F~R 1973 {\em Communications in Mathematical Physics\/}
  {\bf 31} 209--242 {\bf 31} 209--242

\bibitem{D.Bini_etal_1995}
Bini D, Carini P and Jantzen R~T 1995 {\em Classical and Quantum Gravity\/}
  {\bf 12} 2549--2563 {\bf 12} 2549--2563

\bibitem{J.Baker_M.Campanelli_2000}
Baker J and Campanelli M 2000 {\em Physical Review D\/} {\bf 62} 127501 {\bf
  62} 127501 (\textit{Preprint} \eprint{gr-qc/0003031})

\bibitem{A.Coley_etal_2021}
Coley A, Peters J~M and Schnetter E 2021 {\em Classical and Quantum Gravity\/}
  {\bf 38} 17LT01 {\bf 38} 17LT01 (\textit{Preprint} \eprint{2108.04210})

\bibitem{A.Barnes_R.R.Rowlingson_1989}
Barnes A and Rowlingson R~R 1989 {\em Classical and Quantum Gravity\/} {\bf 6}
  949--960 {\bf 6} 949--960

\bibitem{C.Cherubini_etal_2004}
Cherubini C, Bini D, Bruni M and Perjes Z 2004 {\em Classical and Quantum
  Gravity\/} {\bf 21} 4833--4843 {\bf 21} 4833--4843 (\textit{Preprint}
  \eprint{gr-qc/0404075v1})

\bibitem{C.Beetle_L.M.Burko_2002}
Beetle C and Burko L~M 2002 {\em Physical Review Letters\/} {\bf 89} 271101
  {\bf 89} 271101 (\textit{Preprint} \eprint{gr-qc/0210019})

\bibitem{E.Berti_etal_2005}
Berti E, White F, Maniopoulou A and Bruni M 2005 {\em Monthly Notices of the
  Royal Astronomical Society\/} {\bf 358} 923--938 {\bf 358} 923--938 ISSN
  0035-8711 (\textit{Preprint} \eprint{gr-qc/0405146})

\bibitem{R.Penrose_1960}
Penrose R 1960 {\em Annals of Physics\/} {\bf 10} 171--201 {\bf 10} 171--201

\bibitem{S.Weinberg_1972}
Weinberg S 1972 {\em Gravitation and Cosmology: Principles and Applications of
  the General Theory of Relativity\/} (John Wiley and Sons)

\bibitem{R.K.Sachs_1964}
Sachs R~K 1964 {\em Relativity, Groups, and Topology\/} (New York: Gordon and
  Breach) chap Gravitational radiation ed C. DeWitt and B. DeWitt

\bibitem{M.Bruni_etal_1997}
Bruni M, Matarrese S, Mollerach S and Sonego S 1997 {\em Classical and Quantum
  Gravity\/} {\bf 14} 2585--2606 {\bf 14} 2585--2606 (\textit{Preprint}
  \eprint{gr-qc/9609040})

\bibitem{S.Sonego_M.Bruni_1997}
Sonego S and Bruni M 1998 {\em Communications in Mathematical Physics\/} {\bf
  93} 209--218 {\bf 93} 209--218 (\textit{Preprint} \eprint{gr-qc/9708068})

\bibitem{M.Bruni_S.Sonego_1999}
Bruni M and Sonego S 1999 {\em Classical and Quantum Gravity\/} {\bf 16}
  L29--L36 {\bf 16} L29--L36 (\textit{Preprint} \eprint{gr-qc/9906017})

\bibitem{S.A.Teukolsky_1973}
Teukolsky S~A 1973 {\em The Astrophysical Journal\/} {\bf 185} 635--647 {\bf
  185} 635--647

\bibitem{P.Pani_2013}
Pani P 2013 {\em International Journal of Modern Physics A\/} {\bf 28} 1340018
  {\bf 28} 1340018 (\textit{Preprint} \eprint{gr-qc/1305.6759})

\bibitem{U.H.Gerlach_U.K.Sengupta_1978}
Gerlach U~H and Sengupta U~K 1978 {\em Physical Review D\/} {\bf 18} 1789--1797
  {\bf 18} 1789--1797

\bibitem{H.Kodama_M.Sasaki_1984}
Kodama H and Sasaki M 1984 {\em Progress of Theoretical and Experimental
  Physics\/} {\bf 78} 1--166 {\bf 78} 1--166

\bibitem{J.M.Stewart_1990}
Stewart J~M 1990 {\em Classical and Quantum Gravity\/} {\bf 7} 1169--1180 {\bf
  7} 1169--1180

\bibitem{S.W.Goode_1989}
Goode S~W 1989 {\em Physical Review D\/} {\bf 39} 2882--2892 {\bf 39}
  2882--2892

\bibitem{S.Matarrese_etal_1998}
Matarrese S, Mollerach S and Bruni M 1998 {\em Physical Review D\/} {\bf 58}
  043504 {\bf 58} 043504 (\textit{Preprint} \eprint{astro-ph/9707278})

\bibitem{S.Mollerach_S.Matarrese_1997}
Mollerach S and Matarrese S 1997 {\em Physical Review D\/} {\bf 56} 4494--4502
  {\bf 56} 4494--4502 (\textit{Preprint} \eprint{9702234})

\bibitem{R.Maartens_etal_1999}
Maartens R, Gebbie T and Ellis G~F~R 1999 {\em Physical Review D\/} {\bf 59}
  083506 {\bf 59} 083506 (\textit{Preprint} \eprint{astro-ph/9808163})

\bibitem{M.Campanelli_C.O.Lousto_1999}
Campanelli M and Lousto C~O 1999 {\em Physical Review D\/} {\bf 59} 124022 {\bf
  59} 124022 (\textit{Preprint} \eprint{gr-qc/9811019})

\bibitem{A.Garat_R.H.Price_2000}
Garat A and Price R~H 2000 {\em Physical Review D\/} {\bf 61} 044006 {\bf 61}
  044006 (\textit{Preprint} \eprint{gr-qc/9909005})

\bibitem{R.J.Gleiser_etal_2000}
Gleiser R~J, Nicasio C~O, Price R~H and Pullin J 2000 {\em Physics Report\/}
  {\bf 325} 41--81 {\bf 325} 41--81 (\textit{Preprint} \eprint{gr-qc/9807077})

\bibitem{C.G.Tsagas_etal_2008}
Tsagas C~G, Challinor A and Maartens R 2008 {\em Physics Reports\/} {\bf 465}
  61--147 {\bf 465} 61--147 (\textit{Preprint} \eprint{0705.4397})

\bibitem{K.A.Malik_D.Wands_2008}
Malik K~A and Wands D 2009 {\em Physics Reports\/} {\bf 475} 1--51 {\bf 475}
  1--51 (\textit{Preprint} \eprint{0809.4944})

\bibitem{K.A.Malik_D.R.Matravers_2008}
Malik K~A and Matravers D~R 2008 {\em Classical and Quantum Gravity\/} {\bf 25}
  193001 {\bf 25} 193001 (\textit{Preprint} \eprint{0804.3276})

\bibitem{K.D.Kokkotas_B.G.Schmidt_1999}
Kokkotas K~D and Schmidt B~G 1999 {\em Living Reviews in Relativity\/} {\bf 2}
  2 {\bf 2} 2 (\textit{Preprint} \eprint{gr-qc/9909058})

\bibitem{E.Berti_etal_2009}
Berti E, Cardoso V and Starinets A~O 2009 {\em Classical and Quantum Gravity\/}
  {\bf 26} 163001 {\bf 26} 163001 (\textit{Preprint} \eprint{gr-qc/0905.2975})

\bibitem{E.Berti_etal_2018}
Berti E, Yagi K, Yang H and Yunes N 2018 {\em General Relativity and
  Gravitation\/} {\bf 50} 49 {\bf 50} 49 (\textit{Preprint}
  \eprint{gr-qc/1801.03587})

\bibitem{A.Pound_B.Wardell_2021}
Pound A and Wardell B 2020 {\em Black hole perturbation theory and
  gravitational self-force\/} (Springer Singapore) pp 1--119 ed C. Bambi, S.
  Katsanevas, K. D. Kokkotas

\bibitem{C.A.Clarkson_2004}
Clarkson C~A 2004 {\em Physical Review D\/} {\bf 70} 103524 {\bf 70} 103524
  [Erratum: Phys.Rev.D 70, 129902 (2004)] (\textit{Preprint}
  \eprint{astro-ph/0311505})

\bibitem{N.Bartolo_etal_2004}
Bartolo N, Komatsu E, Matarrese S and Riotto A 2004 {\em Physics Report\/} {\bf
  402} 103--266 {\bf 402} 103--266 (\textit{Preprint}
  \eprint{astro-ph/0406398})

\bibitem{D.H.Lyth_etal_2005}
Lyth D~H, Malik K~A and Sasaki M 2005 {\em Journal of Cosmology and
  Astroparticle Physics\/} {\bf 2005} 004 {\bf 2005} 004 (\textit{Preprint}
  \eprint{astro-ph/0411220})

\bibitem{K.Nakamura_2007}
Nakamura K 2007 {\em Progress of Theoretical Physics\/} {\bf 117} 17--74 {\bf
  117} 17--74 (\textit{Preprint} \eprint{gr-qc/0605108})

\bibitem{H.Noh_J-C.Hwang_2004}
Noh H and Hwang J~C 2004 {\em Physical Review D\/} {\bf 69} 104011 {\bf 69}
  104011 (\textit{Preprint} \eprint{0305123})

\bibitem{B.Osano_etal_2007}
Osano B, Pitrou C, Dunsby P, Uzan J~P and Clarkson C 2007 {\em Journal of
  Cosmology and Astroparticle Physics\/} {\bf 2007} 003 {\bf 2007} 003
  (\textit{Preprint} \eprint{gr-qc/0612108})

\bibitem{G.W.Pettinari_etal_2013}
Pettinari G~W, Fidler C, Crittenden R, Koyama K and Wands D 2013 {\em Journal
  of Cosmology and Astroparticle Physics\/} {\bf 2013} 003 {\bf 2013} 003
  (\textit{Preprint} \eprint{1302.0832})

\bibitem{M.Bruni_etal_2014_Mar}
Bruni M, Hidalgo J~C, Meures N and Wands D 2014 {\em The Astrophysical
  Journal\/} {\bf 785} 2 {\bf 785} 2 (\textit{Preprint} \eprint{1307.1478})

\bibitem{E.Villa_C.Rampf_2015}
Villa E and Rampf C 2016 {\em Journal of Cosmology and Astroparticle Physics\/}
  {\bf 2016} 030 {\bf 2016} 030 [Erratum: JCAP 05, E01 (2018)]
  (\textit{Preprint} \eprint{gr-qc/1505.04782})

\bibitem{H.A.Gressel_M.Bruni_2018}
Gressel H~A and Bruni M 2018 {\em Journal of Cosmology and Astroparticle
  Physics\/} {\bf 2018} 016 {\bf 2018} 016 (\textit{Preprint}
  \eprint{1712.08687})

\bibitem{N.Loutrel_etal_2021}
Loutrel N, Ripley J~L, Giorgi E and Pretorius F 2021 {\em Physical Review D\/}
  {\bf 103} 104017 {\bf 103} 104017 (\textit{Preprint}
  \eprint{gr-qc/2008.11770})

\bibitem{J.L.Ripley_etal_2021}
Ripley J~L, Loutrel N, Giorgi E and Pretorius F 2021 {\em Physical Review D\/}
  {\bf 103} 104018 {\bf 103} 104018 (\textit{Preprint}
  \eprint{gr-qc/2010.00162})

\bibitem{M.H-Y.Cheung_etal_2022}
Cheung M~H~Y, Baibhav V, Berti E, Cardoso V, Carullo G, Cotesta R, Del~Pozzo W,
  Duque F, Helfer T, Shukla E and Wong K~W~K 2022

\bibitem{K.Mitman_etal_2022}
Mitman K, Lagos M, Stein L~C, Ma S, Hui L, Chen Y, Deppe N, Hébert F, Kidder
  L~E, Moxon J, Scheel M~A, Teukolsky S~A, Throwe W and Vu N~L 2022
  (\textit{Preprint} \eprint{2208.07380})

\bibitem{M.Bruni_etal_2003}
Bruni M, Gualtieri L and Sopuerta C~F 2003 {\em Classical and Quantum
  Gravity\/} {\bf 20} 535--556 {\bf 20} 535--556 (\textit{Preprint}
  \eprint{gr-qc/0207105})

\bibitem{C.F.Sopuerta_etal_2003}
Sopuerta C~F, Bruni M and Gualtieri L 2004 {\em Physical Review D\/} {\bf 70}
  064002 {\bf 70} 064002 (\textit{Preprint} \eprint{gr-qc/0306027})

\bibitem{C.Pitrou_etal_2015}
Pitrou C, Pereira T~S and Uzan J~P 2015 {\em Physical Review D\/} {\bf 92}
  023501 {\bf 92} 023501 (\textit{Preprint} \eprint{1503.01125})

\bibitem{A.Talebian-Ashkezari_etal_2018}
Talebian-Ashkezari A, Ahmadi N and Abolhasani A~A 2018 {\em Journal of
  Cosmology and Astroparticle Physics\/} {\bf 2018} 001 {\bf 2018} 001
  (\textit{Preprint} \eprint{gr-qc/1609.05893})

\bibitem{S.R.Goldberg_etal_2016}
Goldberg S~R, Clifton T and Malik K~A 2017 {\em Physical Review D\/} {\bf 95}
  043503 {\bf 95} 043503 (\textit{Preprint} \eprint{1610.08882})

\bibitem{A.Passamonti_etal_2005}
Passamonti A, Bruni M, Gualtieri L and Sopuerta C~F 2005 {\em Physical Review
  D\/} {\bf 71} 024022 {\bf 71} 024022 (\textit{Preprint}
  \eprint{gr-qc/0407108})

\bibitem{A.Passamonti_etal_2006}
Passamonti A, Bruni M, Gualtieri L, Nagar A and Sopuerta C~F 2006 {\em Physical
  Review D\/} {\bf 73} 084010 {\bf 73} 084010 (\textit{Preprint}
  \eprint{gr-qc/0601001})

\bibitem{A.Passamonti_etal_2007}
Passamonti A, Stergioulas N and Nagar A 2007 {\em Physical Review D\/} {\bf 75}
  084038 {\bf 75} 084038 (\textit{Preprint} \eprint{gr-qc/0702099})

\bibitem{C.F.Sopuerta_N.Yunes_2009}
Sopuerta C~F and Yunes N 2009 {\em Physical Review D\/} {\bf 80} 064006 {\bf
  80} 064006 (\textit{Preprint} \eprint{gr-qc/0904.4501})

\bibitem{M.Lenzi_C.F.Sopuerta_2021}
Lenzi M and Sopuerta C~F 2021 {\em Physical Review D\/} {\bf 104} 084053 {\bf
  104} 084053 (\textit{Preprint} \eprint{gr-qc/2108.08668})

\bibitem{C.Cherubini_etal_2005}
Cherubini C, Bini D, Bruni M and Perjes Z 2005 {\em Classical and Quantum
  Gravity\/} {\bf 22} 1763--1768 {\bf 22} 1763--1768 (\textit{Preprint}
  \eprint{gr-qc/0408040v1})

\bibitem{StarWars}
\urlprefix\url{https://www.starwars.com}

\bibitem{M.Grasso_etal_2021}
Grasso M, Villa E, Korzy'nski M and Matarrese S 2021 {\em Physical Review D\/}
  {\bf 104} 043508 {\bf 104} 043508 (\textit{Preprint} \eprint{2105.04552})

\bibitem{M.Grasso_E.Villa_2021}
Grasso M and Villa E 2021 {\em Classical and Quantum Gravity\/} {\bf 39} 015011
  {\bf 39} 015011 (\textit{Preprint} \eprint{gr-qc/2107.06306})

\bibitem{T.Buchert_etal_2022}
Buchert T, van Elst H and Heinesen A 2022  (\textit{Preprint}
  \eprint{gr-qc/2202.10798})

\bibitem{J.Yoo_R.Durrer_2017}
Yoo J and Durrer R 2017 {\em Journal of Cosmology and Astroparticle Physics\/}
  {\bf 2017} 16--16 {\bf 2017} 16--16 (\textit{Preprint} \eprint{1705.05839})

\bibitem{A.Z.Petrov_2000}
Petrov A~Z 2000 {\em General Relativity and Gravitation\/} {\bf 32} 1572--9532
  {\bf 32} 1572--9532

\bibitem{A.Barnes_2014}
Barnes A 2014 Einstein spacetimes with constant weyl eigenvalues
  (\textit{Preprint} \eprint{1409.4300})

\bibitem{J.Plebanski_A.Krasinski_2006}
Pleba\'nski J and Krasi\'nski A 2006 {\em An introduction to General Relativity
  and Cosmology\/} (Cambridge University Press)

\bibitem{S.Chandrasekhar_1992}
Chandrasekhar S 1992 {\em The Mathematical Theory of Black Holes\/} (Oxford
  University Press)

\bibitem{P.Szekeres_1965}
Szekeres P 1965 {\em Journal of Mathematical Physics\/} {\bf 6} 1387--1391 {\bf
  6} 1387--1391

\bibitem{A.Lewis_A.Challinor_2006}
Lewis A and Challinor A 2006 {\em Physics Reports\/} {\bf 429} 1--65 {\bf 429}
  1--65 (\textit{Preprint} \eprint{0601594})

\bibitem{F.A.E.Pirani_1957}
Pirani F~A~E 1957 {\em Physical Review\/} {\bf 105}(3) 1089--1099 {\bf 105}(3)
  1089--1099

\bibitem{J.A.Allnutt_1981}
Allnutt J 1981 {\em General Relativity and Gravitation\/} {\bf 13} 1017--1020
  {\bf 13} 1017--1020

\bibitem{W.Kinnersley_1969}
Kinnersley W 1969 {\em Journal of Mathematical Physics\/} {\bf 10} 1195--1203
  {\bf 10} 1195--1203

\bibitem{W.B.Bonnor_W.Davidson_1985}
Bonnor W~B and Davidson W 1985 {\em Classical and Quantum Gravity\/} {\bf 2}
  775--780 {\bf 2} 775--780

\bibitem{I.Robinson_A.Trautman_1962}
Robinson I and Trautman A 1962 {\em Proceedings of the Royal Society London
  A\/} {\bf 265} 463--473 {\bf 265} 463--473

\bibitem{A.Nerozzi_etal_2005}
Nerozzi A, Beetle C, Bruni M, Burko L~M and Pollney D 2005 {\em Physical Review
  D\/} {\bf 72} 024014 {\bf 72} 024014 (\textit{Preprint}
  \eprint{gr-qc/0407013})

\bibitem{A.Nerozzi_etal_2006}
Nerozzi A, Bruni M, Re V and Burko L~M 2006 {\em Physical Review D\/} {\bf 73}
  044020 {\bf 73} 044020 (\textit{Preprint} \eprint{gr-qc/0507068})

\bibitem{S.W.Goode_J.Wainwright_1982}
Goode S~W and Wainwright J 1982 {\em Physical Review D\/} {\bf 26}(12)
  3315--3326 {\bf 26}(12) 3315--3326

\bibitem{N.Meures_M.Bruni_2011_Jan}
Meures N and Bruni M 2012 {\em Monthly Notices of the Royal Astronomical
  Society\/} {\bf 419} 1937 {\bf 419} 1937 (\textit{Preprint}
  \eprint{1107.4433})

\bibitem{N.Meures_M.Bruni_2011_Jun}
Meures N and Bruni M 2011 {\em Physical Review D\/} {\bf 83} 123519 {\bf 83}
  123519 (\textit{Preprint} \eprint{1103.0501})

\bibitem{Planck_CMB_2018}
Collaboration P 2020 {\em Astronomy \& Astrophysics\/} {\bf 641} A5 {\bf 641}
  A5 (\textit{Preprint} \eprint{1907.12875})

\bibitem{M.Bruni_etal_2014_Sep}
Bruni M, Hidalgo J~C and Wands D 2014 {\em The Astrophysical Journal Letters\/}
  {\bf 794} L11 {\bf 794} L11 (\textit{Preprint} \eprint{1405.7006})

\bibitem{C.B.Collins_J.M.Stewart_1971}
Collins C~B and Stewart J~M 1971 {\em Monthly Notices of the Royal Astronomical
  Society\/} {\bf 153} 419--434 {\bf 153} 419--434

\bibitem{K.Rosquist_R.T.Jantzen_1985}
Rosquist K and Jantzen R~T 1985 {\em Physics Letters A\/} {\bf 107} 29--32 {\bf
  107} 29--32

\bibitem{A.Harvey_D.Tsoubelis_1977}
Harvey A and Tsoubelis D 1977 {\em Physical Review D\/} {\bf 15} 2734--2737
  {\bf 15} 2734--2737

\bibitem{A.Harvey_etal_1979}
Harvey A, Tsoubelis D and Wilsker B 1979 {\em Physical Review D\/} {\bf 20}
  2077--2078 {\bf 20} 2077--2078

\bibitem{B.Fornberg_1988}
Fornberg B 1988 {\em Mathematics of computation\/} {\bf 51} 699--706 {\bf 51}
  699--706

\bibitem{Sciama}
\urlprefix\url{http://www.sciama.icg.port.ac.uk/}

\end{thebibliography}

\appendix

\section{Finite Differencing test} \label{sec: append FDtest}

To understand the limitations of the FD schemes, we test them in 3 simple cases: a polynomial, a sinusoidal and an exponential, with both the $4^{th}$ and $6^{th}$ order schemes. 

\begin{figure}
    \centering
    \includegraphics[width=0.8\linewidth]{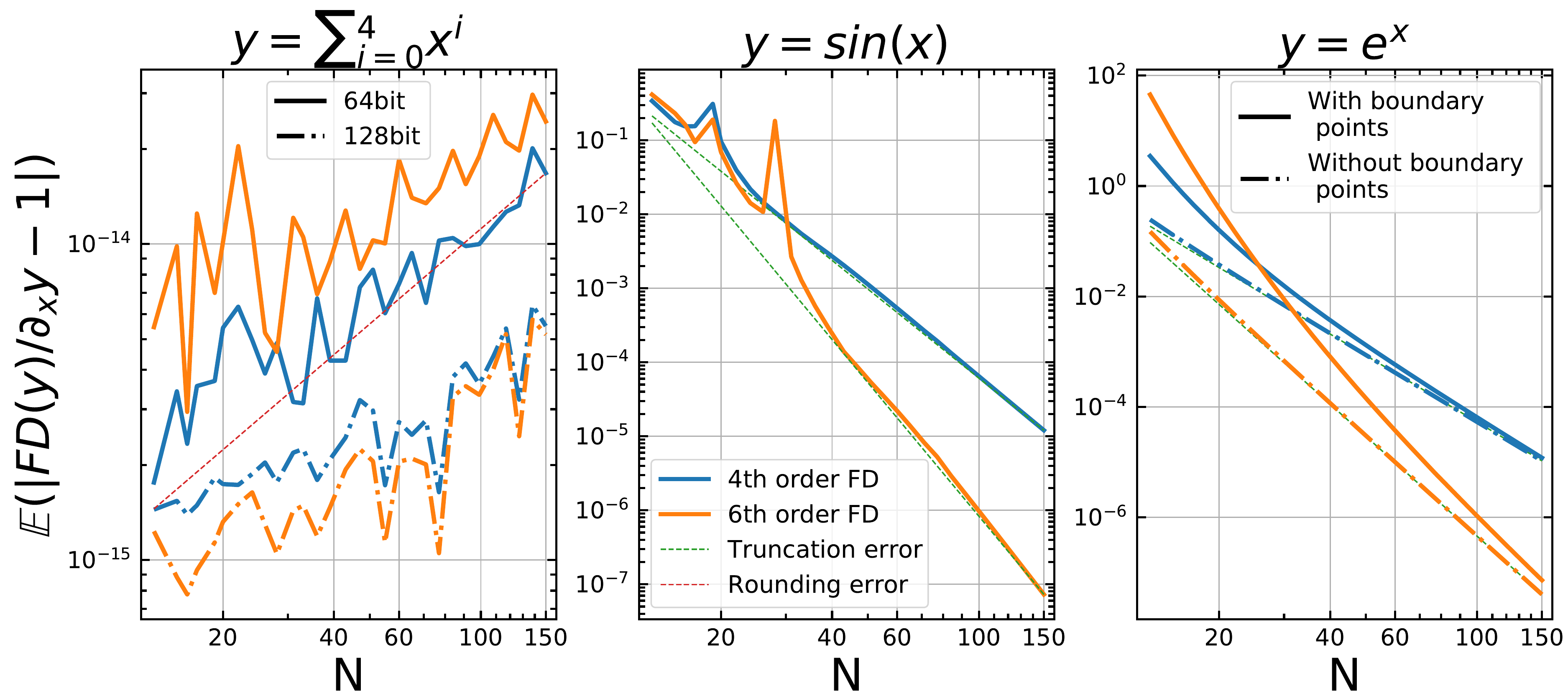}
    \caption{Average relative error of the FD scheme applied to a polynomial, sinusoidal and exponential as a function of the $N$ points in the data arrays. The FD uses a $4^{th}$ (blue) or $6^{th}$ (orange) order scheme. The dash-dotted lines on the left plot distinguish floating point precision, and on the right, they distinguish cases where the edge points calculated with forward and backward FD schemes are included or cut out.}
    \label{fig: FDtest}
\end{figure}

As seen in Fig.~(\ref{fig: FDtest}), the polynomial case is no challenge to the scheme, having a relative error of $10^{-14}$, and a rounding error increasing with a $N^1$ slope. This error does not originate from the FD approximation but from computing limitations, so here the higher order FD has a slightly higher error due to the additional operations. This error is reduced by increasing computing precision, as seen when we go from 64bit to 128bit. 

The sinusoidal case shows convergence according to the expected truncation error, so this benefits from increasing the FD order. 

\begin{figure}
    \centering
    \includegraphics[width=0.5\linewidth]{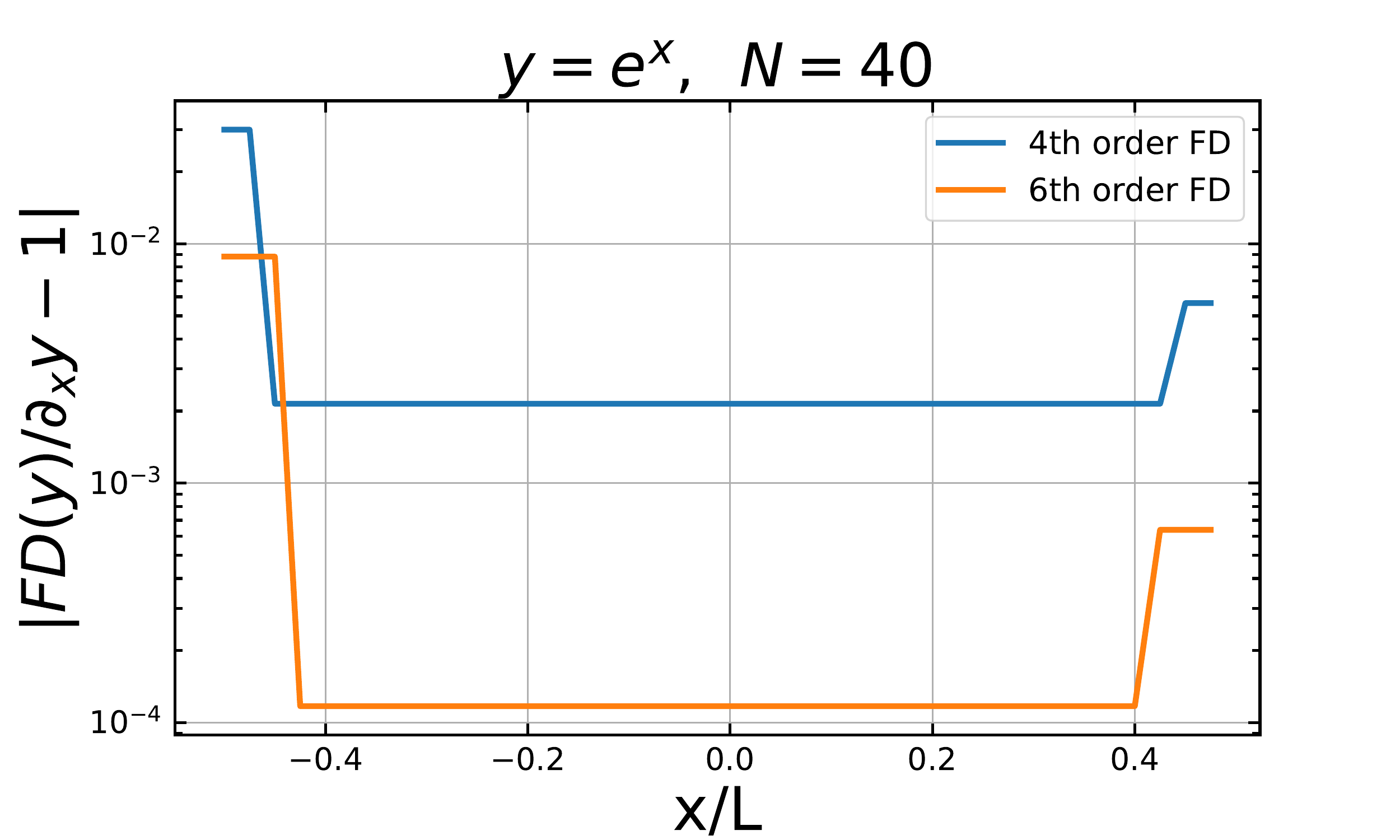}
    \caption{Relative error of the FD scheme applied to an exponential distribution, for $4^{th}$ and $6^{th}$ order schemes and applied to a data array of 40 points.}
    \label{fig: FDscheme}
\end{figure}

The same could be said of the exponential, however here the boundary points need to be considered. 
Unless periodic boundary conditions can be applied, a combination of forward and backward FD schemes were used. 
Fig.~(\ref{fig: FDscheme}) shows that the relative error of these different schemes can significantly differ for an exponential distribution. 
Indeed the left-most points (calculated with a forward scheme) and right-most points (backward scheme) have larger errors than the central ones computed with the centred scheme. 
The forward and backward schemes are of the same convergence order as the centred scheme but start with a higher relative error. 
In the right panel of Fig.~(\ref{fig: FDtest}) we then consider the case where these edge points are included, full lines, and when they are cut off, dash-dotted lines. 
In the first case, the order of convergence is higher at low resolution and tends towards the expected value at high resolution. 
This is because, as resolution is increased, more points are computed with the centred scheme, whereas the same number of points are computed with the forward and backward scheme throughout.
Then, excluding the boundary points, reduces the initial relative error and allows it to explicitly follow the expected convergence. 
For the examples considered above, the edge points have then been excluded.

\section{Analytic expressions} \label{sec: analyticsolution}

The metric $g_{\alpha\beta}$, the extrinsic curvature $K_{\alpha\beta}$ and the energy-momentum tensor $T_{\alpha\beta}$ are passed to our codes in order to retrieve ${}^{(4)}R$, ${}^{(3)}R$, $E^2$, and $B^2$. The analytic expressions of $g_{\alpha\beta}$ have been presented in Section~\ref{sec: Example spacetimes}, then $K_{\alpha\beta}$ can easily be obtained from it's time derivative $K_{\alpha\beta}=-\frac{1}{2}\frac{\partial}{\partial t}({g}_{\alpha\beta})$ in the synchronous gauge, and we will present below the analytical expressions of $T_{\alpha\beta}$ for the different metrics. 
For the $\Lambda$-Szekeres, Bianchi II Collins-Stewart, and Bianchi IV Harvey and Tsoubelis metrics, the coordinates are comoving with the fluid, so the fluid velocity is $u^\alpha = \{1,\;0,\;0,\;0\}$, and because this matches the normal to the hypersurfaces, we will drop the frame specific notation $\rho^{\{n\}}=\rho^{\{u\}}=\rho$. 
For the test-metric, $T_{\alpha\beta}$ is build from Einstein's equations;  here, for simplicity, we choose to express this $T_{\alpha\beta}$ in the normal frame  $n^\alpha$.
Finally, for the Bianchi VI Rosquist and Jantzen metric, although it describes a tilted perfect fluid, the different terms of $T_{\alpha\beta}$ are again expressed in the frame $n^\alpha$: then,  in this frame the fluid $T_{\alpha\beta}$  no longer appears as that of a perfect fluid.

Then with Maple \cite{Maple}, we have derived the expressions of ${}^{(4)}R$, ${}^{(3)}R$, $E^2$, and $B^2$ listed below. Here, an overhead dot, e.g.\ $\dot{v}$ means time derivative of said variable $v$.

\subsection{The $\Lambda$-Szekeres models of Barrow and Stein-Schabes}\label{sec: analyticsolution_szekeres}

To help compute time derivatives of the metric in Section~\ref{sec: Ex Szekeres} we provide here the Hubble scalar
\begin{equation}
    H = \frac{\dot{a}}{a} = H_0\sqrt{\Omega_{m0}a^{-3}+\Omega_{\Lambda 0}}.
\end{equation}
The hypergeometric function in Eq.~(\ref{eq: Szekeres_F}) is the result of the following integral
\begin{equation}
    \int \frac{\sinh{(\tau)}}{\cosh{(\tau)}} d\tau = \frac{3}{5}\cosh{(\tau)} \frac{\sinh{(\tau)}^{5/3}}{\cosh{(\tau)}} {}_{2}F_{1}\left(\frac{5}{6},\; \frac{3}{2};\; \frac{11}{6};\; -\sinh{(\tau)}^2 \right).
\end{equation}

The $\Lambda$-Szekeres metric is a solution of Einstein's equations with dust and $\Lambda$: the energy-momentum tensor takes the form $T_{\alpha\beta} = \rho u_\alpha u_\beta$, with the energy density provided by Eq.~(\ref{eq: SzekeresBackground}) and Eq.~(\ref{eq:szekdelta}). Then the Maple \cite{Maple} results for this spacetime are:

\begin{equation}
\begin{aligned}
    {}^{(3)}R &= -\frac{4v_1}{a^2Z}, \\
    {}^{(4)}R &= {}^{(3)}R +
6 \left(H^2+\frac{ \ddot{a}}{a}\right)
+\frac{2}{Z}\left(4 H \dot{Z}+\ddot{Z}\right), \\
    E^2 &= \frac{(v_1 + v_2)^{2}}{6 a^{4} Z^{2}},\\
     B^2 &= 0,
\end{aligned}
\end{equation}
with $v_1 = \partial_x\partial_x Z = \partial_y\partial_y Z$, and $v_2 = a\dot{a}\dot{Z}+a^2\ddot{Z}$, and we use the property $\partial_x\partial_y Z = \partial_y\partial_x Z = 0$.
In particular, here the magnetic part of the Weyl tensor is null, $B_{\alpha\beta}=0$, because $\partial_x \dot{Z}=\partial_y \dot{Z}=0$.

Then we also have the invariants $I = E^2/2$ and $J = (E^2/6)^{3/2}$, confirming that $I^3 = 3^3 J^2$: therefore this is a spacetime of special type. 
With the null vector base obtained by following the methodology of Section~\ref{sec: invariants}, we find the Weyl scalars $\Psi_\alpha=\{3\Psi_2,\;0,\;\Psi_2,\;0,\;3\Psi_2\}$,
with $\Psi_{2}^2 = E^2/24$. 
These are not built with the principal null direction and, given their expressions, one can see that they can be further simplified.
Indeed, the complex scalars are $K=N=0$ and $L = 3 \Psi_{2}^2$, confirming that we find this spacetime to be of Petrov type D, and we know that in this case only $\Psi_2$ is non zero, if built with the principal null directions.
We demonstrate this through two tetrad rotations, first to make $\Psi_0$ vanish, such that the new scalars are $\tilde{\Psi}_\alpha=\{0,\;0,\;-2\Psi_2,\;3i\Psi_2,\;3\Psi_2\}$, then to make the new $\tilde{\Psi}_4$ vanish, hence the final scalars are $\hat{\Psi}_\alpha=\{0,\;0,\;-2\Psi_2,\;0,\;0\}$. 
This indeed leaves only $\hat{\Psi}_2\neq0$, and it has gained a factor of $-2$ through these rotations. 
This is the same result as Eq.~(42) of \cite{N.Meures_M.Bruni_2011_Jun}.

As expected we have identified the spacetime to be of type D; however, should $E^2=0$, which is the case if $\beta_+=0$, then it would reduce to an FLRW metric, i.e.\ it would be of type O.

The Maple ``PetrovType()" function will identify this spacetime as type I unless it is also provided with the following definition $Z(x, y, z, t)=1+\beta_+(z)\mathcal{F}(t)+\mathcal{A}\beta_+(z)(x^2+y^2)$ (without needing to define $\beta$, $\mathcal{F}$ or $\mathcal{A}$), it then finds this spacetime to be of type D. For the other spacetimes, the classification we make by computing the invariants corroborates the classifications made by this function, where we only need to provide the metric as information.

\subsection{A non-diagonal inhomogeneous test metric}\label{sec: analyticsolution_testmetric}

With $\gamma=At(-2 + A^2t^2)$ the determinant of the spatial metric of Section~\ref{sec: Ex test metric}, we find:
\begin{equation}
\begin{aligned}
    {}^{(3)}R &= \frac{At^2}{2 \gamma}\Big(t (\partial_z A)^{2} (2+3 A^{2} t^{2}) - 4 \gamma \partial_z\partial_z A\Big), \\
    {}^{(4)}R &= \frac{A}{2 \gamma^{2}}\Big((2+3 A^{2}t^{2})(- 2 A + t^{3}(\partial_z A)^{2}) - 4t^{2}\gamma \partial_z\partial_z A\Big),\\
    E^{2\{n\}} &=\frac{1}{96 \gamma^{4}}\Big(2 A^4 c_{2}^2 \left(3 A^2 t^2+2\right)+ c_{2}^2 t^3 (\partial_z A)^{2}\left(8 A^3+ t (\partial_z A)^{2}\left(A^2 t^2+3\right)\right) \\
    &+4 \gamma t^2 \partial_z\partial_z A\left(4 A^3 c_{2}+\left(A^2 t^2+3\right) \left(c_{2} t(\partial_z A)^{2} +\gamma  \partial_z\partial_z A\right)\right)\Big) \\
    B^{2\{n\}} &= \frac{5 A^3 t^3}{32 \gamma^4}(\partial_z A)^{2} (2 + A^{2} t^{2})^{2},
\end{aligned}
\end{equation}
where we simplify these expressions with the following substitution $c_2 = 2- 3 A^{2} t^{2}$. 

Then using Einstein's equations we find $T_{\alpha\beta}$ for a non perfect fluid \cite{G.F.R.Ellis_etal_2012}:
\begin{equation}\label{eq:nonperfectfluid}
    T_{\alpha\beta} = \rho^{\{n\}} n_\alpha n_\beta + p^{\{n\}} \gamma_{\alpha\beta} + 2q_{(\alpha}^{\{n\}} n_{\beta)} + \pi_{\alpha\beta}^{\{n\}},
\end{equation}
with the energy density $\rho^{\{n\}}$, pressure $p^{\{n\}}$, energy flux $q_{\alpha}^{\{n\}}$, and anisotropic pressure $\pi_{\alpha\beta}^{\{n\}}$ all expressed in the normal frame $n^\alpha$. They can be identified following their definitions:
\begin{equation}
\begin{aligned}
    \rho^{\{n\}} &=T^{\alpha\beta}n_{\alpha}n_{\beta}= \frac{At}{4 \kappa \gamma^{2}}\Big(3 A^{2}\gamma + t^{2}(\partial_z A)^{2}(2+3 A^{2} t^{2}) -4 t\gamma \partial_z\partial_z A\Big), \\
    p^{\{n\}} &= \gamma_{\alpha\beta}T^{\alpha\beta}/3= \frac{A}{12\kappa \gamma^{2}}\Big(4A + 3 A^{5} t^{4} + (2+3 A^{2} t^{2})(2A - t^{3}(\partial_z A)^{2}) + 4t^{2}\gamma \partial_z\partial_z A\Big), \\
    q_{\alpha}^{\{n\}} &= -\gamma_{\alpha}^{\beta} T_{\beta\mu} n^{\mu}= \frac{At \partial_z A}{4\kappa \gamma^{2}}\Big(0, \;\; At(-6+A^{2} t^{2}), \;\; 2+A^{2} t^{2}, \;\; 2-7 A^{2} t^{2}\Big), \\
    \pi_{\alpha\beta}^{\{n\}} &= \gamma_{\alpha \mu} \gamma_{\beta \nu} T^{\mu \nu}
- \frac{1}{3}\gamma_{\alpha \beta} \gamma_{\mu \nu} T^{\mu \nu} = \frac{1}{12\kappa\gamma^2}\begin{pmatrix}
    0 & 0 & 0 & 0 \\
    0 & \pi_{xx} & \pi_{xy} & \pi_{xz} \\
    0 & \pi_{xy} & \pi_{yy} & \pi_{yz} \\
    0 & \pi_{xz} & \pi_{yz} & \pi_{zz} \\
    \end{pmatrix}, \\
\end{aligned}
\end{equation}
with the factorised components of the anisotropic pressure expressed as below, with $c_1 = t^{2}(\partial_z A)^{2}(2-3 A^2 t^2) + 2 t\gamma \partial_z\partial_z A$:
\begin{equation}
\begin{aligned}
    \pi_{xx} &=  16 A^{3}t + (3+A^2 t^2)c_1\\
    \pi_{xy} &= A^2 (4 + 3 A^{4} t^{4}) + 4Atc_1\\
    \pi_{xz} &= A^2 (4 + 3 A^{4} t^{4}) + Atc_1\\
    \pi_{yy} &= 2A^{3}t(2+3 A^{2} t^{2}) + (3+A^2 t^2) c_1\\
    \pi_{yz} &= 6At(2 - A^{4} t^{2}) + 3c_1\\
    \pi_{zz} &= 2A^{3}t(2+3 A^{2} t^{2})+ (3-2 A^2 t^2) c_1.
\end{aligned}
\end{equation}

Then, in computing the invariants to determine the Petrov type (too long to be included here),
we find that $I$ and $J$ do not satisfy the requirements for this spacetime to be special, so it is of Petrov type I. Note that, should $A$ be a constant along $z$, then ${}^{(3)}R=B^{2\{n\}}=0$, and
\begin{equation}
    I^3-27J^2 = \frac{A^{12} c_{2}^6}{2^{15} \gamma^{10}}.
\end{equation}
Then, at the point in time where $c_2=0$ this spacetime would be of type O, with $E^{2\{n\}}=0$.

\subsection{Bianchi II Collins-Stewart} \label{sec: analyticsolution_BianchiII}

Here $\gamma$ is the $\gamma$-law index of the perfect fluid so the energy-stress tensor takes the form:
\begin{equation}
    T_{\alpha\beta} = \rho( (\gamma - 1) g_{\alpha\beta} + \gamma u_\alpha u_\beta)
\end{equation}
with $\rho$ given in Section~\ref{sec: Ex Bianchi II}. Then we obtain the following expressions:
\begin{equation}
\begin{aligned}
    {}^{(3)}R &= -\frac{s^2}{8 \gamma ^2 t^2} \\
    {}^{(4)}R &= \frac{3 \gamma ^2-36 \gamma +44}{8 \gamma ^2 t^2} + {}^{(3)}R \\
    E^2 &= \frac{(3 \gamma -2)^2 (5 \gamma -6)^2}{384 \gamma ^4 t^4} \\
    B^2 &= \frac{-3 (\gamma -2) (3 \gamma -2)^3}{128 \gamma ^4 t^4}. \\
\end{aligned}
\end{equation}
The only non-zero Weyl scalar is 
\begin{equation}
    \Psi_2 = \frac{(3\gamma - 2)(6-5\gamma + 3 i s)}{48 \gamma^2 t^2},
\end{equation}
directly obtained with the scheme in Section~\ref{sec: invariants}, without further frame rotations. The invariants are then $I=3\Psi_{2}^2$, $J=-\Psi_{2}^3$ and $K=L=N=0$, therefore this spacetime is of Petrov type D for both dust and radiation.

\subsection{Bianchi VI tilted model} \label{sec: analyticsolution_BianchiVI}

The spacetime from Section~\ref{sec: Ex Bianchi VI} has the following metric determinant $g=-k^{2}t^{2(1+2s)}$, and we find the following expressions:
\begin{equation}
\begin{aligned}
    {}^{(3)}R &= \frac{-2}{k^2 t^2}, \\
    {}^{(4)}R &= {}^{(3)}R + \frac{1}{2 t^{2}}\Big(m^{2} c_{1}^{2} + 4 q^{2} + 12 s^{2}\Big),\\
    E^{2\{n\}} &= \frac{1}{24 k^{4} t^{4}}\Big[
    16 
    + 16 k^{2} q^{2}\Big(-2 + k^{2} (q^{2} + 3c_{2}^2)\Big)\\
    &+ k^{2} m^{2} c_{1}^{2} \Big( 16 +
    4 k^{2} m^{2} c_{1}^{2} 
    + k^{2} (11 q^{2} - 18 q c_{2} + 3c_{2}^2) 
    \Big)\Big],\\
     B^{2\{n\}} &= \frac{9 m^{2} c_{1}^{2} + 16c_{2}^2}{8 k^{2}t^{4}},
\end{aligned}
\end{equation}
where we use the parameter substitutions $c_{1} = (q - s + 1)$ and $c_{2} = (s - 1)$.

Although this spacetime follows a $\gamma$-law perfect fluid in a tilted frame, we work with the normal frame $n^\alpha = \{1,\;0,\;0,\;0\}$, meaning that the stress-energy tensor takes the non-perfect fluid form, Eq.~(\ref{eq:nonperfectfluid}), with the following quantities:
\begin{equation}
\begin{aligned}
    \rho^{\{n\}} &= \frac{-1}{4 \kappa k^{2} t^{2}}\Big(4+ k^{2} (m^{2} c_{1}^{2} + 4 q^{2} - 4 s (s + 2))\Big),\\
    p^{\{n\}} &= \frac{1}{12 k^{2} \kappa t^{2}}\Big(4- k^{2} (3 m^{2} c_{1}^{2} + 12 q^{2} + 4 s (5 s - 2))\Big), \\
    q_{\alpha}^{\{n\}} &= \frac{-1}{2 \kappa t}\Big(0, \;\; m^{2} c_{1} + 4 q, \;\; m c_{1} e^{x} / k t^{c_{1}}, \;\; 0\Big), \\
    \pi_{\alpha\beta}^{\{n\}} &= \frac{1}{6\kappa}\begin{pmatrix}
    0 & 0 & 0 & 0 \\
    0 & \pi_{xx} & \pi_{xy} & 0 \\
    0 & \pi_{xy} & \pi_{yy} & 0 \\
    0 & 0 & 0 & \pi_{zz} \\
    \end{pmatrix}, \\
\end{aligned}
\end{equation}
with the factorised components to the anisotropic pressure:
\begin{equation}
\begin{aligned}
    \pi_{xx} &= -8 - 8 c_{2} k^{2} s + m^{2} (4 - k^{2} (- 6 c_{2} q + 3 c_{1}^{2} m^{2}  + 3 q^{2} + 11 s^{2} - 14 s + 3)),\\
    \pi_{xy} &= \frac{m e^{x}}{k t^{c_{1}}}(4 - k^{2} (3 c_{1}^{2} m^{2} + 3 q^{2} + 6 q + 5 s^{2} - 8 s + 3)), \\
    \pi_{yy} &= \frac{e^{2 x}}{k^{2}t^{2 c_{1}}}(4 - k^{2} (3 c_{1}^{2} m^{2} + s (4 c_{1} + 8 q))),\\
    \pi_{zz} &= \frac{4t^{2 (q + s - 1)} }{k^{2}e^{2 x}}(1 + k^{2} s (3 q + s - 1) ).
\end{aligned}
\end{equation}

Computing the invariants, we find the spacetime to be of Petrov type I, in particular for $\gamma=1.22$, $I^3 - 27 J ^2 \simeq 61.05 t^{12}$.

\subsection{Bianchi IV vacuum plane wave} \label{sec: analyticsolution_PlaneWave}

As described by the title of this spacetime from Section~\ref{sec: Ex Bianchi IV plane wave}, this  is a vacuum solution, so $T_{\alpha\beta} = {}^{(4)}R =  0$, then we find ${}^{(3)}R = -2/t^2$, and $E^2 = B^2 = 1/2t^4$ and the complex scalars are all null: $I=J=K=L=N=0$. 
Therefore this spacetime is of type N, as can be established by the Weyl scalars in the null principal direction $\Psi_\alpha=\{0,\;0,\;0,\;0,\;t^{-2}\}$.

\end{document}